\documentclass[a4paper,11pt]{article}
\pdfoutput=1 

\usepackage{jheppub} 

\usepackage{youngtab}
\usepackage{hyperref}
\usepackage{fancybox}
\usepackage{float}
\usepackage{amsfonts}
\usepackage{amsmath}
\usepackage{amsbsy}
\usepackage{graphicx}
\usepackage{amssymb}
\usepackage{amsthm}
\usepackage{bm}
\usepackage{comment}
\usepackage{epsfig}
\usepackage{latexsym}
\usepackage{pdflscape}

\usepackage{color}

\makeatletter
\def\@fpheader{\relax}
\makeatother

\DeclareSymbolFont{AMSa}{U}{msa}{m}{n}
\DeclareSymbolFont{AMSb}{U}{msb}{m}{n}
\DeclareMathSymbol{\fieldR}{\mathalpha}{AMSb}{"52}

\newcommand{\beq}{\begin{eqnarray}}
\newcommand{\eeq}{\end{eqnarray}}
\newcommand{\bea}{\begin{eqnarray}}
\newcommand{\eea}{\end{eqnarray}}
\newcommand{\be}{\begin{equation}}
\newcommand{\ee}{\end{equation}}
\newcommand{\bq}{\begin{equation}}
\newcommand{\eq}{\end{equation}}

\newcommand{\nn}{\nonumber}

\def\eg{{\it e.g. }}
\def\ie{{\it i.e. }}

\def\6{\partial}

\def\6{\partial}

\begin{document} 

\title {Boson stars and solitons confined in a Minkowski box}


\author[a]{Oscar J. C. Dias,}
\affiliation[a]{STAG research centre and Mathematical Sciences, University of Southampton, UK}

\author[a]{Ramon Masachs,}

\author[a]{and Paul Rodgers}

\emailAdd{ojcd1r13@soton.ac.uk}
\emailAdd{ramonmasachs@gmail.com}
\emailAdd{pwr1u17@soton.ac.uk}

\begin{abstract}
{We consider the static charged black hole bomb system, originally designed for a (uncharged) rotating superradiant system by Press and Teukolsky. A charged scalar field confined in a Minkowski cavity with a Maxwell gauge field has a quantized spectrum of normal modes that can fit inside the box. Back-reacting non-linearly these normal modes, we find the hairy solitons, a.k.a boson stars (depending on the chosen $U(1)$ gauge), of the theory. The scalar condensate is totally confined inside the box and, outside it, we have the Reissner-Nordstr\"om solution. The Israel junction conditions at the box surface layer determine the stress tensor that the box must have to confine the scalar hair. Some of these horizonless hairy solutions exist for any value of the scalar field charge and not only above the natural critical charges of the theory (namely, the critical charges for the onset of the near-horizon and superradiant instabilities of the Reissner-Nordstr\"om black hole).
However, the ground state solutions have a non-trivial intricate phase diagram with a main and a secondary family of solitons (some with a Chandrasekhar mass limit but others without) and there are a third and a fourth critical scalar field charges where the soliton spectra changes radically. Most of these intricate properties are not captured by a higher order perturbative analysis of the problem where we simply back-react a normal mode of the system. }
\end{abstract}

\maketitle
\flushbottom

\section{Introduction}\label{sec:intro}

Confining gravitational boxes became notorious in the context of superradiant instabilities when Press and Teukolsky introduced the black hole bomb system \cite{Press:1972zz} (see also \cite{Cardoso:2004nk}). A scalar wave impinging on a rotating black hole may extract energy from the black hole provided its frequency $\omega$ satisfies the superradiant bound $\omega < m_{\varphi} \Omega_{H}$, where $m_{\varphi}$ is the azimuthal quantum number and $\Omega_{H}$ is the angular velocity of the horizon. 
In standard conditions this scalar wave would then disperse to the asymptotic region and die-off. 
However, when surrounded by a reflecting cavity, the scalar wave undergoes multiple superradiant amplifications and reflections and an instability builds up. 
Similar superradiant instabilities occur for  charged scalar fields confined around Reissner-Nordstr\"om black holes (RN BHs), whereby the superradiant frequency bound now reads $\omega < e \mu$, where $e$ is the charge of the scalar field and $\mu$ is the chemical potential of the black hole \cite{Denardo:1973pyo}. 
In both black hole bomb systems, the onset or zero-mode of the superradiant instability signals a bifurcation to a novel family of hairy black holes: the solution outside the box   
is described by the Kerr or RN solution but, inside the box, there is also a non-trivial  scalar field floating above the horizon. The scalar field cloud is stationary because either centrifugal effects or Coulomb repulsion balance the system against gravitational collapse.  
 
Consider now this very same gravitational box confining a charged scalar field but, this time, it is simply placed in Minkowski background (eventually also with a radial electric field) with no event horizon. As is well known, at linear order in perturbation theory, the scalar field frequencies that can fit inside the box radius are quantized; these are the normal modes of the system. Interestingly, beyond linear order in perturbation theory, there is nothing impeding us from back-reacting a normal mode to higher orders while keeping the solution regular everywhere and still confined as a stationary configuration inside the box \cite{Dias:2018yey}. In this case we have an asymptotically Minkowski soliton or boson star confined in a box.  

In \cite{Dias:2018yey},  we have constructed some of the  above static charged hairy solitons (boson stars) and hairy black holes in perturbation theory (with the amplitude of the scalar field and the ratio between the horizon and box radius as expansion parameters). By construction, such solutions are valid only for small energy and charge and they are perturbatively connected to the Minkowski box solution. In the present manuscript we complete the analysis initiated in \cite{Dias:2018yey} and solve the full nonlinear Einstein-Maxwell-scalar field equations to find the exact numerical solution, that describes the solitons (boson stars) of the theory confined in a box, also in the non-perturbative regime. At low energies/charges our solutions are well described by the perturbative predictions of \cite{Dias:2018yey}. However, at intermediate and large energies (when compared to the box lengthscale) the phase diagram of solutions develops an intricate structure that was not anticipated at all by the perturbative analysis. In particular, we will find a {\it main branch} of boson stars (solitons) that, in the small charge regime, is described within perturbation theory but that has (at higher charges) a Chandrashekhar mass limit and multi-branched structure that was not captured within perturbation theory. Additionally, we also find a {\it secondary family} of boson stars that is not captured at all by perturbation theory.  Finally, we will find that the properties of boson stars have a non-trivial dependence on the electric scalar field charge $e$, most of which were not anticipated by the perturbative analysis. 

Some key properties of asymptotically flat caged solitons turn out to be similar to those observed in asymptotically anti-de Sitter solitons \cite{Gentle:2011kv}, \cite{Basu:2010uz,Bhattacharyya:2010yg,Dias:2011tj,Arias:2016aig,Markeviciute:2016ivy,Markeviciute:2018cqs,Dias:2016pma}. We thus identify features that seem to, or might be  universal to charged scalar condensates (or other bosonic field condensates) confined in a potential well. 

\section{Summary of phase diagram}\label{sec:summary}

For clarity, in this section we summarize our main results since some of the plots that we will present in section~\ref{sec:phasediag} are elaborated. We find charged scalar boson stars (a.k.a. solitons) confined inside a gravitational box in an asymptotically Minkowski background. These are regular, static, horizon-free solutions to Einstein-Maxwell theory coupled to a scalar field that vanishes outside a box but not inside it (the latter has an Israel stress tensor that supports the pressure of the scalar field and keeps the solution stationary). 

As discussed in the perturbative analysis of \cite{Dias:2018yey}, the properties of charged hairy solutions on a box depend on the charge of the scalar field $e$. In particular, one identifies the following four critical scalar charges (two of them, $e_{\gamma}$ and $e_c$, are not captured within  perturbation theory  \cite{Dias:2018yey}):

\begin{description}

\item $\bullet$ $e=e_{\hbox{\tiny NH}}=\frac{1}{2 \sqrt{2}}\sim 0.354$. This is the charge above which scalar fields can trigger a violation of the near horizon $AdS_{2}$ Breitenl\"ohner-Freedman bound of the extremal RN black hole whose horizon radius approaches, from below, the box radius. This violation renders (near-)extremal RN BHs unstable. This is the so-called near horizon scalar condensation instability first studied in the context of AdS black holes and the holographic gravity/condensed matter correspondence programme \cite{Gubser:2008px,Hartnoll:2008vx,Hartnoll:2008kx,Dias:2010ma}. We ask the reader to see Section III.B of \cite{Dias:2018zjg} for a detailed analysis that gets this critical charge. As far as we could perceive, this charge does not play a relevant role in the discussion of the solitons of the theory.

\item   $\bullet$ $e=e_{\hbox{\tiny S}}=\frac{\pi }{\sqrt{2}}\sim 2.221$. This is the critical charge that saturates the superradiant bound $\omega=e\mu$ for an extremal RN BH (which has $\mu=\sqrt{2}$) when we take the frequency to be the lowest frequency, $\omega=\pi$, that can fit inside a box of (dimensionless) unit radius in Minkowski spacetime \cite{Press:1972zz,Denardo:1973pyo}. We ask the reader to see Section III.A of \cite{Dias:2018yey} for a more detailed analysis that leads to this critical charge. 

\item  $\bullet$ $e=e_\gamma$ and  $e=e_c$ with $e_{\hbox{\tiny NH}}<e_\gamma<e_c<e_{\hbox{\tiny S}}$. In the present manuscript we find that the system has a third and a fourth critical charges, that we find within numerical error to be $e_\gamma \sim 1.13$ and $e_c\simeq 1.8545\pm 0.0005$ (unlike the other two charges, we are not aware of a heuristic analysis that allows to capture analytically these two critical values without solving the full equations of motion). These charges are not captured by the perturbative analysis of \cite{Dias:2018yey}. We will find that our system has (at least) two distinct families of ground state solitons (there is then an infinite tower of excited soliton families that, in the perturbative regime, correspond to the backreaction of excited normal modes with higher radial overtones). One $-$ that we call the {\it main} soliton family $-$ can be seen as the backreaction of the  charged ground state normal mode of a scalar field in a Minkowski box \cite{Dias:2018yey} since it exist for small energies/charges up to a Chandrashekhar limit. On the other hand, the second family $-$ that we denote as the {\it secondary} soliton family $-$ exists only for intermediate or large energies/charges and thus it is not captured in the perturbative analysis of \cite{Dias:2018yey}.     
The secondary soliton family exists only for $e_\gamma<e<e_c$. In a phase diagram of soliton solutions, as we approach the critical charge $e_\gamma$ from above, the secondary soliton family ceases to exist because it no longer fits inside the box.
On the other hand, as we approach  the critical charge $e_c$ from below, the main and secondary families of solitons approach each other and they connect precisely at $e=e_c$.  

\end{description}



In the present manuscript, we solve numerically the full Einstein-Maxwell equations for a charged scalar field confined in a covariant box with dimensionless radius $R=1$. Therefore, our analysis is now fully nonlinear and not restricted to small energies/charges. We focus our attention on soliton solutions (and leave the study of hairy BHs for a later publication \cite{DaveyDiasRodgers:2021}) since the phase diagram of these solutions is already very intricate. 
We will recover the main soliton family of solutions of \cite{Dias:2018yey} for small energies/charges (which exists for any $e>0$), thereby confirming and setting the regime of validity of the perturbative analysis, but also solutions with intermediate and large energies with a Chandrasekhar mass limit and multi-branched structure not captured by \cite{Dias:2018yey}. In particular, we will find the  secondary family of solitons and reveal the existence of the critical charges $e_{\gamma}$ and $e_c$.\footnote{The perturbative analysis of  \cite{Dias:2018yey} also finds that hairy BHs that are perturbatively connected to a boxed Minkowski spacetime exist only for $e\geq e_{\hbox{\tiny S}}$; by construction, the zero horizon radius limit of such hairy BHs is one of our hairy solitons. Perturbative theory \cite{Dias:2018yey} does not capture the existence of hairy BHs with $e< e_{\hbox{\tiny S}}$. In a companion manuscript  \cite{DaveyDiasRodgers:2021}, some of the authors will show that a full nonlinear analysis finds that hairy BHs also exist in the range $e_{\hbox{\tiny NH}} \leq e<e_{\hbox{\tiny S}}$ but, typically, they are no longer necessarily connected to the solitons of the theory in their zero entropy limit.
}

By Birkhoff's theorem, outside the box (in particular, in the asymptotic region), our solutions are necessarily described by the RN solution \cite{WILTSHIRE198636,inverno:1992}.\footnote{Birkhoff's theorem for Einstein-Maxwell theory states that the unique spherically symmetric solution of the Einstein-Maxwell equations with non-constant area radius function $r$ (in the gauge \eqref{fieldansatz}) is the Reissner-Nordstr\"om solution. If $r$ is constant then the theorem does not apply since one has  the Bertotti-Robinson ($AdS_2\times S^2$) solution.} It follows that we cannot use the Arnowitt-Deser-Misner (ADM) mass $M$ and charge $Q$ (measured by a Gauss law at infinity \cite{Arnowitt:1962hi}) to distinguish the several solutions of the theory. Instead, we need to resort to the Brown-York quasilocal mass $\mathcal{M}$ and charge $\mathcal{Q}$ \cite{Brown:1992br}, measured at the box location, to display our solutions in a phase diagram of the theory. These quantities satisfy their own first law of thermodynamics that we use to (further) check our solutions. Dimensionless quasilocal mass and charge are given in units of the box radius $L$, $\mathcal{M}/L$ and  $\mathcal{Q}/L$, respectively. A natural reference in this quasilocal phase diagram is the extremal RN BH 1-parameter family with horizon inside the box. However, above a certain scalar charge $e$, the solitons have a mass/charge that can be very close to the ones of the extremal RN. For this reason we will find useful to plot  $\Delta\mathcal{M}/L$ {\it vs} $\mathcal{Q}/L$ where $\Delta\mathcal{M}=\mathcal{M}-\mathcal{M}\big|_{\rm ext\, RN}$ is the mass difference between the hairy solution and the extremal RN that has the {\it same} $\mathcal{Q}/L$. 
So, in this phase diagram the horizontal line with $\Delta\mathcal{M}=0$ identifies the extremal RN BH solution. Its dimensionless horizon $R_+=r_+/L$ can fit inside the box of radius $L$ if $R_+\leq 1$ (which corresponds to $\mathcal{Q}/L\leq 2^{-1/2}$) and non-extremal RN BHs exist above this line.
 But the horizon of non-extremal RN BHs fit inside the box ($R_+\leq 1$) only if these solutions are to the left of the red dashed line that will  be displayed in our plots. We will find that this line represents the maximal quasilocal charge that solutions that fit inside the box can have (with or without scalar condensate). 

\begin{figure}[th]
\centerline{
 \includegraphics[width=.505\textwidth]{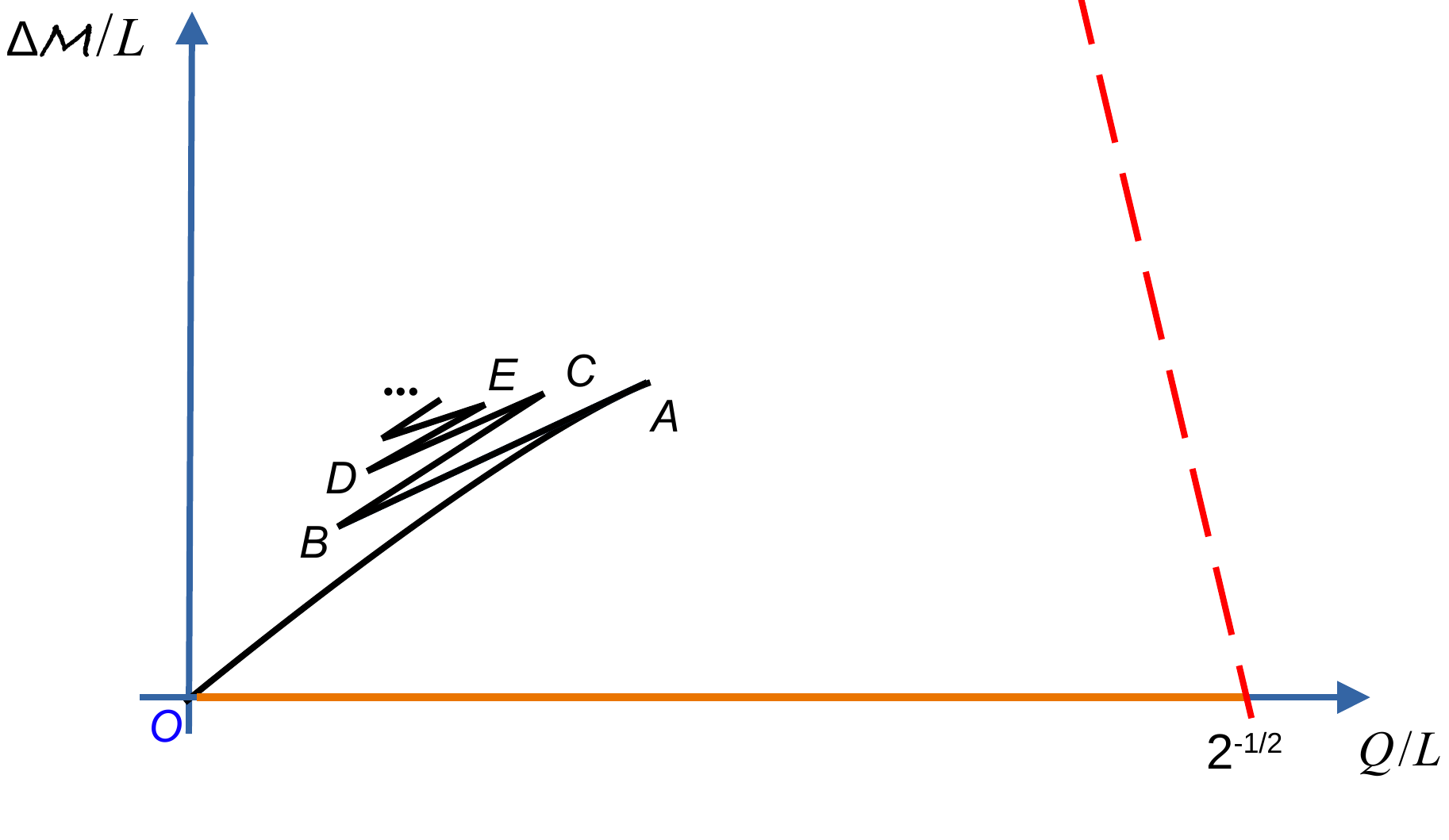}
\hspace{0.3cm}
\includegraphics[width=.50\textwidth]{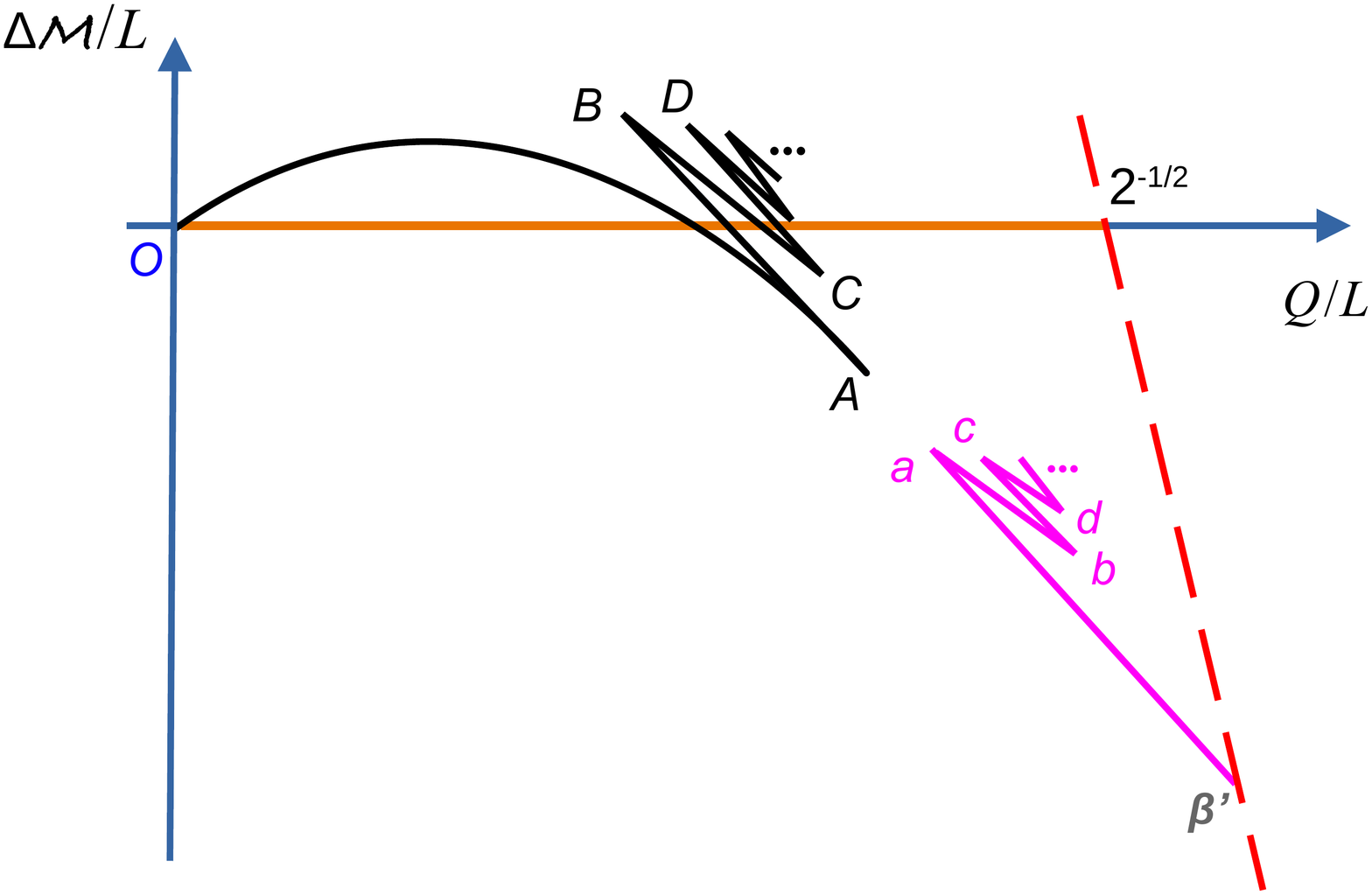}
}
\centerline{
\includegraphics[width=.505\textwidth]{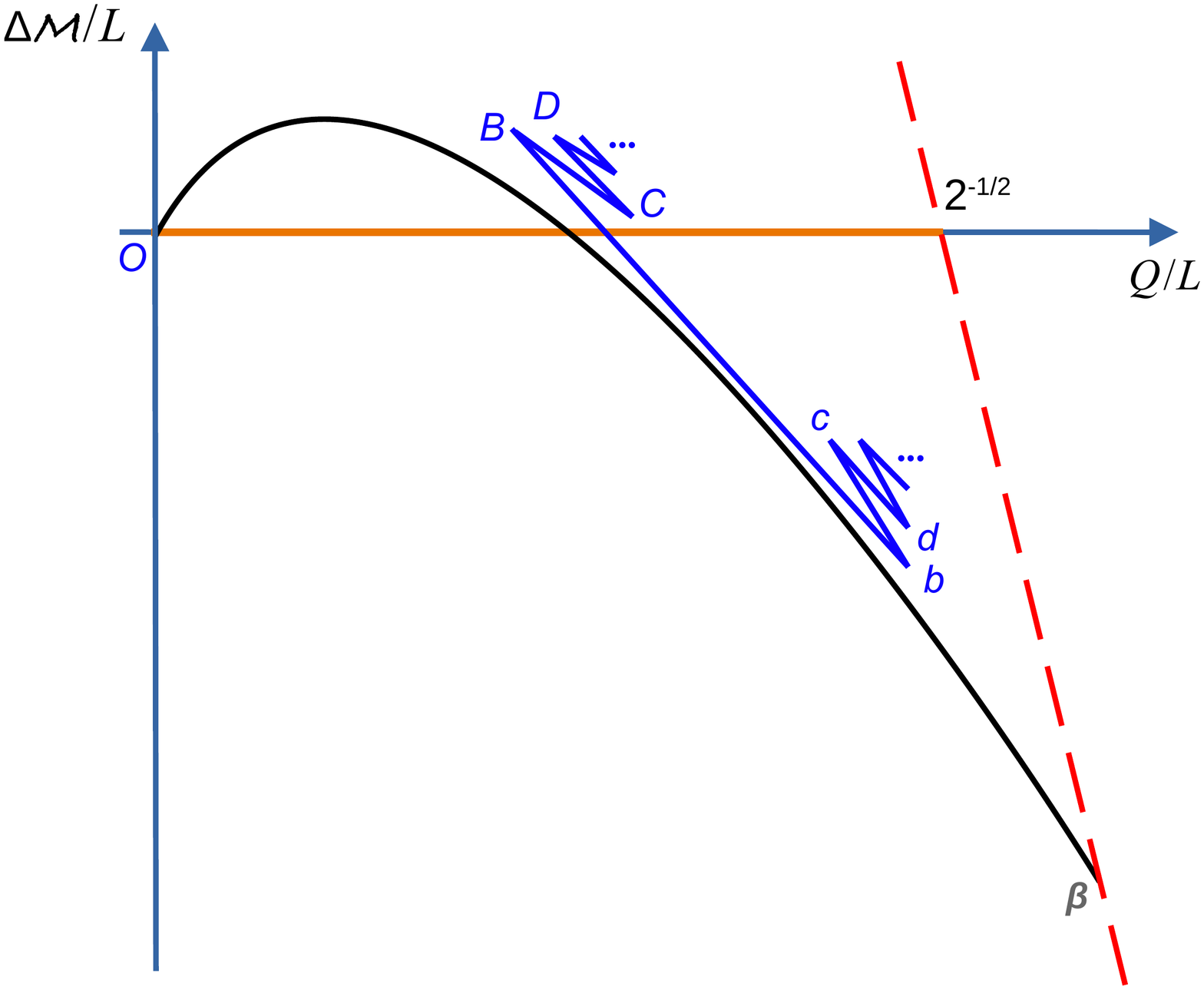}
\hspace{0.3cm}
\includegraphics[width=.505\textwidth]{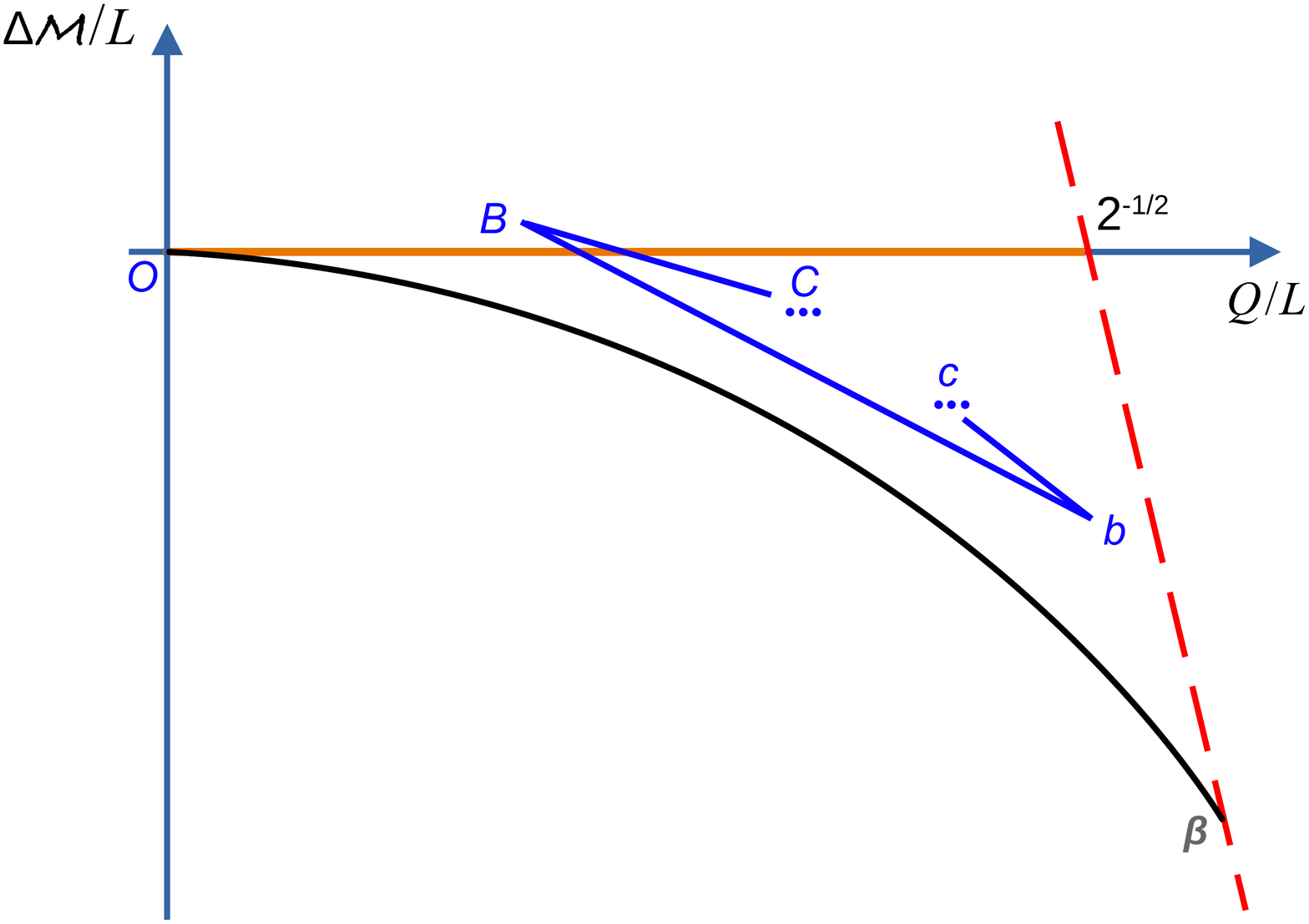} }
\caption{Sketch of the quasilocal phase diagram for solitons as we span relevant windows of scalar field charge $e$.  The critical charges are such that $0< e_{\hbox{\tiny NH}} < e_c < e_{\hbox{\tiny S}}$.  The quantity $\Delta{\cal M}$ is the quasilocal mass difference between the soliton and an extremal RN BH that has the $\textit{same}$ quasilocal charge $\mathcal{Q}/L$. Hence the orange line at $\Delta{\cal M} = 0$ describes the extremal RN solution that must have $\mathcal{Q}/L\leq 2^{-1/2}$ to fit inside the box.
The red dashed line represents the maximal quasilocal charge of solutions that can fit inside the box. It intersects the extremal RN line at  $\mathcal{Q}/L=2^{-1/2}$. Non-extremal RN BHs confined in the box have $\Delta\mathcal{M}>0$ and fill the  triangular region bounded by $\mathcal{Q}=0$ and by the orange and red dashed lines. The main soliton family is always given by black curves that start at $O$. The secondary soliton family is given either by magenta or blue curves. {\bf Top-left panel:} case $e<e_{\gamma}$.  {\bf Top-right panel:} case $e_{\gamma}<e<e_c$. {\bf Bottom-left panel:} case $e_c<e<e_{\hbox{\tiny S}}.$ {\bf Bottom-right panel:} case $e > e_{\hbox{\tiny S}}.$}
\label{FIG:Summary_sketch}
\end{figure}

Using this phase diagram, a summary of our main findings is (see sketch in Fig.~\ref{FIG:Summary_sketch}):
\begin{enumerate}

\item $e<e_\gamma\sim 1.13$. Later, we will give data for the case $e=0.23$ (section~\ref{subsec:lowerGamma}; Figs.~\ref{FIGe0.23:MassCharge}-\ref{FIGe0.23:phiK}).
Solutions with $e<e_\gamma$ are qualitatively similar and  blind to the critical charge $e_{\hbox{\tiny NH}}$. Here, we {\it sketch} a phase diagram that highlights the main properties of these solutions in the top-left panel of Fig.~\ref{FIG:Summary_sketch}. We simply have the `{\it main soliton family}' (or `{\it perturbative soliton family}' represented by the black line $OABCDE\cdots$). Nearby $O$, the properties of the branch $OA$ for small charge were already captured by the perturbative analysis of \cite{Dias:2018yey} but this soliton then develops an intricate series of cusps $A, B, C, D, E, \cdots$ that could not be anticipated by the analysis of \cite{Dias:2018yey}. In particular, this main soliton family has a Chandrasekhar limit at $A$. As we move from the main branch $OA$, along the sequence of secondary zig-zagged branches $AB$, $BC$, $CD$, ..., we find that the Kretschmann curvature invariant at the origin of the soliton is growing without bound. For this reason, it becomes increasingly hard to follow this solution beyond a certain point (say, point $E$) but we gathered enough evidence to predict that this solution might well develop an infinite number of cusps as it approaches a singular limit where the curvature invariants blow up. At a physical level, nothing wrong happens at the cusps. Actually, these cusps are simply smooth turning points if we plot $\mathcal{M}/L$, $\mathcal{Q}/L$, or  the values of the gravitational field $f\equiv g_{tt}$, electric potential $A_t$ or scalar field $\phi$ at the origin as a function of the derivative $\epsilon\equiv \phi'(R=1)$ of the scalar field at the box radius (see Appendix). Depending on which of these functions we look at we can have a curve with damped oscillations or a spiral curve with a series of turning points $A, B, C, D, E, \cdots$ (see e.g. Figs.~\ref{FIGe0.23:fA}-\ref{FIGe0.23:phiK} for $e=0.23$). 

As $e$ increases from $e=0$ to $e=e_\gamma\sim 1.13$, the qualitative features of the solutions do not change significantly. These main solitons always have more quasilocal mass than the extremal RN BH with the same quasilocal charge and their slope in the $\mathcal{Q}$-$\Delta\mathcal{M}$ is positive but decreases as $e$ grows.  This discussion is best illustrated in Fig.~\ref{FIGseveral:MassCharge} where we plot the main soliton family for different values of the electric scalar field charge $e$. In particular, we give three cases, $e=0.23, 0.5$ and $e=1$ that have $e<e_\gamma$.

\item $e_\gamma<e<e_c\simeq 1.8545\pm 0.0005$. Later, we will give data for the case  $e=1.854$ (section~\ref{subsec:GammaC}; Figs.~\ref{FIGe1.854:MassCharge}-\ref{FIGe1.854:phiK}). Here, we summarize the analysis with a {\it sketch} of the phase diagram in the top-right panel of Fig.~\ref{FIG:Summary_sketch}, that emphasizes the main properties of these solutions. As for $e<e_\gamma$, we have the `{\it main soliton family}' (or `{\it perturbative soliton family}' represented by the black line $OABCDE\cdots$). But, unlike for $e<e_\gamma$, after a gap $Aa$ in $\mathcal{Q}/L$, we now also have a `{\it secondary soliton family}' $\beta'abcd\cdots$ (that we might also call the `{\it non-perturbative soliton family}'). 

The main soliton family $OABCDE\cdots$ for $e_\gamma<e<e_c$ has similar properties to the ones already found in the $e<e_\gamma$ case. The only minor difference is that, as $e$ grows well above $e_\gamma$, at a certain point the main soliton family bends downwards as $\mathcal{Q}/L$ grows. And at a certain critical charge, $e\sim 0.5$, we can have portions of the main soliton in the neighbourhood  of its Chandrasekhar point $A$ with less quasilocal mass than the extremal RN BH with same $\mathcal{Q}/L$: see the sequence of solitons in  Fig.~\ref{FIGseveral:MassCharge} for different values of $e$ in this range.

The  top-right panel of Fig.~\ref{FIG:Summary_sketch} also sketches the secondary family $\beta'abcd\cdots$ of solitons  (magenta line). This family only exists above a critical charge $\mathcal{Q}/L$ (point $a$) that depends on $e$ and extends all the way up to the maximal charge that can fit inside the box (represented by the point $\beta'$ on the red dashed line). Thus, this family is not captured by the perturbative analysis of \cite{Dias:2018yey}. As we move from $\beta'$ to cusp $a$ and continue through the cusps $b$, $c$, $d, \cdots$ one finds that the Kretschmann curvature at the origin of the soliton is growing without bound and we find evidence that the solution will terminate on a singular solution where the curvature diverges, probably after going through an infinite number of cusps.

For scalar charge $e=e_{\gamma}$, point $a$ coincides with point $\beta'$ (see  top-right panel of Fig.~\ref{FIG:Summary_sketch}).
Just above $e_{\gamma}$, point $a$ is very close to $\beta'$ and the gap $Aa$ between the main and secondary solitons is very large. The secondary soliton that we found with smallest $e$ has $e=1.144$. We take this and the findings summarized in the right plot of Fig.~\ref{FIGseveral:MassCharge} to extrapolate that $e_{\gamma}\sim 1.13$. For our purposes it is not necessary to determine $e_{\gamma}$ with higher accuracy. As $e$ keeps increasing away from $e_{\gamma}$, the gap $Aa$ decreases very quickly and, quite importantly, it goes to zero precisely at $e=e_c$ where the main and secondary solitons merge.
This evolution of the $\mathcal{Q}$-$\Delta\mathcal{M}$ phase diagram with $e$ is best illustrated in  the right panel of Fig.~\ref{FIGseveral:MassCharge} where we display the main and secondary family of solitons for different values of $e$ in the window $e_\gamma<e<e_c$. Note that, consistent with the description given above, secondary solitons exist only in the $\mathcal{Q}$-$\Delta\mathcal{M}$ region bounded by the closed curve $a_c\beta_c\gamma$ (see auxiliary dashed gray curve of  the right panel of Fig.~\ref{FIGseveral:MassCharge}). Just above $e_{\gamma}$, \eg for $e=1.15$, the secondary soliton family is described by a very short segment  (red diamonds) very close to point $\gamma$ (\ie $\beta'$ is close to $\gamma$). That is to say, secondary solitons with $e<e_{\gamma}$ do not exist because they would not fit inside the box with radius $R=1$ (the red dashed line $\beta_c\gamma$). On the other hand just below $e_c$, \eg $e=1.854$, the main soliton (black disks) family is very close to the secondary soliton family (magenta triangles that almost coincide with the limiting curve $a_c\beta_c$, \ie $\beta'\to \beta_c$). Precisely at $e=e_c$, these two families merge at point $A\equiv a \equiv a_c$.

\begin{figure}[h]
\centerline{
\includegraphics[width=.49\textwidth]{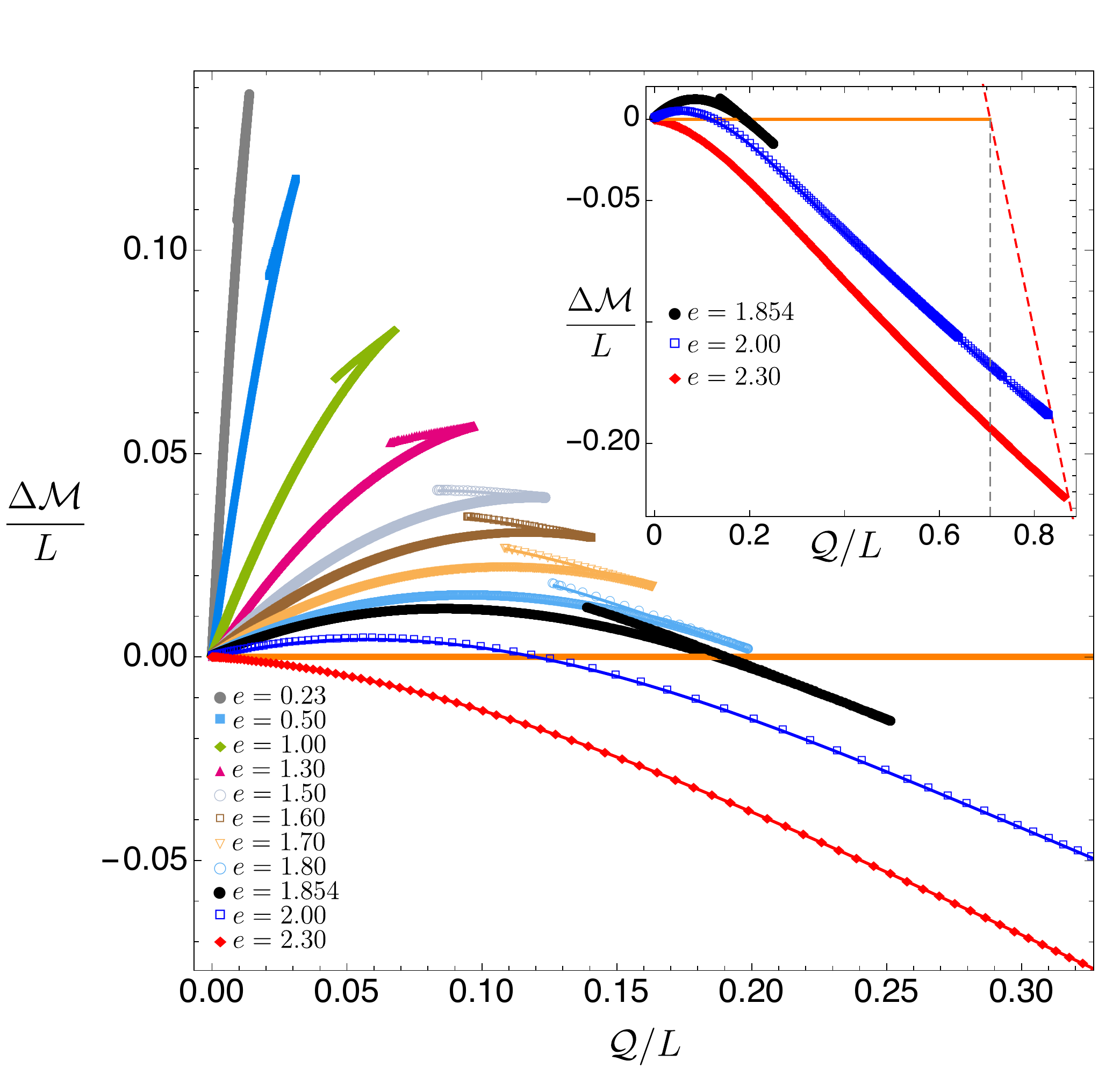}
\hspace{0.3cm}
\includegraphics[width=.48\textwidth]{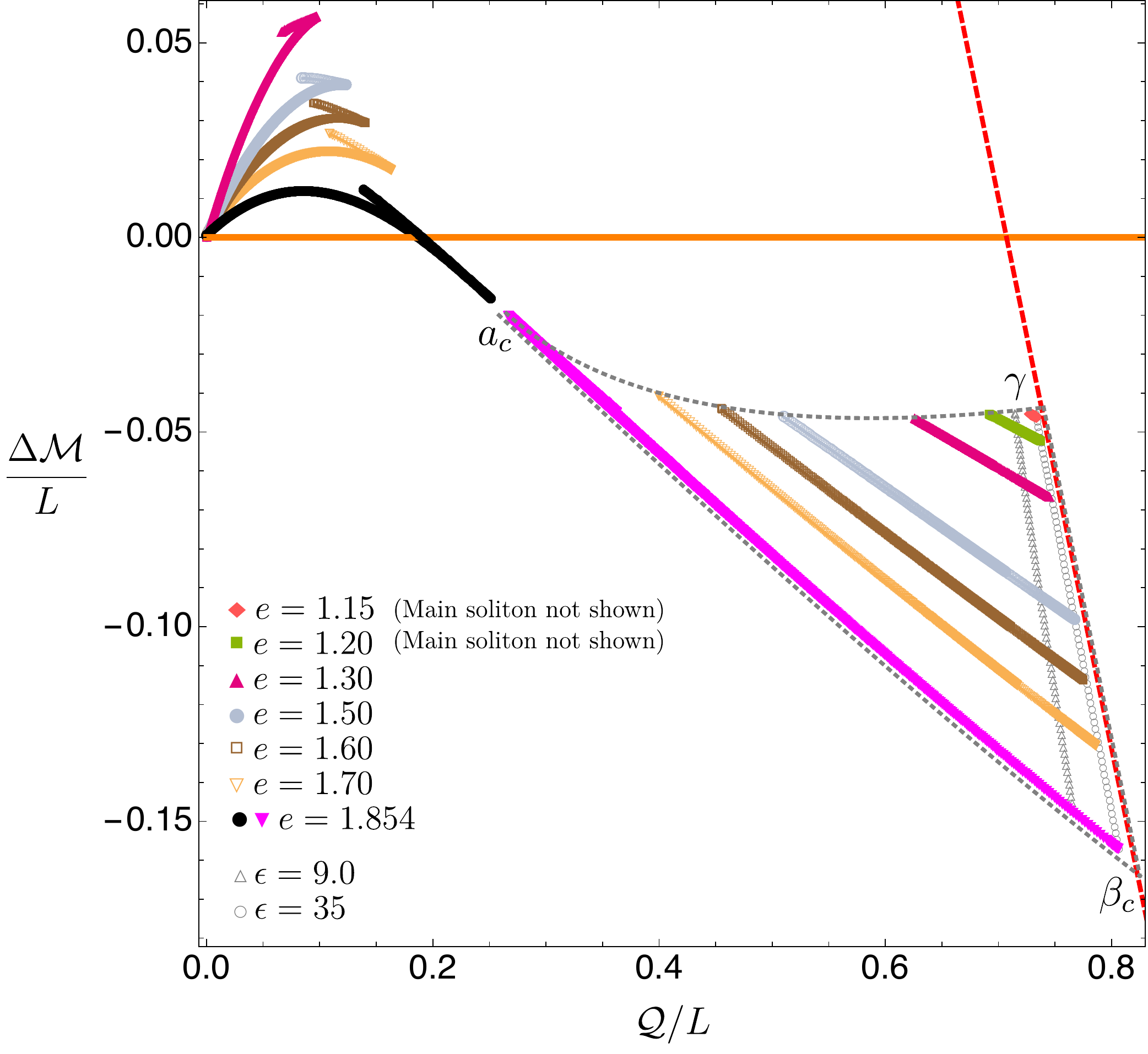}
}
\caption{Quasilocal phase diagram for solitons with different scalar condensate electric charges. The quantity $\Delta{\cal M}$ is the mass difference between the soliton and an extremal RN BH that has the $\textit{same}$ charge $\mathcal{Q}/L$. Hence the orange line at $\Delta{\cal M} = 0$ describes the extremal RN solution. The dashed vertical line in the inset plot describes the line $\mathcal{Q}=2^{-1/2}$ which is the maximum local charge that an extremal RN BH can have whilst fitting inside a box with radius $R=1$. The red dashed line represents the maximal quasilocal charge of solutions that can fit inside the box. {\bf Left panel:} here we give several  examples of  the  main soliton  family only. Notice the change in appearance of the curves as we increase the scalar field charge above $e_c \simeq 1.8545$ (see also inset plot).  {\bf Right panel:} here we concentrate our attention only in  cases with  $e_{\gamma}<e<e_c$ but, this time, we present both the main and secondary solitons. The secondary solitons only exist in the region bounded by the auxiliary gray dashed closed curve $a_c \beta_c \gamma$. The secondary soliton curve with $e$ just above $e_\gamma$ is close to point $\gamma$, while the soliton with $e=1.854$, just below $e_c$, is the magenta curve (very close to $a_c\beta_c$). Note that the gap in $\mathcal{Q}/L$ between the main soliton and the secondary one starts very large at $e=e_\gamma$ but then decreases and goes to zero precisely at $e=e_c$. We also present a secondary soliton curve with constant scalar field condensate, $\epsilon=35$, to illustrate how we can march in $e$ to obtain different solutions and further conclude that secondary solitons are bounded by the auxiliary lines $a_c\gamma$ and $a_c\beta_c$.}
\label{FIGseveral:MassCharge}
\end{figure} 

\item $e_c<e< e_{\hbox{\tiny S}}$. Later, we will give data for the case $e=1.855$ that fits in this window  (section~\ref{subsec:CandS}; Figs.~\ref{FIGe1.855:MassCharge}-\ref{FIGe1.855:phiK}).
Precisely at $e=e_c$ we find that the previously discussed main and secondary soliton families merge: point $A$ of top-right panel of Fig.~\ref{FIG:Summary_sketch} merges with point $a$ at $e=e_c$. This happens in a curious way since above $e_c$ the branch $OA$  merges to the  branch $a\beta'$ (of the top-right panel) and they now form the main soliton family of solitons that, this time, is cusp-free: this is the black line curve $O\beta$ in the bottom-left panel of Fig.~\ref{FIG:Summary_sketch}. On the other hand, the old (\ie top-right panel) sequence of secondary branches/cusps $ABCD\cdots$ of the main soliton family is now connected to the old sequence of secondary branches/cusps $abcd\cdots$ of the secondary soliton family: this is the blue line $\cdots DCBbcd\cdots$ in the bottom-left  panel of Fig.~\ref{FIG:Summary_sketch}. So, precisely at $e=e_c$,  one has 4 families of solitons ``bifurcating'' from the merger $A\equiv a$, \ie the black and blue curves of the bottom-left panel intersect at $A\equiv a$ (equivalently, approaching $e_c$ from below, the black and magenta curves of the top-right panel intersect at $A\equiv a$). Then, as we increase $e$ above $e_c$ the black and blue curves disconnect and their distance increases as $e$ grows. This merging and interactions  between the black/magenta and black/blue curves as we approach $e=e_c$ from bellow/above is observed not only on the quasilocal  phase diagram but also undoubtedly     
confirmed when we analyse \eg the plots for the Kretchsmann curvature invariant at the origin, $K\big|_{R=0}$, as a function of the scalar condensate $\epsilon\equiv \phi'(R=1)$: compare latter Fig.~\ref{FIGe1.854:phiK} (for $e=1.854\lesssim e_c$) with Fig.~\ref{FIGe1.855:phiK} (for $e=1.855\gtrsim e_c$).  

The sharp transition of the properties of the main soliton family when $e$ changes from a value below $e_c$ into one above $e_c$ is also illustrated in Fig.~\ref{FIGseveral:MassCharge} where we plot the main soliton family for different values of the electric scalar field charge $e$. In this plot
the black disk curve has $e=1.854<e_c$ while the blue square ($e=2$) and red diamond ($e=2.3$) curves have  $e>e_c$. The inset plot of this figure clearly shows that the Chandrasekhar limit seen for  $e<e_c$ is no longer present for $e>e_c$.   

\item $e \geq e_{\hbox{\tiny S}}=\frac{\pi }{\sqrt{2}}\sim 2.221$. Later, we will give data for the case $e=2.3$ that fits in this window  (section~\ref{subsec:higherS}; Figs.~\ref{FIGe2.3:MassCharge}-\ref{FIGe2.3:phiK}).
Solitons with charges in this range have properties that are very similar to those displayed by solitons with $e_c<e< e_{\hbox{\tiny S}}$. In particular, as sketched in the bottom-right panel of Fig.~\ref{FIG:Summary_sketch}, we still have similar main soliton (black curve $O\beta$) and secondary soliton (blue curve $\cdots CBbc\cdots$) families (these curves become more separated apart as $e$ grows). But charges in this range also leave their unique footprint. Indeed, the main difference that justifies distinguishing the cases $e_c<e< e_{\hbox{\tiny S}}$ and  $e \geq e_{\hbox{\tiny S}}$ is that, for the latter case, the main soliton family $O\beta$ always has smaller quasilocal mass than the extremal RN BH with the same charge, no matter how small $\mathcal{Q}/L$ is. That is to say, these solitons always have $\Delta \mathcal{M}<0$, with the slope of the black curve $O\beta$ at $\mathcal{Q}=0$ being zero precisely at $e=e_{\hbox{\tiny S}}$ and negative for $e>e_{\hbox{\tiny S}}$ (so, this slope at the origin for $e<e_{\hbox{\tiny S}}$ is positive). Moreover, for $e \geq e_{\hbox{\tiny S}}$ and only in this case, extremal RN BHs are unstable all the way down to $\mathcal{Q}=0$ \cite{Dias:2018zjg}. These two facts suggest that for this regime of $e$, as confirmed in the perturbative analysis of  \cite{Dias:2018yey}, the zero horizon radius limit of the hairy black holes of the theory  should be our soliton and thus they can be constructed perturbatively for arbitrarily small scalar condensate amplitude and horizon radius. 

The secondary family of solitons (the magenta curve $\cdots CBbc\cdots$ in the bottom-right plot of Fig.~\ref{FIG:Summary_sketch}) still exists for $e>e_{\hbox{\tiny S}}$ but it is further separated from the main soliton family (black curve $O\beta$). It becomes increasingly much more difficult to follow this family beyond the cusps $B, b$ for values of $e>e_c$. So it is hard to explore more energetic soliton families for values of $e$ much larger than  $e_c$. 
However, one real possibility (we found evidence for this) is that there is a sequence of critical charges $e_{c_2}, e_{c_3}\cdots$. Here, $e_{c_2}$ would be the charge above which the cups $C$ and $c$ would merge (very much like the cusps $A$ and $a$ merged at $e=e_c\sim 1.8545$) and form a {\it closed} line family in the phase diagram ($CBbc$ with $c\equiv C$) together with a {\it third} open family of solitons that would have higher mass and that would correspond to the extension to higher $e$ of the open curve $\cdots DCcd \cdots$ in the bottom-left plot of Fig.~\ref{FIG:Summary_sketch}. If so, as the charge grows we could have not only two ground state families of solitons but a sequence of them: first 3 families (one of them closed), then 4 families (two of them closed), etc. For this reason we leave the curve $\cdots Bb \cdots$ incomplete a bit beyond the cusps $B$ and $b$ in the bottom-right panel of Fig.~\ref{FIG:Summary_sketch}.

\end{enumerate}

Once we have found the hairy solitons that are confined inside the Minkowski box we can study the properties that the box must have to support the pressure exerted by the scalar condensate that is inside it. As mentioned above, the box must have an Israel stress tensor (that accounts for the extrinsic curvature jump at the box) that prevents the scalar field to expand all the way to the asymptotic region where it would die-off. In section \ref{sec:Boxstructure}, we compute this Israel stress tensor and find that there are boxes with an energy density and pressure that obey several or all forms of the energy conditions \cite{Wald:106274}. We will also compute the ADM mass and charge of our solitons as measured at the asymptotic boundary \cite{Arnowitt:1962hi}. These ADM charges include the contributions from the hairy soliton and the box and we will find that, quite often, the ADM mass can be negative. Recall that Schwarzschild and RN BHs with negative ADM mass are singular solutions, but our solitons with negative mass are regular everywhere. 

Many of the above physical features observed in boson stars (solitons) confined in a box in an asymptotically flat background are similar to those observed in asymptotically anti-de Sitter solitons \cite{Gentle:2011kv} (see also \cite{Basu:2010uz,Bhattacharyya:2010yg,Dias:2011tj,Arias:2016aig,Markeviciute:2016ivy,Markeviciute:2018cqs,Dias:2016pma}). In this case, the AdS boundary conditions act as a natural gravitational box with radius inversely proportional to the cosmological length that provide confinement. Therefore, we have good reasons to expect that many of the features that we identify in the present study are universal properties of charged scalar fields subject to some sort of confining mechanism.  

\section{Einstein-Maxwell gravity with a confined scalar field}\label{sec:Model}

\subsection{Theory and setup}

Consider Einstein-Maxwell gravity in four dimensions coupled to a charged scalar field with action:
\begin{align}\label{action}
S=\frac{1}{16 \pi G_N}\int{\mathrm d^4 x\sqrt{g}\left({\cal R}-\frac{1}{2}F_{\mu\nu}F^{\mu\nu}-2D_{\mu}\phi(D^{\mu}\phi)^{\dagger}+V(|\phi|)\right)},
\end{align}
where ${\cal R}$ is the Ricci scalar,  $A$ is the Maxwell gauge potential, $F=\mathrm d A$, and  $D_{\mu}=\nabla_{\mu}-i q A_{\mu}$ is the gauge covariant derivative of the system. We consider the potential $V(|\phi|)= m^2\phi\phi^{\dagger}$ with $m$ the mass of the scalar field. For concreteness we will take $m=0$ but solitons with $m>0$ should also exist. We fix Newton's constant $G_N \equiv 1$. 

We are interested in solitonic solutions of \eqref{action} that are static, spherically symmetric and asymptotically flat. Using reparametrisations of the time and radial coordinates, $ t\to \tilde{t}=t+H(t,r) $ and $r\to \tilde{r}(r)$, we work in the `radial/Schwarzschild gauge where we fix the radius of a round $S^2$ to be the areal radius $r$ and there is no cross term $dt dr$. An {\it ansatz} with the desired symmetries is
\begin{align}\label{fieldansatz}
\mathrm d s^2=-f(r)\mathrm d t^2+g(r)\mathrm d r^2+r^2\mathrm d \Omega_2^2, \qquad A_{\mu}\mathrm d x^{\mu}=A_t(r)\mathrm d t, \qquad \phi=\phi^{\dagger}=\phi(r),
\end{align} 
with $\mathrm d \Omega_2^2$ being the metric for the unit 2-sphere (expressed in terms of the polar and azimuthal angles $x=\cos\theta$ and $\varphi$).  We choose to work with the static {\it ansatz} \eqref{fieldansatz} where the scalar field is real. Horizonless solutions in this gauge are usually called solitons. However, we can also perform a $U(1)$ gauge transformation with gauge parameter $\chi=-\omega t/q$,
\begin{align}\label{eq:U1gauge}
\phi=|\phi|e^{i\varphi} \to |\phi|e^{i(\varphi +q \,\chi)} \,,\qquad A_t \to A_t +\nabla_t \chi,
\end{align} 
to rewrite the scalar field as $\phi=|\phi|e^{-i\omega t}$, in which case we would be in a frame where the scalar field oscillates in time with a frequency $\omega$.\footnote{However, since the energy-momentum tensor of the scalar field only depends on $\phi \phi^\dagger$ and $\partial \phi (\partial \phi)^\dagger$, in the new gauge the gravitational and Maxwell fields would still be invariant under the action of the Killing vector field $\partial_t$.}  Horizonless solutions in this gauge are usually called boson stars. The solitons and boson starts of the theory are therefore the same objects since they differ only by a $U(1)$ gauge transformation.

We introduce a spherical confining box of radius $L$ in our system. However, the system has the scaling symmetry:
\begin{align}\label{1scalingsym}
\begin{split}
\{t,r,x,\varphi\}\to\{\lambda_1 t,\lambda_1 r,x,\varphi\} , &\qquad \{f,g,A_t,\varphi\}\to \{f,g,A_t,\varphi\}, \\
\{q,L, r_+,m\}&\to \left\{\frac{q}{\lambda_1},\lambda_1 L, \lambda_1 r_+,\frac{m}{\lambda_1}\right\}
\end{split}
\end{align}
which  leaves the equations of motion invariant and rescales the line element and the gauge field 1-form as $\mathrm{d}s^2\to \lambda_1^2 \mathrm{d}s^2$ and $A_t \mathrm{d}t \to \lambda_1 \,A_t \mathrm{d}t$. We use this scaling symmetry to work with dimensionless coordinates and measure thermodynamic quantities in units of $L$ (effectively this sets $L\equiv 1$),
\begin{equation}\label{adimTR}
T=\frac t L\,,\qquad R=\frac r L\,; \qquad R_+=\frac{r_+}{L}\,, \qquad e=q L\,, \qquad m_\phi=m L \,.
\end{equation}
The box is now at $R=1$.

The equations of motion for the fields $f(R),\ A_t(R),\ g(R)$ and $\phi(R)$ obtained from extremising the action \eqref{action}  with $m=0$ can be found in \cite{Dias:2018yey}. We have a system of three second order ODEs for $f,A_t,\phi$ and $g$ is given by an algebraic relation of the other three functions.
To have a well-posed boundary value problem we need to specify the boundary conditions at the origin and asymptotic boundary of our spacetime. Moreover we must also impose junction conditions at the timelike hypersurface $\Sigma$ at $R=1$ where the box is located. Our solitons have vanishing scalar field at and outside this box, $\phi(R\geq 1)=0$.

The system is described by three second order ODEs which means that there are six arbitrary integration constants  when we do a Taylor expansion around the origin, $R=0$. Regularity, requires that we impose Dirichlet boundary conditions whereby we set three of the above integration constants to zero in order to eliminate terms that would diverge at this boundary \cite{Dias:2015nua}. We are thus left with only three constants $f_0,A_0,\phi_0$ (say) such that the regular fields have the Taylor expansion around the origin:
\begin{equation}\label{BCorigin}
f(0)=f_0+\mathcal O(R^2),\qquad A_t(0)=A_0+\mathcal O(R^2),\qquad  \phi(0)=\phi_0+\mathcal O(R^2).
\end{equation}
Take now the asymptotic boundary of our spacetime, $R\to\infty$. Outside the box the scalar field vanishes, $\phi=0$, and the solutions of the equation of motion are: $f^{out}(R)=c_f-\frac{M_0}{R}+\frac{\rho^2}{2 R^2},$ $A_t^{out}(R)=c_A+\frac{\rho}{R}$ and $g^{out}(R)=c_f/f^{out}(R)$ (henceforward, the superscript~$^{out}$  represents fields outside the box). Here, $c_f,M_0,c_A$ and $\rho$ are arbitrary integration constants which are not constrained, \ie  we have an asymptotically flat solution for any value of these constants. But the theory has a second scaling symmetry,
\begin{align}\label{2scalingsym}
\{T,R,x,\varphi\}\to\{\lambda_2 T, R,x,\varphi\} , \quad \{f,g,A_t,\varphi\}\to \{\lambda_2^{-2} f,g,\lambda_2^{-1} A_t,\varphi\}, \quad \{e, R_+\}\to \{e, R_+\},
\end{align}
that we use to set $c_f=1$ so that $f |_{r\to\infty}=1$ (and $g^{out}=1/f^{out}$).
Outside the box the solution to the equations of motion is then 
\begin{equation}\label{BCinfinity}
f^{out}(R)\big|_{R\geq 1}=1-\frac{M_0}{R}+\frac{\rho^2}{2 R^2}\,,\qquad A_t^{out}(R)\big|_{R\geq 1}=c_A+\frac{\rho}{R}\,,\qquad \phi^{out}(R)\big|_{R\geq1}=0\,,
\end{equation}
which is the Reissner-Nordstr\"om solution as required by Birkhoff's theorem for the Einstein-Maxwell theory \cite{WILTSHIRE198636,inverno:1992}. However, \eqref{BCinfinity} has three free integration constants, $M_0,c_A,\rho$, which will be determined only after we have the solution inside the box. 

Our solutions are asymptotically flat. Therefore, some of the parameters  in \eqref{BCinfinity} are related to the ADM  conserved charges  \cite{Arnowitt:1962hi}.
Namely, the  adimensional ADM mass and electric charge of the system are given by (setting $G_N\equiv 1$):\footnote{
Note that the Maxwell term in action \eqref{action} is $\frac{1}{2}F^2$, not the perhaps more common $F^2$ term. It follows that the extremal RN BH satisfies the  ADM relation $M=\sqrt{2}|Q |$, instead of $M=|Q|$ that holds when the Maxwell term in the action is $F^2$.}
\begin{align}\label{eq:ADM}
\begin{split}
&M/L=\lim_{R\to \infty}\frac{R^2 f'(R)}{2\sqrt{f(R)  g(R)}}=\frac{M_0}{2},\\
&Q/L=\lim_{R\to \infty}\frac{R^2 A_t'(R)}{2 f(R) g(R)}=-\frac{\rho}{2}.
\end{split}
\end{align}
These ADM conserved charges measured by a Gauss law at the asymptotic boundary include the contribution from the energy-momentum content of the box that confines the scalar hair. 

In these conditions, solitons of the theory are a 1-parameter family of solutions that we can take to be, \eg $f_0$ as defined in \eqref{BCorigin}, or the value of the (interior) derivative of the scalar field at the box,  $\epsilon\equiv \phi^{\prime\:in} \big|_{R=1}$ as discussed in the next subsection.

As mentioned in section \ref{sec:summary}, it follows from Birkhoff's theorem that  in the asymptotic region our solutions are necessarily described by the RN solution and  we cannot use the ADM mass $M$ and charge $Q$  to distinguish the several solutions of the theory. Instead, we need to resort to the Brown-York quasilocal mass $\mathcal{M}$ and charge $\mathcal{Q}$, measured at the box to display our solutions in a phase diagram of the theory \cite{Brown:1992br}.
From section II.C of \cite{Dias:2018yey} (which we ask the reader to visit for details), the Brown-York quasilocal mass and charge contained inside a 2-sphere with radius $R=1$ are ($G_N\equiv 1$)
\begin{eqnarray}\label{BYmassCharge}
{\cal M}/L&=& R\left( 1-\frac{1}{\sqrt{g(R)}}\right)\Big|_{R=1}. \nonumber\\
{\cal Q}/L  &=& \frac{R^2A_t'(R)}{2 \sqrt{g(R)} f(R)}\Big|_{R=1}.
\end{eqnarray}
To complete the thermodynamic description of our solutions we still need to define the chemical potential of the soliton which is  given by the value of the gauge potential at the box, 
\begin{align}\label{eq:QLchp}
\mu=A_t\big|_{R=1},
\end{align}
and the quasilocal quantities must satisfy the quasilocal first law of thermodynamics \cite{Dias:2018yey}: 
\begin{align}\label{1stlawsoliton}
\mathrm d {\cal M}=\mu \,\mathrm d {\cal Q}\,. 
\end{align}
We will use this law as a non-trivial check of our solutions.

As explained before, for reference we will often compare the soliton families of solutions against extremal RN BHs. RN BHs confined in a box can be parametrized using the dimensionless horizon radius $R_+$ and the chemical potential $\mu$, and their quasilocal mass and charge are  \cite{Dias:2018yey}
\begin{align}\label{quasilocalRN}
{\cal M}/L \big|_{RN}= 1 - \frac{\sqrt{2}(1-R_+)}{\sqrt{2-(2-\mu^2)R_+}}, \qquad \mathcal{Q} /L \big|_{RN}= \frac{\mu R_+}{\sqrt{2} \sqrt{2-(2-\mu^2)R_+}}.
\end{align}
where $0<R_+\leq 1$ (for the horizon to be confined inside the box) and $0\leq \mu\leq \mu_{\rm ext}$, with extremality reached at $\mu_{\rm ext}=\sqrt{2}$. Note that at extremality one has ${\cal M}/L = R_+$ and ${\cal Q}/L =R_+/\sqrt{2}$. On the other hand, for any $\mu$, when $R_+=1$ one has ${\cal M}/L = 1$ and ${\cal Q}/L =2^{-1/2}$. 

\subsection{Junction conditions and Israel stress tensor at the box}\label{subsec:JCisrael}

Above, we discussed the boundary conditions at the origin and asymptotic boundaries.
However, solitons are solutions that glue an interior spacetime ($R<1$; with superscript $^{in}$) with the known RN exterior background solution \eqref{BCinfinity} ($R>1$; with superscript $^{out}$) . So all we need to do is to find the interior solution. But for that we must specify appropriate physical conditions at the outer boundary of our numerical integration domain, namely at $R=1$.

The scalar field must vanish at and outside the box, \ie for $R\geq 1$ but its derivative when approaching the box from the interior, \ie as $R\to 1^-$, does not vanish (unless we have the trivial RN solution) and we will call this quantity $\epsilon$:\footnote{Note that our theory has the symmetry $\phi\to -\phi$ so we can focus our attention only on the case $\epsilon>0$.}
\begin{equation}\label{def:epsilon} 
\phi^{in} \big|_{R=1}=\phi^{out} \big|_{R=1}=0,\qquad \phi^{out}(R)=0, \qquad \phi^{\prime\:in} \big|_{R=1}\equiv \epsilon, 
\end{equation}
\ie for $R\leq 1$ the scalar field is forced to have the Taylor expansion $\phi\big|_{R=1^-}=\epsilon (R-1)+\mathcal{O}(R-1)^2$. 
We are forcing a jump in the derivative of the scalar field normal to the cavity  timelike hypersurface $\Sigma$. The latter is defined by $\frak{f}(R)=R-1=0$ and  has outward unit normal $n_{\mu}=\partial_\mu\frak{f}/|\partial \frak{f}|$ ($n_\mu n^\mu=1$).  Naturally, this forcing condition on the scalar field has consequences: we need to impose junction conditions at $\Sigma$ on the other fields. Ideally, we would like to have a smooth crossing, whereby the gravitational and gauge fields and their normal derivatives are continuous at $\Sigma$. But this is not possible when we have a non-vanishing scalar field inside the box. It follows that the Israel junction conditions require a non-vanishing jump in the extrinsic curvature across the box.

It is a good idea to set some notation to discuss this issue further.
Adopting the viewpoint of an observer in the interior region, the layer surface $\Sigma$ is parametrically described by $R=1$ and $T=T^{in}(\tau)=\tau$ and the induced line element and gauge 1-form of the shell $\Sigma$ read
\begin{eqnarray}\label{SigmaIn}
{\mathrm d}s^2|_{\Sigma^{in}}&=&h_{ab}^{in} \,{\mathrm d}\xi^a {\mathrm d}\xi^b =- f^{in}|_{R=1}{\mathrm d}\tau^2+{\mathrm d}\Omega^2_2\,,\nonumber \\
 A_t |_{\Sigma^{in}}&=& a_{a}^{in} {\mathrm d}\xi^a = A_t^{in}|_{R=1}{\mathrm d}\tau\,,
\end{eqnarray}
where $\xi^a$ describe coordinates in $\Sigma$, $h_{ab}^{in}$ is the induced metric in $\Sigma$ and $a_a^{in}$ is the induced gauge potential in $\Sigma$. 
On the other hand, from the perspective of an observer outside the cavity, $\Sigma$ is parametrically described by $R=1$ and $T=T^{out}(\tau)=N \tau$ (so,  $N$ is a reparametrization freedom parameter) so that the induced line element and gauge 1-form are
\begin{eqnarray}\label{SigmaOut}
{\mathrm d}s^2 |_{\Sigma^{out}}&=&h_{ab}^{out} \,{\mathrm d}\xi^a {\mathrm d}\xi^b =- N^2 f^{out}|_{R=1}{\mathrm d}\tau^2+{\mathrm d}\Omega^2_2\,,\nonumber \\
 A_t |_{\Sigma^{out}}&=& a_{a}^{out} {\mathrm d}\xi^a = N A_t^{out}|_{R=1}{\mathrm d}\tau\,,
\end{eqnarray}

The junction conditions required to join smoothly two backgrounds at a timelike hypersurface $\Sigma$ were studied by Israel \cite{Israel:1966rt,Israel404,Kuchar:1968,Barrabes:1991ng} built on previous work of Lanczos and Darmois. A solution is smooth at $\Sigma$ if and only if: 1) the induced metric $h_{ab}$ and induced gauge potential $a_a$  are continuous (\ie ${\mathrm d}s^2|_{\Sigma^{in}}={\mathrm d}s^2|_{\Sigma^{out}}$ and $A |_{\Sigma^{in}}=A |_{\Sigma^{out}}$), and 2)  the extrinsic curvature $K_{ab}$ (essentially the normal derivative of the induced metric) and the normal derivative of the induced gauge field, $f_{aR}$, are continuous.  Denoting, as we have been doing, the solution inside (outside) $\Sigma$ by the superscript $^{in}$ ($^{out}$), the Israel junction conditions are
\begin{subequations}\label{IsraelJunctionConditions}
\begin{align}
& a_{a}^{in}\big|_{R=1}=a_{a}^{out}\big|_{R=1}\,, \label{eq:Israeljunction1} \\
&h_{ab}^{in}\big|_{R=1}=h_{ab}^{out}\big|_{R=1}\,; \label{eq:Israeljunction2} \\
& f_{aR}^{in}\big|_{R=1}=f_{aR}^{out}\big|_{R=1}\,,\label{eq:Israeljunction3} \\
&K_{ab}^{in}\big|_{R=1}=K_{ab}^{out}\big|_{R=1}; \label{eq:Israeljunction4}
\end{align}
\end{subequations}
where $h_{ab}=g_{ab}-n_a n_b$ is the induced metric at $\Sigma$ and  $K_{ab}=h_a^{\phantom{a}c}\nabla_c n_b$ is the extrinsic curvature.

In the absence of the scalar condensate, we can set $N=1$ and all the junction conditions \eqref{IsraelJunctionConditions} are satisfied. However, our hairy solitons are continuous but not differentiable at $R=1$. Namely, they satisfy the 3 conditions \eqref{eq:Israeljunction1}-\eqref{eq:Israeljunction3} but not \eqref{eq:Israeljunction4}. Since the latter  extrinsic curvature condition  is not obeyed, our hairy solitons are singular at $\Sigma$. But this singularity simply signals the presence  of a Lanczos-Darmois-Israel surface stress tensor ${\cal S}_{ab}$ at the hypersurface layer proportional to the difference of the extrinsic curvature across  the hypersurface. This  Lanczos-Darmois-Israel surface stress tensor induced in $\Sigma$ is  \cite{Israel:1966rt,Israel404,Kuchar:1968,Barrabes:1991ng}
\begin{align}\label{eq:inducedT}
{\cal S}_{ab}=-\frac{1}{8 \pi}\Big([K_{ab}]-[K] h_{ab}\Big),
\end{align}
where $K$ is the trace of the extrinsic curvature and $[K_{ab}]\equiv K_{ab}^{out}\big|_{R=1}-K_{ab}^{in}\big|_{R=1}$. This surface tensor is the pull-back of the energy-momentum tensor integrated over a small region around the hypersurface $\Sigma$ \ie it is obtained by integrating the appropriate Gauss-Codazzi equation  \cite{Israel:1966rt,Israel404,Kuchar:1968,Barrabes:1991ng,MTW:1973} and it is also given by the difference between the Brown-York surface tensor just outside and inside the surface layer  \cite{Brown:1992br} (see also discussion in \cite{Dias:2018yey}).
Essentially, \eqref{eq:inducedT} describes the energy-momentum tensor of the cavity (the ``internal structure" of the box) that we have to build to confine the scalar field inside. With our explicit construction of the hairy solutions of the system we will be able to compute this  Lanczos-Darmois-Israel stress tensor.
Note that since the two Maxwell junction conditions \eqref{eq:Israeljunction1}-\eqref{eq:Israeljunction2} are obeyed,  our solitons will have a surface layer that has no electric charge.

Our strategy is now clear. To find the soliton solution inside the box, we integrate numerically the Einstein equation in the domain $R\in [0,1]$ subject to the boundary conditions \eqref{BCorigin}  at the origin and, at the box, we impose $\phi(1^{-})=0$ and use the scaling symmetry \eqref{2scalingsym} to set $f(1^{-})=1$. With this information we can already read univocally the quasilocal charges \eqref{BYmassCharge} of the system.
We then impose the three junction conditions \eqref{eq:Israeljunction1}-\eqref{eq:Israeljunction3} at the box to match the interior solution with the outer solution  described by the RN solution \eqref{BCinfinity}. This allows to find the parameters $M_0,C_A,\rho$ in  \eqref{BCinfinity} as a function of the reparametrization freedom parameter  $N$  introduced in \eqref{SigmaOut}. 
The Israel stress tensor ${\cal S}_a^b$ is just a function of $N$ and, if $\phi^{in}\neq 0$, there is no choice of $N$ that kills all the components of ${\cal S}_a^b$ (there are two non-vanishing components, ${\cal S}_t^t$ and ${\cal S}_\theta^\theta={\cal S}_\varphi^\varphi$). The fact that we have arbitrary freedom to select $N$ simply reflects the freedom we have in the choice of the energy-momentum content of the box needed to contain the scalar condensate inside it. We will show that there are choices that preserve some or all the energy conditions \cite{Wald:106274}. Once we make a choice for $N$, we can also compute the ADM mass and charge \eqref{eq:ADM} of the solution which includes the contribution from the box.
\subsection{Numerical schemes}\label{subsec:method}

The solitons we search for are a 1-parameter family of solutions. 
We will generate these solutions numerically following one of two routes that differ on the choice made for the marching parameter along the family: 1) we march varying the value of the scalar condensate quantity $\epsilon\equiv \phi'(R=1)$, or  
2) we  march changing the value of the function $f$  at the origin, $f_0\equiv f(0)$. In both cases, the marching parameter that we give as an input to our numerical code appears as a boundary condition.

We will use these two marching approaches because numerical convergence of each one is different at different regions of the parameter space. The first marching strategy is chosen essentially because $\epsilon\equiv \phi'(R=1)$ is the expansion parameter of the perturbative construction of \cite{Dias:2018yey} thus we can straightforwardly use the perturbative solution of  \cite{Dias:2018yey} as a seed for the numerical scheme. However, we will find that away from the perturbative regime, $\epsilon$ no longer uniquely parametrizes the family of solutions (see plots in Appendix). This is because for a given $\epsilon$ there will be more than one soliton (the two or more solitons differ in their charges $\mathcal{M}$ and $\mathcal{Q}$). This happens in the neighborhood of the cusps of Fig.~\ref{FIG:Summary_sketch}. For this reason, it is good to use an alternative parametrization where we march along the soliton branch using the value of the function $f=g_{tt}$ at the origin:  we find that physical/thermodynamic quantities are a monotonic function of $f_0\equiv f(0)$ as we move along any of the soliton families even when we cross any of the cusps sketched in Fig.~\ref{FIG:Summary_sketch}.

When we use $\epsilon$ as a marching parameter we find convenient to introduce the field redefinitions,
\begin{align}\label{functions_epsilon}
f(R) = \tilde{q}_{1}(R), \qquad  A(R) = \tilde{q}_{2}(R), \qquad  \phi(R) = \left(1-R^2 \right) \tilde{q}_{3}(R). 
\end{align}  
These redefinitions automatically impose the condition $\phi(1)=0$ when we search for smooth functions $\tilde{q}_{1,2,3}$. Additionally, we also impose the normalization condition $ \tilde{q}_{1}(1)=1$. All other boundary conditions discussed in the previous two subsections follow from requiring that the equations of motion are also valid at $R=0$ and $R=1$.

On the other hand, when we use the marching parameter $f_0$, it is useful to introduce the field redefinitions 
\begin{align}\label{functionsf0}
&f(R) = \left(1- R^2 \right) f_0 + R^2 \left(1-\left(1-R^2\right) q_{1}(R)\right), \\
&A(R) = q_{2}(R), \\ \nn
&\phi(R) = \left(1-R^2 \right) q_{3}(R). \nn
\end{align}
which have the advantage of introducing explicitly the marching parameter $f_0$ in the problem and  that smooth functions $q_1$ and $q_3$ automatically satisfy the boundary conditions
$f(0)=f_0$, $f(1)=1$ and $\phi(1)=0$. The other boundary conditions for $q_{1,2,3}$ are derived boundary conditions in the sense that they follow directly from the equations of motion evaluated at the boundaries \cite{Dias:2015nua}.
 
 To solve numerically our boundary value problem, we use a standard Newton-Raphson algorithm and discretise the coupled system of three ODEs using pseudospectral collocation (with Chebyshev-Gauss-Lobatto nodes along the $R$). The resulting algebraic linear systems are solved by LU decomposition. These numerical methods are described in detail in the review \cite{Dias:2015nua}. Since we are using pseudospectral collocation, our results should display exponential convergence with the number of grid points. We check this is indeed the case and the thermodynamic quantities that we display have, typically,  8 decimal digit accuracy. We further use the quasilocal first law \eqref{1stlawsoliton} (typically, obeyed within an error smaller than $10^{-3}\%$) to check our solutions.

\section{Phase diagram for charged solitons confined in a Minkowski box}\label{sec:phasediag}

As discussed in our summary of results (section~\ref{sec:summary}), a system with a scalar field with charge $e$ confined inside a box has two natural critical charges: $e_{\hbox{\tiny NH}}=\frac{1}{2 \sqrt{2}}\sim 0.354$ and  $e_{\hbox{\tiny S}}=\frac{\pi }{\sqrt{2}}\sim 2.221$. Moreover, we also find the existence of two other important critical charges, $e_{\gamma}\sim 1.13$ and $e_c \sim 1.8545\pm 0.0005$, that we could not predict using heuristic or perturbative  analysis. These charges satisfy the relations $e_{\hbox{\tiny NH}}<e_{\gamma}<e_c< e_{\hbox{\tiny S}}$. The charge $e_{\hbox{\tiny NH}}$ plays no special role in the discussion of the phase diagram of solitons of the system. Thus, we do not discuss it further, and in the next subsections, we describe the properties of solitons in the following 4 windows of scalar charge: 1) $e\leq e_{\gamma}$, 2) $e_{\gamma}<e\leq e_c$, 3)  $e_c<e\leq e_{\hbox{\tiny S}}$, and 4) $e> e_{\hbox{\tiny S}}$. For concreteness we will display results for a particular value of $e$ for each one of these windows: 1) $e=0.23$ (section~\ref{subsec:lowerGamma}), 2) $e=1.854$ (section~\ref{subsec:GammaC}), 3) $e=1.855$  (section~\ref{subsec:CandS}), and 4) $e=2.3$ (section~\ref{subsec:higherS}).  Altogether, these results (and others not presented) will allow to extract the conclusions summarized in section~\ref{sec:summary}.

It follows from an analysis of the RN quasilocal charges \eqref{quasilocalRN} that, in the quasilocal $\mathcal{Q}-\mathcal{M}$ plot, the region that represents RN BHs whose horizon radius fits inside the box is the triangular surface bounded by the  lines $\mathcal{Q}=0$,  $\mathcal{M}/L=1$ and 
$\mathcal{M}=\sqrt{2}\mathcal{Q}$. 
When plotting the adimensional quasilocal mass $\mathcal{M}/L$ as a function of the dimensionless quasilocal charge $\mathcal{Q}/L$ we will often find that soliton curves are very close to the extremal RN curve that we use for reference. Therefore, we will also plot $\Delta{\cal M}$ as a function of $\mathcal{Q}$, where $\Delta{\cal M}$  is the quasilocal mass difference between the soliton and the extremal RN BH that has the $\textit{same}$ quasilocal charge. For reference, in our plots the orange curve describes this extremal RN BH solution which must have quasilocal charge $\mathcal{Q} \leq 2^{-1/2}$ to fit inside the covariant box with dimensionful radius $L$ (and thus adimensional radius $R=1$).

In the $\mathcal{Q}-\Delta\mathcal{M}$ plane, non-extremal RN BHs exist in the triangular region with boundaries $\mathcal{Q}=0$, $\Delta\mathcal{M}=0$ and $\Delta\mathcal{M}=1-\sqrt{2}\mathcal{Q}/L$. The latter curve is
\begin{equation}\label{redDashed}
\left(\mathcal{Q}/L,\Delta\mathcal{M}/L\right)=\left(L^{-1} \mathcal{Q}\big|_{\rm ext\,RN},1-L^{-1}\mathcal{M}\big|_{\rm ext\,RN} \right)=\left( \frac{R_+}{\sqrt{2}}, 1-R_+\right)
\end{equation}
where $\mathcal{M}\big|_{\rm ext\,RN}$ and $\mathcal{Q}\big|_{\rm ext\,RN}$ are given by 
 \eqref{quasilocalRN} with $\mu=\mu_{\rm ext}=\sqrt{2}$.
 In our $\mathcal{Q}-\Delta\mathcal{M}$ plots, the red dashed line is the parametric curve \eqref{redDashed} with $R_+$ {\it allowed to take also values above 1}.   
It turns out that the most charged solitons that we will find seem to approach this dashed red line (in the limit where scalar condensate amplitude $\epsilon$ approaches infinity).  In this sense, for a given quasilocal mass (smaller than 1), this red dashed line represents the maximal quasilocal charge that solutions that can fit inside the box can have.

To test our numerical results and to set the regime of validity of the perturbative analysis of \cite{Dias:2018yey}, in many of our plots we will also display the perturbative prediction of \cite{Dias:2018yey} (which is valid only for small solitons) using green curves.

\subsection{$e< e_{\gamma}\sim 1.13$ \label{subsec:lowerGamma}}

We illustrate the properties of solitons in this range using the case $e=0.23$ (other examples can be found in Fig.~\ref{FIGseveral:MassCharge}). 
As discussed in our summary section~\ref{sec:summary}, for this range of $e$ the system only has the main  family of solitons.\footnote{We cannot exclude the existence of a second family but if it exists, it is not connected to the secondary soliton family present for $e>e_\gamma$ and that we discuss in the following subsections. Also note that beyond the ground state, there is also an infinite tower of excited solitons that are the nonlinear backreaction of the normal mode radial overtones of the system.}

\begin{figure}[t]
\centerline{
\includegraphics[width=.49\textwidth]{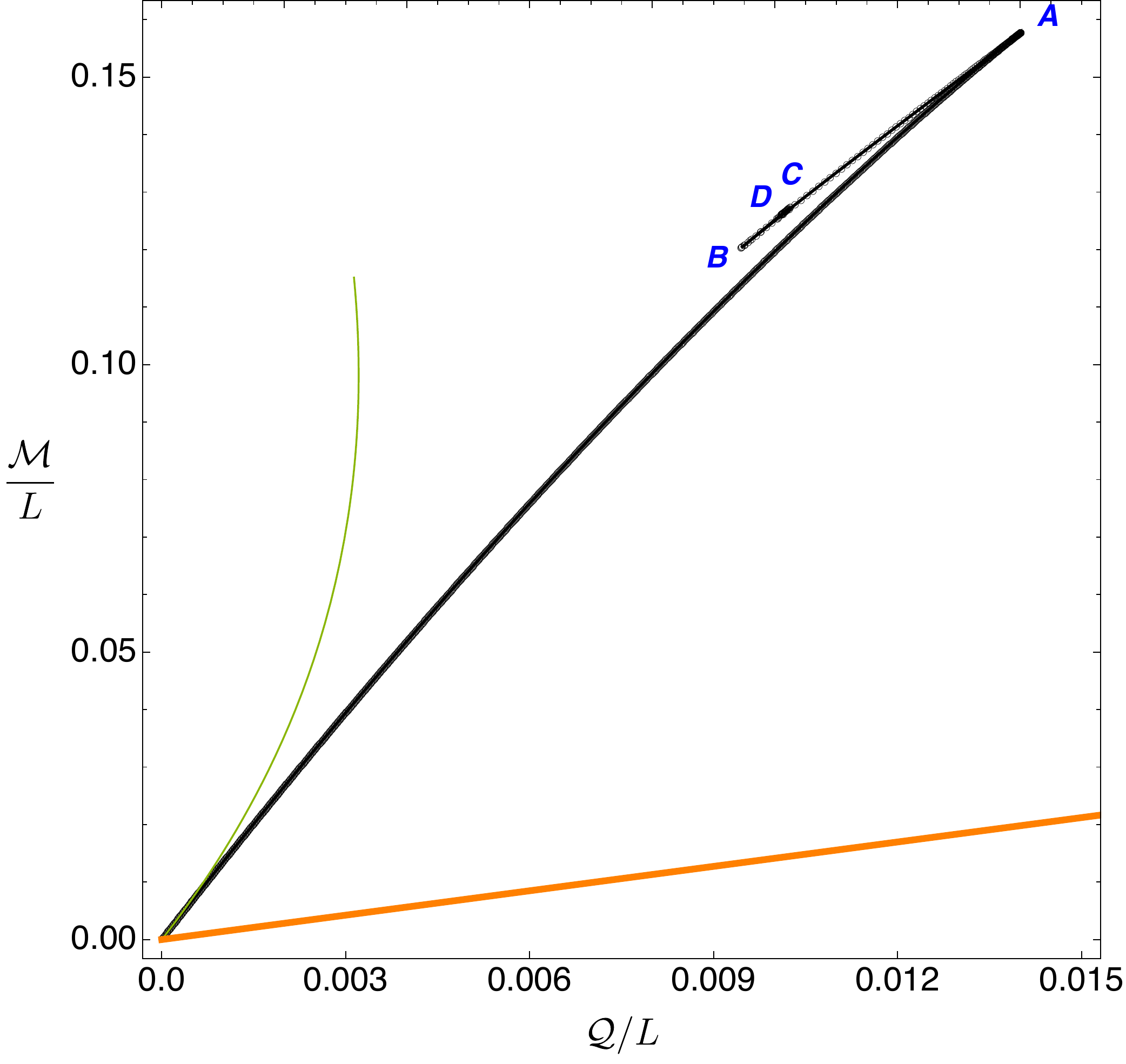}
\hspace{0.3cm}
\includegraphics[width=.505\textwidth]{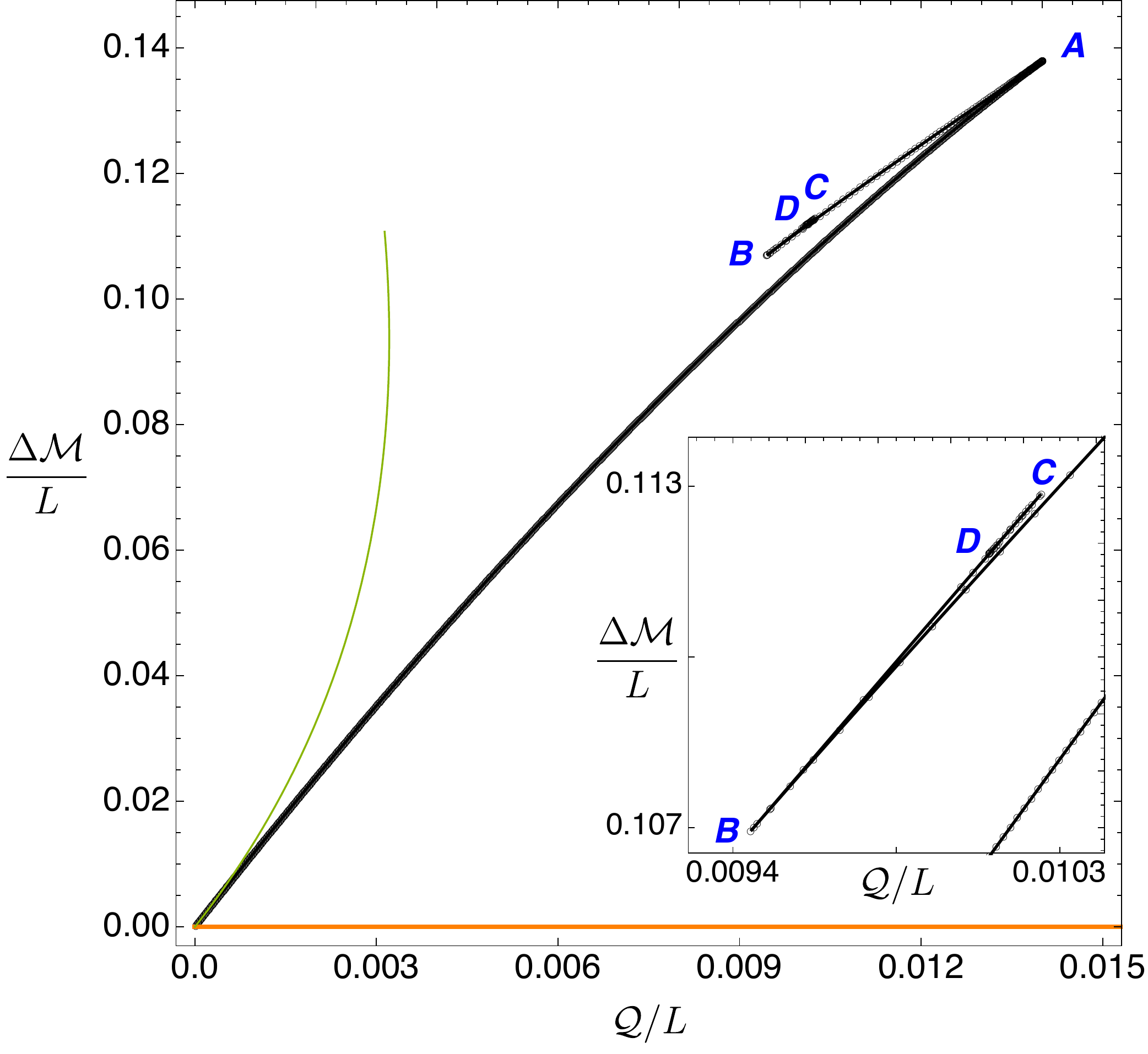}
}
\caption{Main soliton  family with $e=0.23$ ($e<e_{\gamma}$). The orange curve describes the extremal RN BH solution, and the green curve is the perturbative prediction of \cite{Dias:2018yey}. 
 {\bf Left panel:}  Quasilocal mass $\mathcal{M}/L$ as a function of the quasilocal charge $\mathcal{Q}/L$.  {\bf Right panel:} $\Delta{\cal M}$ is the mass difference between the soliton and an extremal RN BH that has the $same$ quasilocal charge. For larger $e$ this quantity will distinguish better some solitons from the extremal RN BH solution. The inset plots reveals a zig-zag structure along the cusps $A, B, C, D,\cdots$}
\label{FIGe0.23:MassCharge}
\end{figure}

 In Fig.~\ref{FIGe0.23:MassCharge} we plot the dimensionless quasilocal mass $\mathcal{M}/L$ as a function of the dimensionless quasilocal charge $\mathcal{Q}/L$ of the main soliton branch. As a first observation, note that our full nonlinear numerical construction of solitons (black curve) agrees well with the perturbative construction of \cite{Dias:2018yey} at small values of the charge $\mathcal{Q}$ (green curve). However, for larger charges the elaborated zig-zagged structure that is developed is not captured by the perturbative analysis.

\begin{figure}[t]
\centerline{
\includegraphics[width=.49\textwidth]{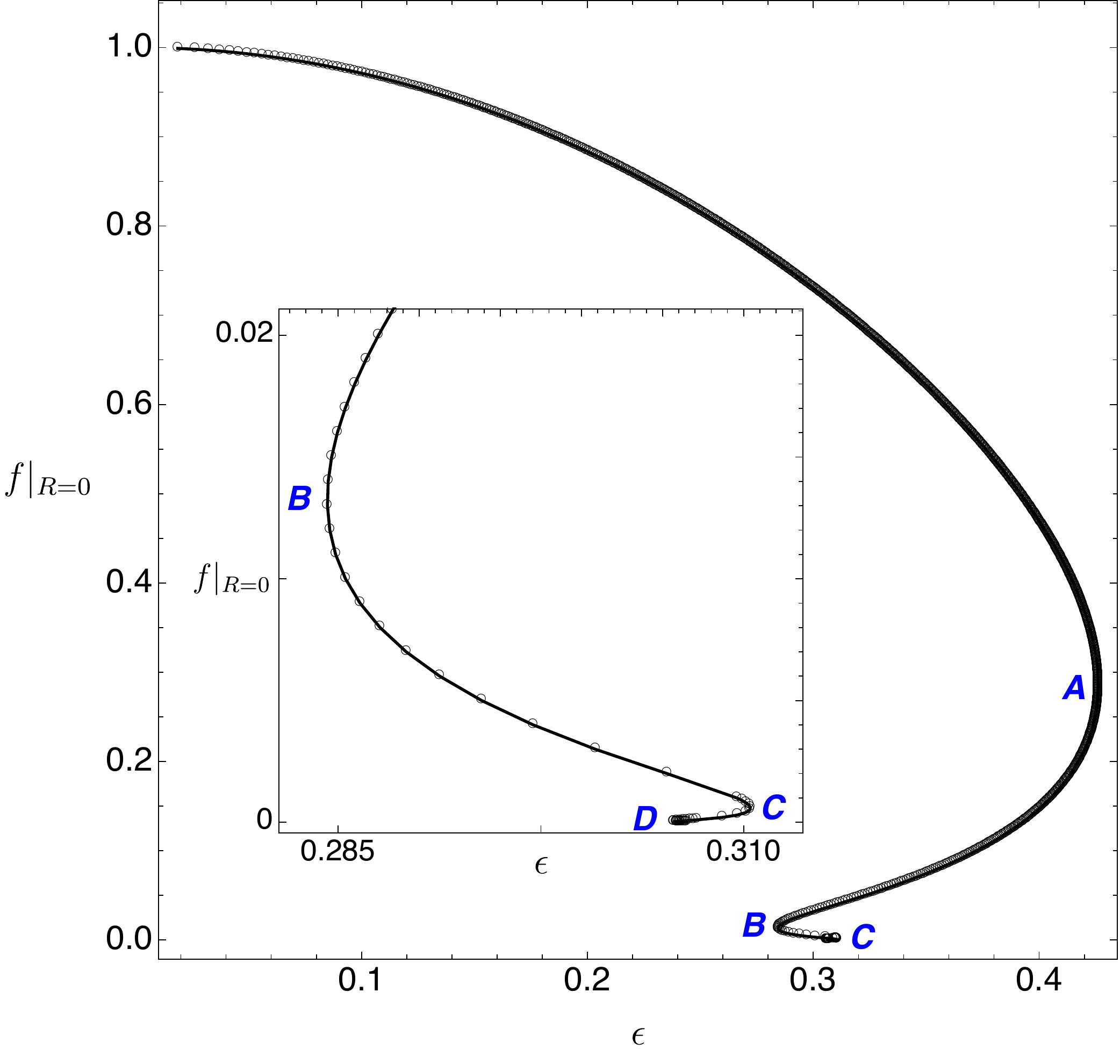}
\hspace{0.3cm}
\includegraphics[width=.49\textwidth]{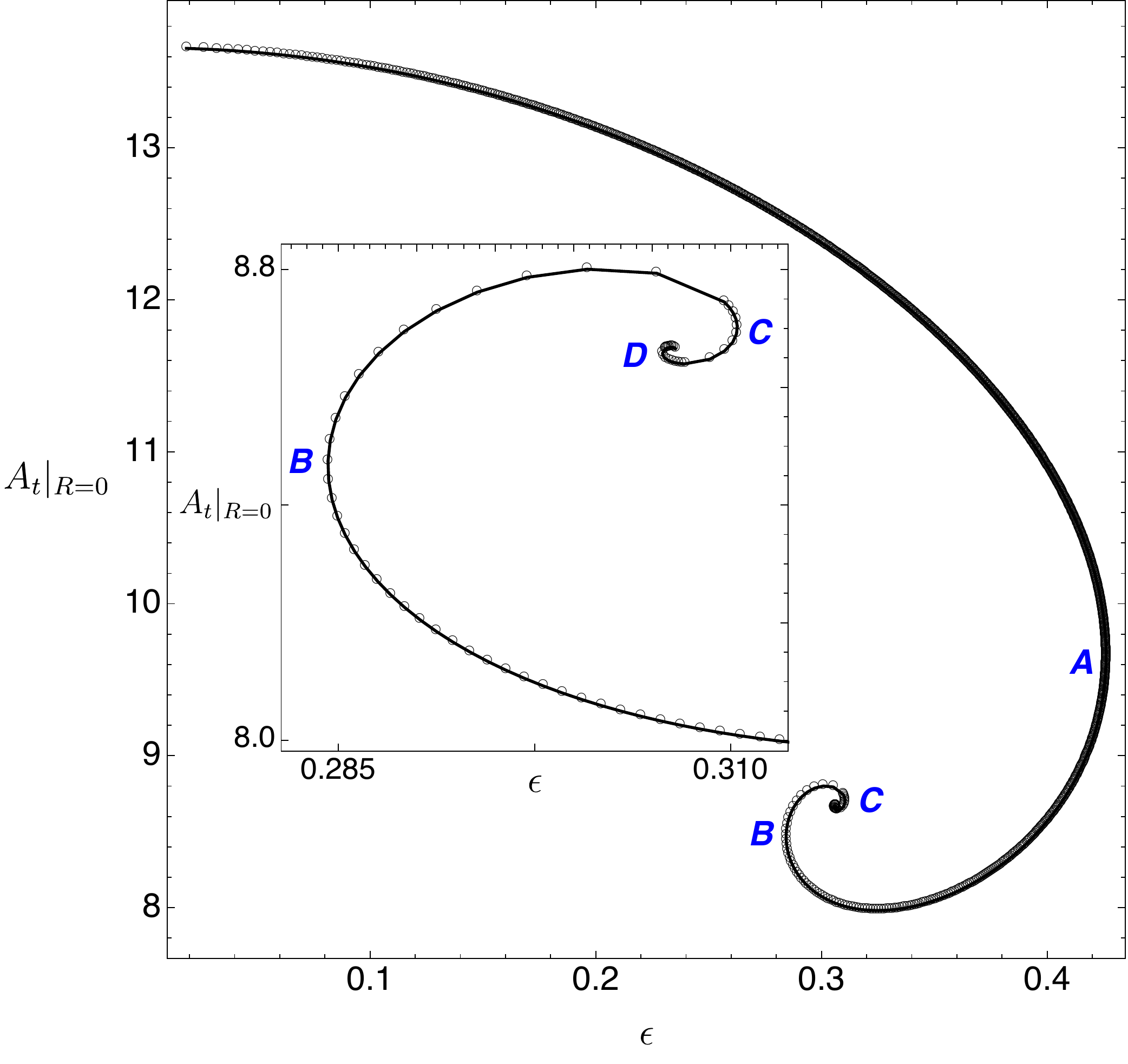}
}
\caption{Main soliton  family with $e=0.23$ ($e<e_{\gamma}$). {\bf Left panel:} As we ramp up the value of the amplitude $\epsilon\equiv \phi'(1)$ the core value of the metric function $f$ is (very) slowly approaching zero. The inset plot zooms in the damped oscillations around the turning points $B,C,D$.  {\bf Right panel:}  Contrast this with the core value of the gauge field $A$ which is spiralling inward indefinitely as $\epsilon$ is ramped up. The inset plot zooms in the spirals  around the turning points $B,C,D$. }
\label{FIGe0.23:fA}
\end{figure} 

\begin{figure}[t]
\centerline{
\includegraphics[width=.48\textwidth]{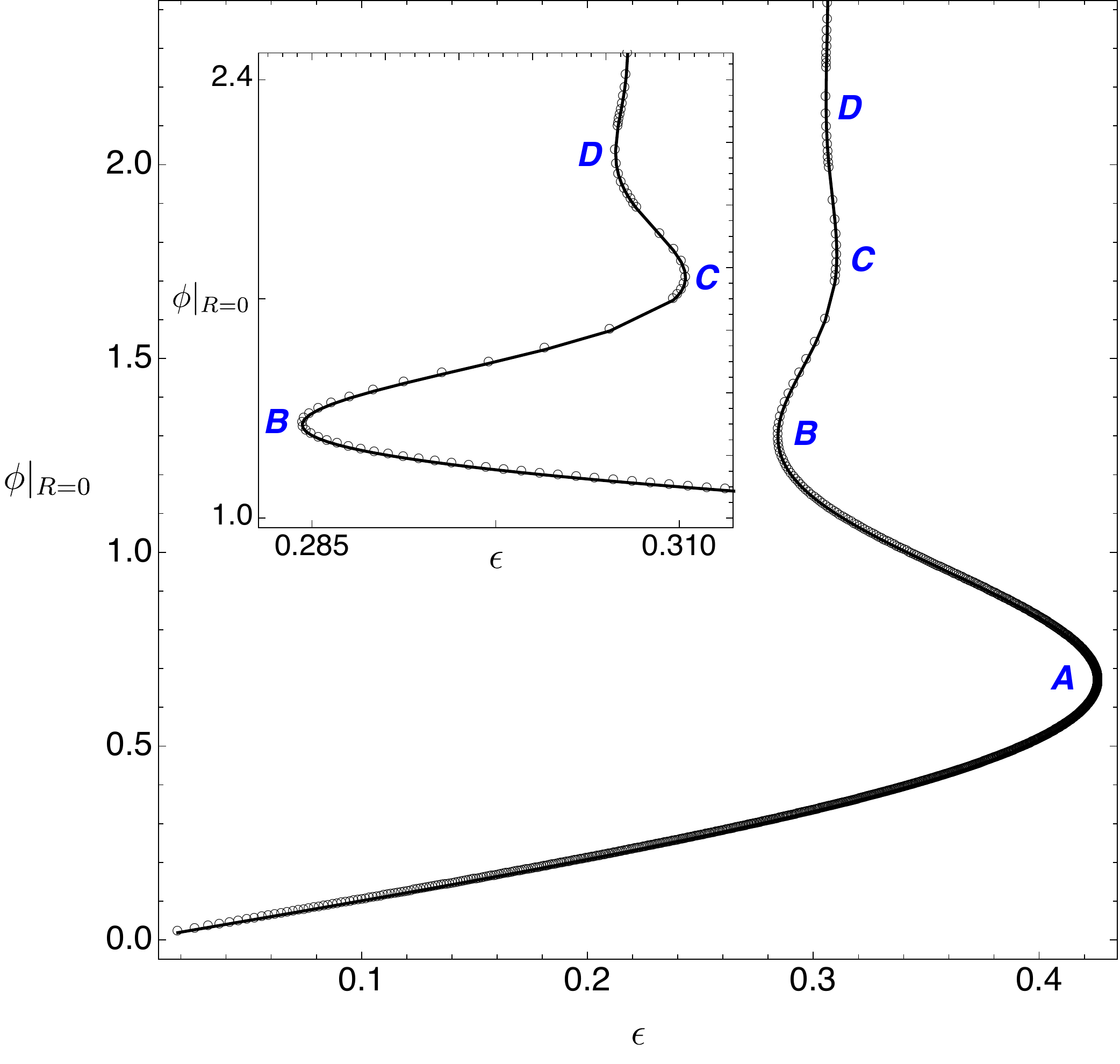}
\hspace{0.3cm}
\includegraphics[width=.515\textwidth]{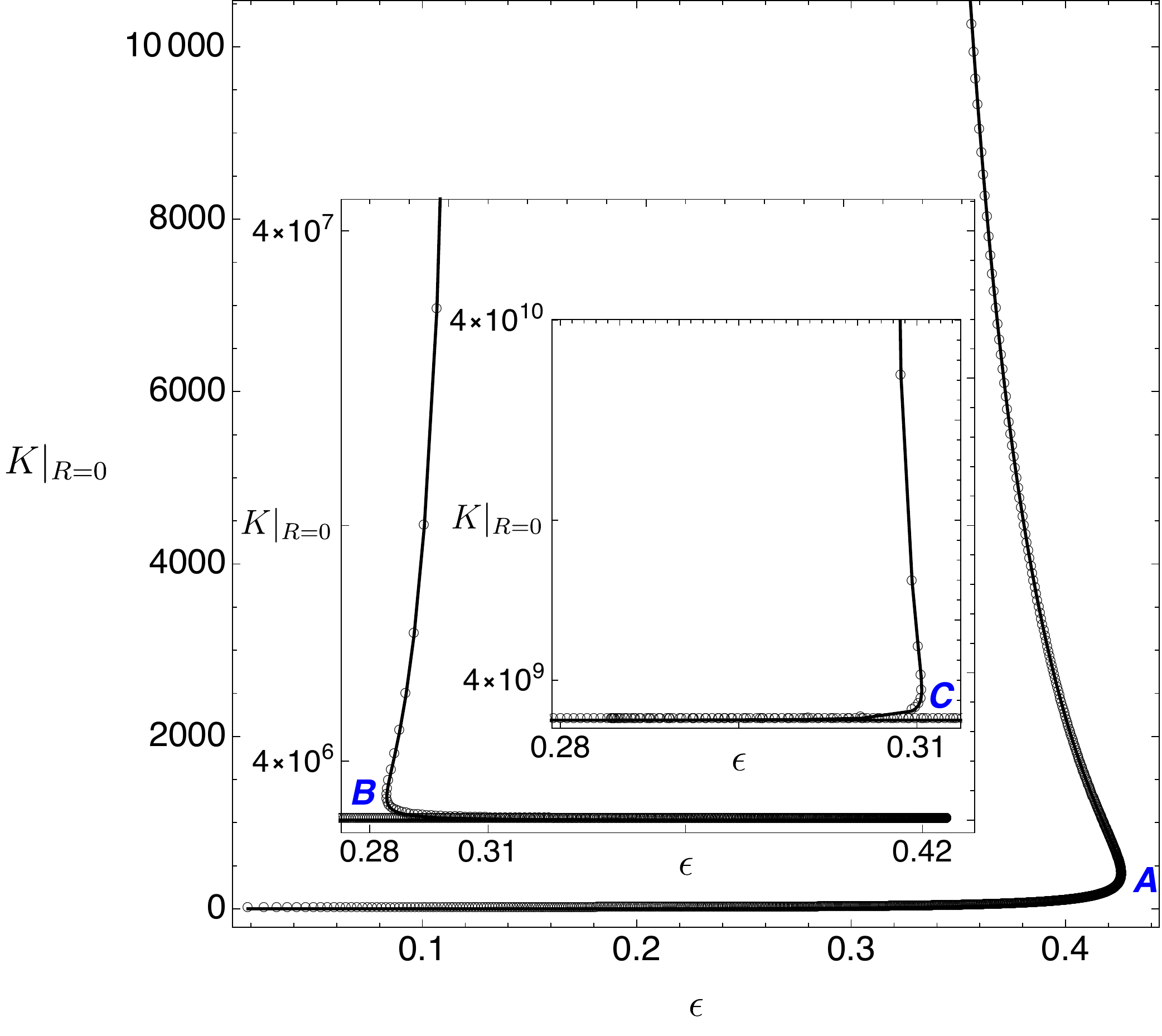}
}
\caption{Main soliton family with $e=0.23$ ($e<e_{\gamma}$).  {\bf Left panel:} As we ramp up the value of the amplitude $\epsilon\equiv \phi'(1)$ the core value of the scalar field $\phi$ is increasing seemingly without bound. The inset plot zooms in the damped oscillations around the turning points $B,C,D$. {\bf Right panel:}  Kretschmann invariant at the origin as a function of $\epsilon$. The turning points $A$, $B$, $C$, $D$ are seeing a substantial increase in the curvature as we go from one to the next. This is despite the fact that they lie close to each other in the phase diagram. The inset plots zoom in around the point $B$ and then around the point $C$ to be able to show more clearly the large growth that $K$ experiences (note the different scales in which plot).}
\label{FIGe0.23:phiK}
\end{figure} 

The zig-zagged structure in the quasilocal charge-mass plot translates into a damped oscillatory behaviour, when we plot the fields $f$ and $\phi$ (or even the Kretschmann) at the origin as a function of the scalar condensate amplitude $\epsilon\equiv \phi'(1)$ (see Figs.~\ref{FIGe0.23:fA}-\ref{FIGe0.23:phiK}), or into a spiral motion in the $A(0)-\epsilon$ plane (see right panel of Fig.~\ref{FIGe0.23:fA}); see also plots in the Appendix. Altogether, this reveals the existence of a  a Chandrasekhar limit at $\mathcal{Q}=\mathcal{Q}_{Ch} \approx 0.0094 L$ (point $A$ in Figs.~\ref{FIGe0.23:MassCharge}-\ref{FIGe0.23:phiK}). We can understand this limit as follows. For very small $e$, the soliton is essentially an almost neutral boson star and the latter is expected to have a Chandrasekhar limit at a critical mass. Expanding in $e$,  the soliton charge $\mathcal{Q}$ can be measured by the gauge field sourced by the above weakly charged boson star and so it should be of order $e$, \ie the Chandrasekhar quantities $(\mathcal{M}_{Ch},\mathcal{Q}_{Ch})$ should grow linearly with $e$ (for small $e$) and then keep growing with $e$.
We find this is indeed the case (as partially illustrated in Fig.~\ref{FIGseveral:MassCharge}) as long as $e<e_c$ (to be discussed later; as we cross $e_c$ into $e>e_c$, $\mathcal{Q}_{Ch}$ jumps {\it discontinuously} to the maximum charge that can fit inside the box).

At the Chandrasekhar limit (point $A$ in Fig.~\ref{FIGe0.23:MassCharge}) the family of solitons reaches a regular cusp and afterwards, it develops a sequence of zig-zagged branches $AB$,  $BC, \cdots$  splitted by new regular cusps $B, C, D,\cdots$ As we move along these, the parameter  $f_0=f(0)$ is increasingly approaching zero, which signals that one is approaching a naked singularity (see left panel of Fig.~\ref{FIGe0.23:fA}). This conclusion is further corroborated by the fact that the  Kretschmann curvature at the origin, $K\big|_{R=0}=R_{abcd}R^{abcd}\big|_{R=0}$, is growing unboundedly as we follow the path $OABCD\cdots$. It becomes increasingly harder to follow this soliton family beyond point $D$ but the above results suggest that the soliton might well go through an infinite sequence of cusps (in the charge-mass plot) and an infinite sequence of damped oscillations or spirals in the plots of Figs.~\ref{FIGe0.23:fA}-\ref{FIGe0.23:phiK}, before it finally reaches a naked singularity configuration where $f\big|_{R=0}\to 0$ and $K\big|_{R=0}\to \infty$.

The main soliton family of solutions has similar properties to the ones displayed in  Figs.~\ref{FIGe0.23:MassCharge}-\ref{FIGe0.23:phiK} for any $e<e_c$, although it becomes harder to capture higher order cusps as $e$ increases. 

\subsection{$e_{\gamma}<e<e_c\simeq 1.8545\pm 0.0005$ \label{subsec:GammaC}}

\begin{figure}[t]
\centerline{
\includegraphics[width=.49\textwidth]{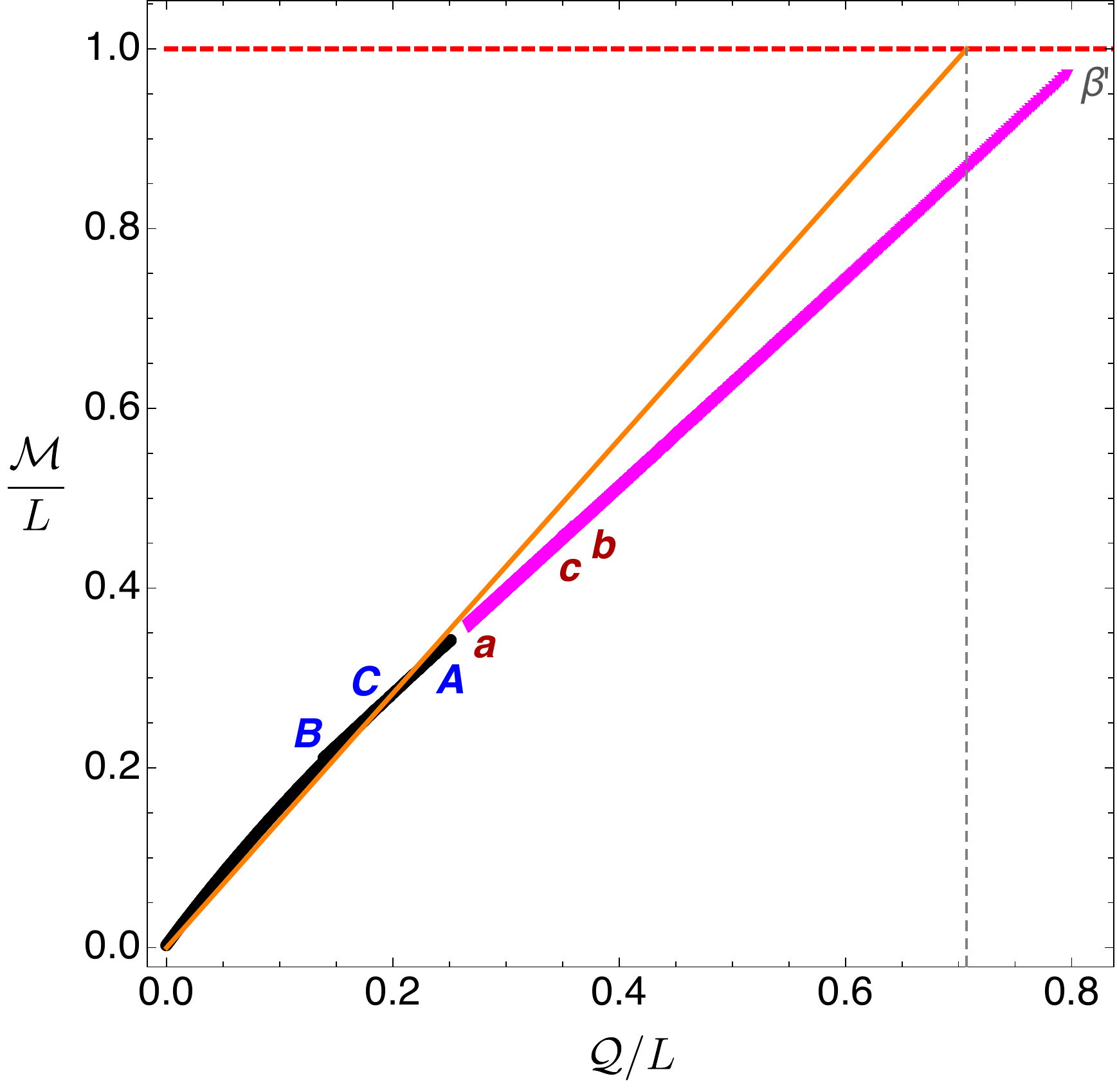}
}
\centerline{
\includegraphics[width=.505\textwidth]{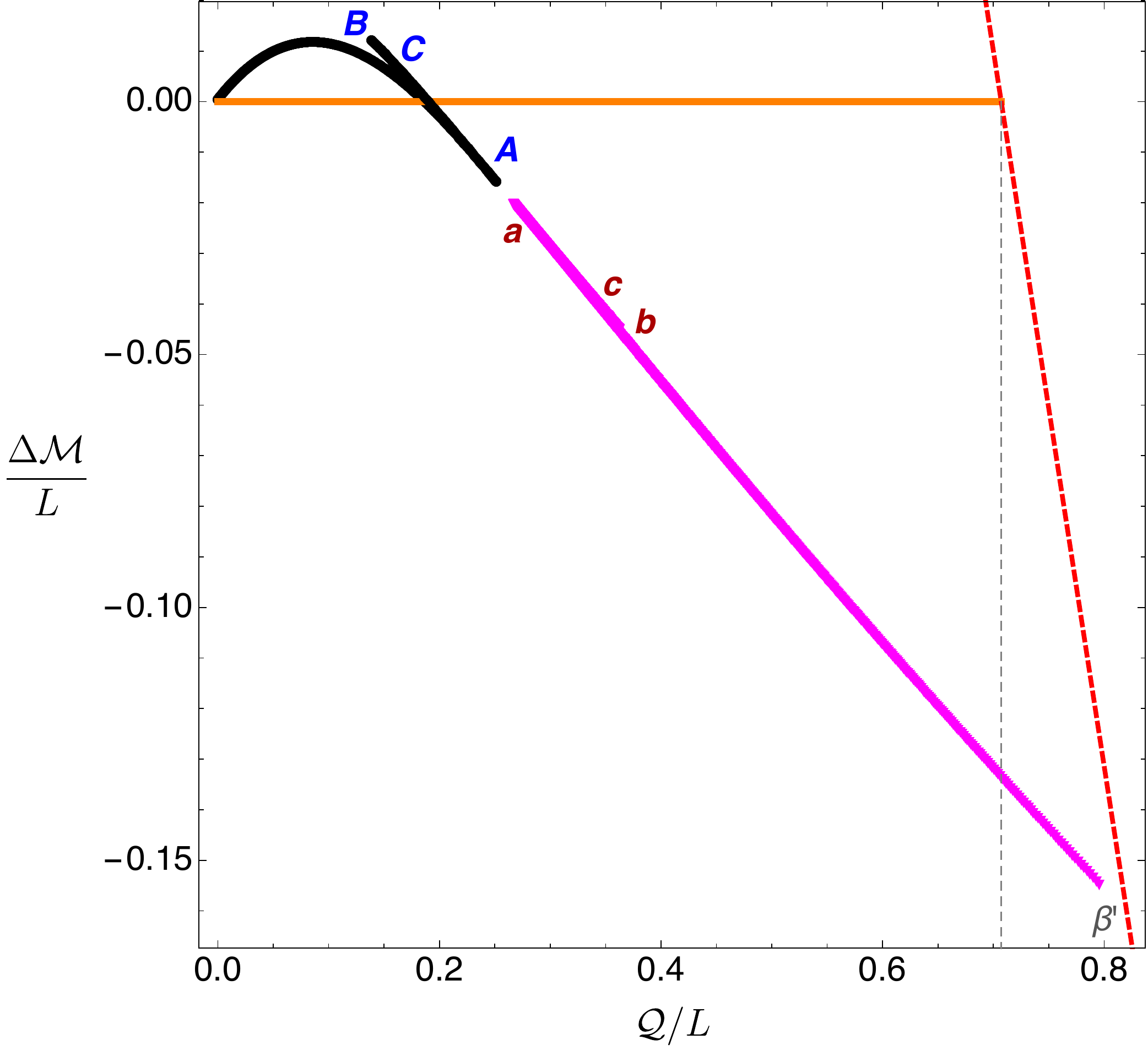}
\hspace{0.3cm}
\includegraphics[width=.505\textwidth]{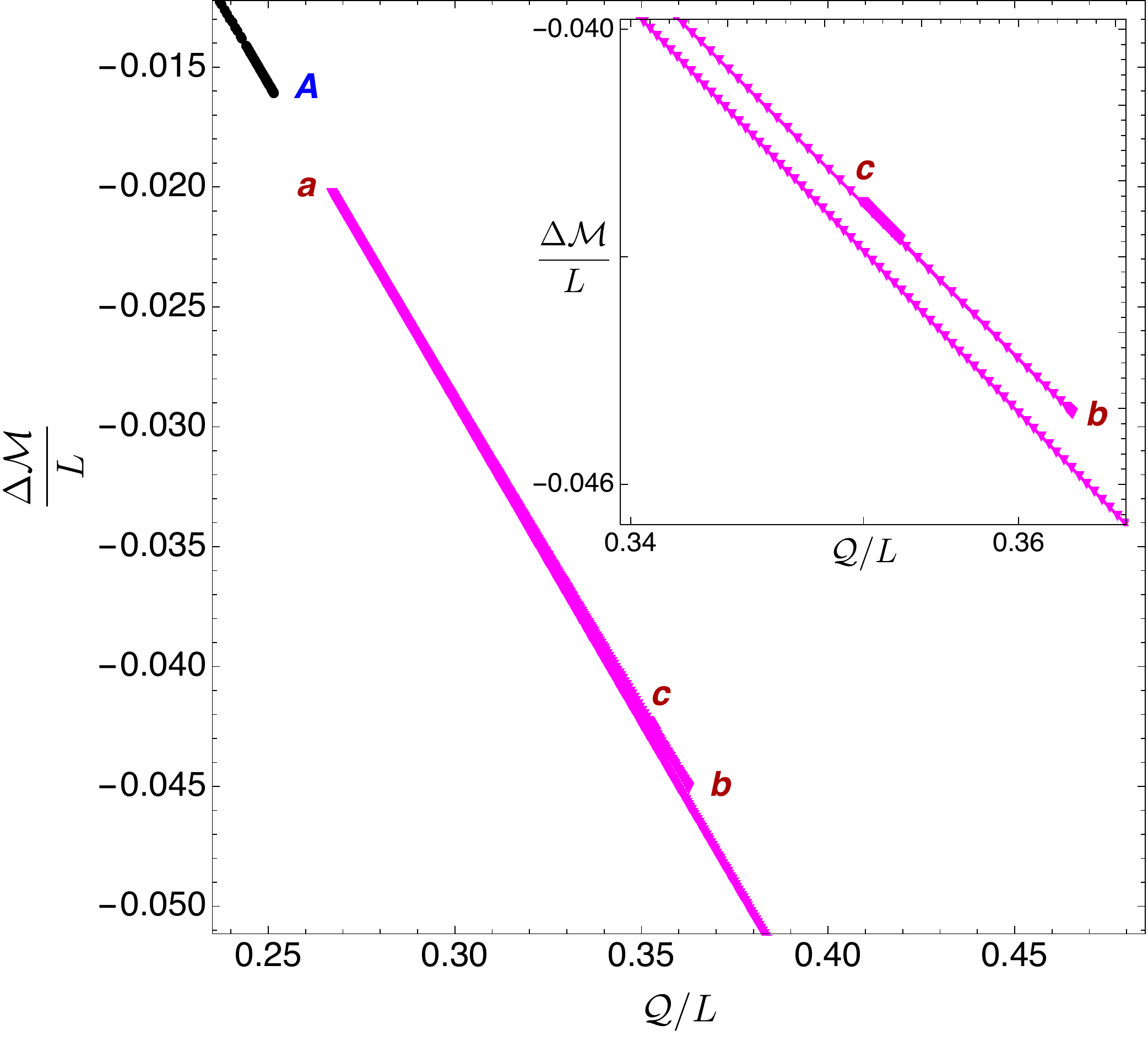} }
\caption{Soliton families with $e=1.854$ ($e_{\gamma}<e<e_c$). {\bf Top panel:} quasilocal mass {\it vs} charge. The main family (black curve $OABC\cdots$) looks to be on top of the extremal RN BH (orange line) which is why we find it useful to plot the quantity $\Delta{\cal M}$ which is shown on the bottom panel. {\bf Bottom-left panel:}  the main black soliton curve starts more massive ($\Delta\mathcal{M}>0$) than the extremal RN but then becomes less massive ($\Delta\mathcal{M}<0$) above a certain charge. The secondary soliton family (magenta curve) eventually hits the dashed red line at $\beta'$ and  extends to a finite lower charge $\mathcal{Q}$ where it develops a series of cusps $a,b,c,\cdots$. {\bf Bottom-right panel:} zoom of the left plot to amplify the gap $Aa$ between the main and secondary families (magenta) and the cusp structure $abc \cdots$ of the secondary family.}
\label{FIGe1.854:MassCharge}
\end{figure} 

\begin{figure}[t]
\centerline{
\includegraphics[width=.49\textwidth]{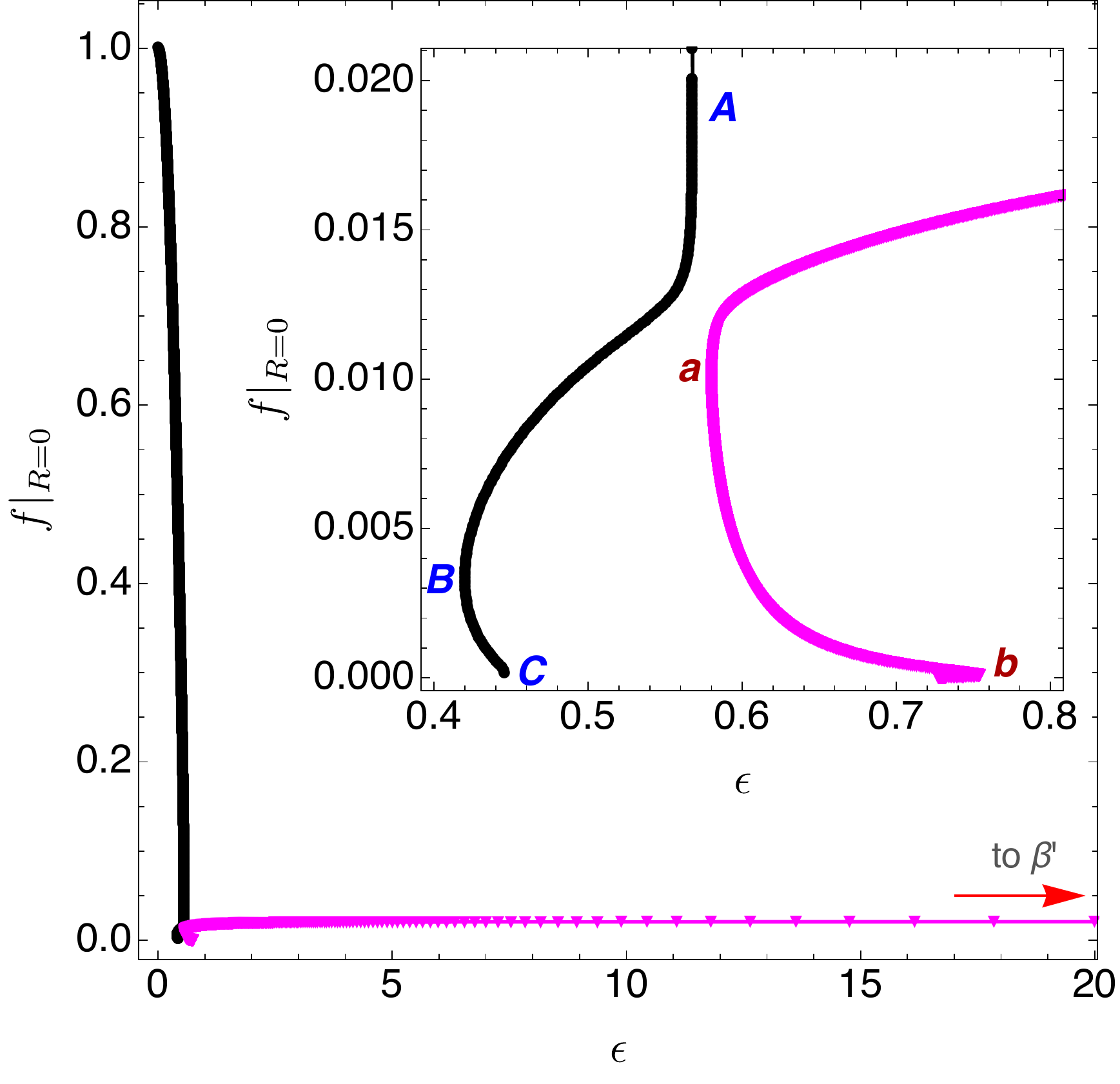}
\hspace{0.3cm}
\includegraphics[width=.49\textwidth]{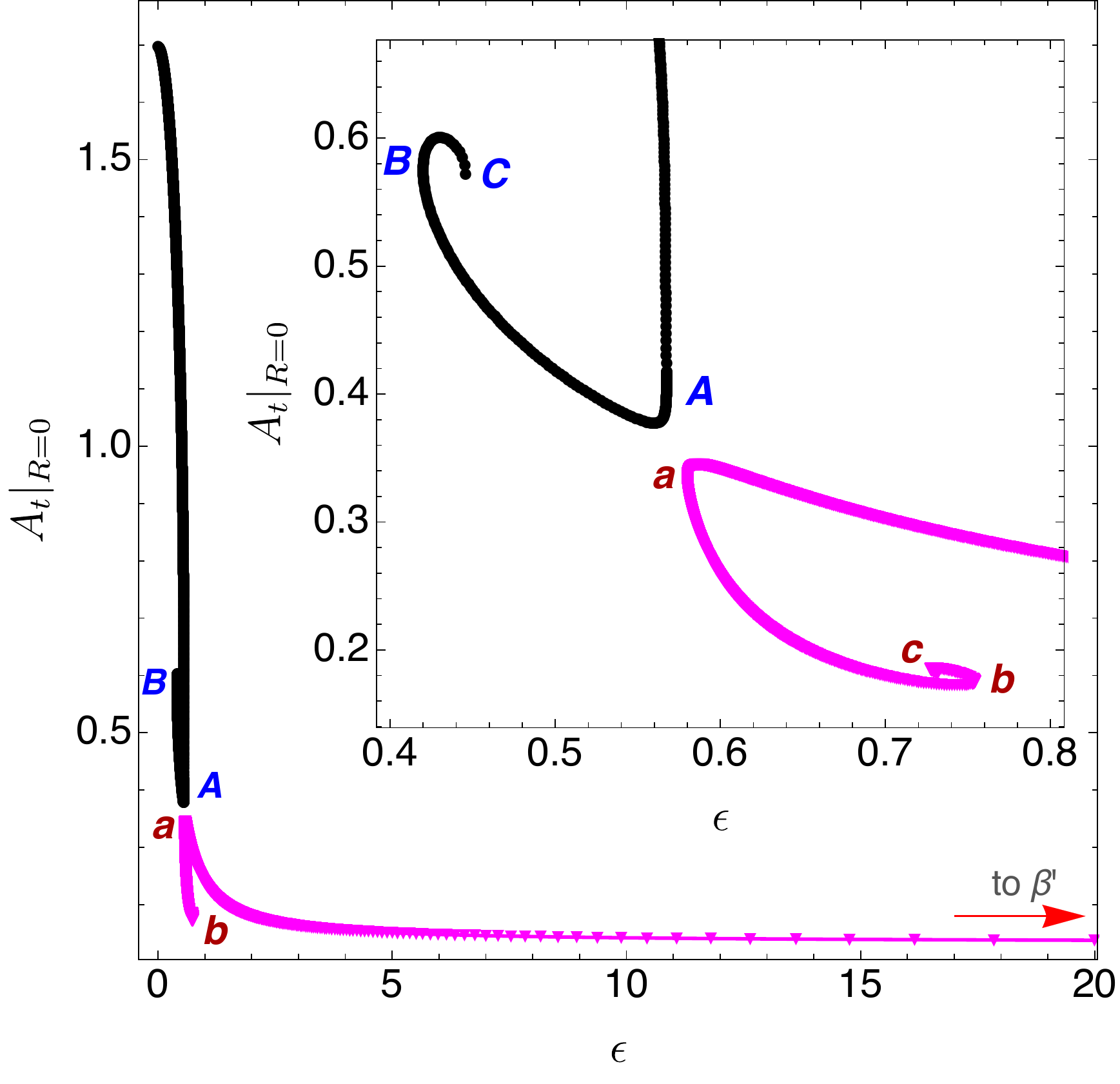}
}
\caption{Soliton families with $e=1.854$ ($e_{\gamma}<e<e_c$). As in previous plots, the black curve represents the main family and the magenta line represents the secondary family. In both figures, the inset plots zooms in around the region where the two soliton families approach the most. As $e\to e_c^-$, the gap between $A$ and $a$ decreases until these two points merge precisely at $e=e_c$. {\bf Left panel:} The function $f$ evaluated at the origin of the soliton is plotted as a function of the amplitude $\epsilon$. The two families seem to be on top of each other for a small range of $\epsilon$ but the inset plot shows that this is not the case and a small gap is still present while $e\neq e_c$. {\bf Right panel:} The gauge field $A_t$ evaluated at the origin as a function of $\epsilon$. Both families spiral inward towards a singularity in one of their endpoints: this corresponds to the cusp structure observed in the $\mathcal{Q}-\Delta\mathcal{M}$ plot. }
\label{FIGe1.854:fA}
\end{figure} 

\begin{figure}[t]
\centerline{
\includegraphics[width=.48\textwidth]{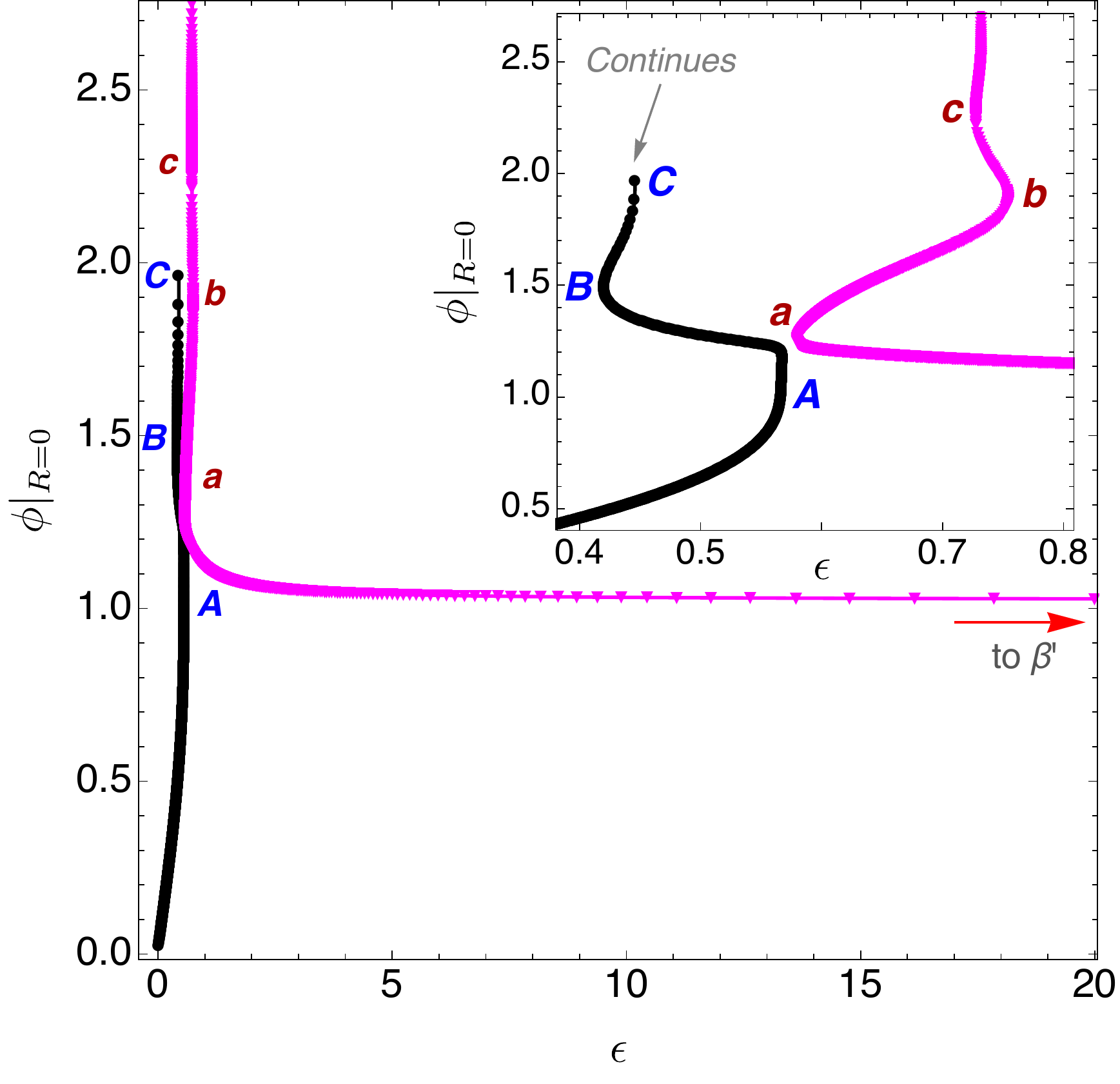}
\hspace{0.3cm}
\includegraphics[width=.515\textwidth]{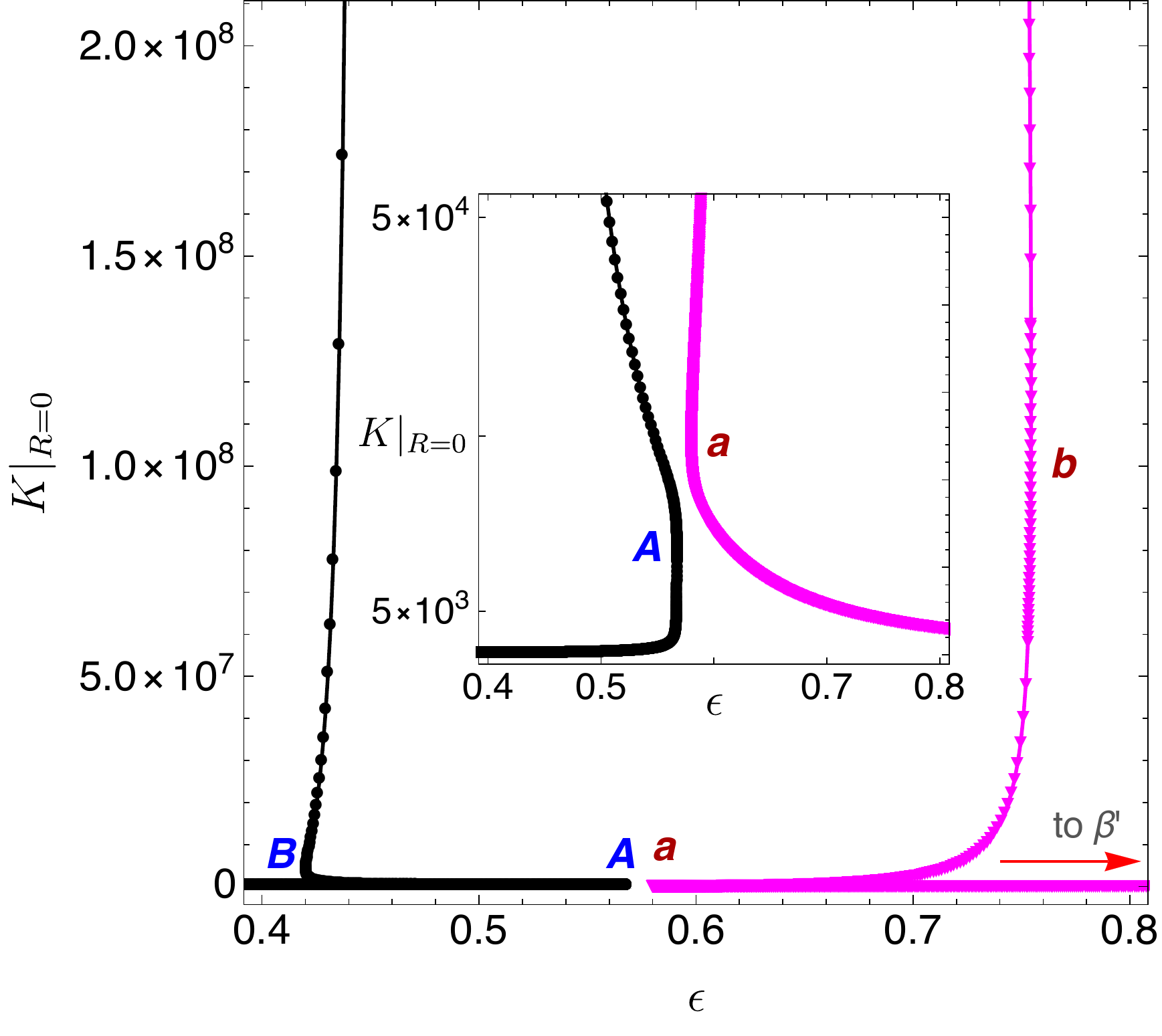}
}
\caption{Soliton families with $e=1.854$ ($e_{\gamma}<e<e_c$) with the main (black disks) and secondary (magenta triangles) families. In both figures, the inset plots zooms in around the region where the two soliton families approach the most. As $e\to e_c^-$, the gap between $A$ and $a$ decreases until these two points merge precisely at $e=e_c$. {\bf Left panel:}  The core value of the scalar field is plotted as a function of $\epsilon$. {\bf Right panel:}  The value of the Kretschmann at the origin is plotted as a function of $\epsilon$. Both families have diverging Kretschmann as one moves along the damped oscillation trenchs.}
\label{FIGe1.854:phiK}
\end{figure} 

We now describe the solitons of the system for a scalar field charge in the window  $e_{\gamma}<e<e_c\simeq 1.8545\pm 0.0005$. We choose to discuss the particular case $e = 1.854 \lesssim e_c$ for reasons that will soon be clear (further examples all the way down to $e\to e_{\gamma}\sim 1.13$ can be found in the right panel of Fig.~\ref{FIGseveral:MassCharge}). For the plots associated with this scalar charge, the black curve represents what we call the `main family' (as mentioned in the introduction this family is perturbatively connected to a Minkowski box) and the magenta curve represents a distinct `secondary family' which is \textit{not} perturbatively connected to a Minkowski box but still describes genuine regular solitonic solutions to the theory. 

The main soliton family for these values of the charge is qualitatively similar to what we had for $e = 0.23$: in the quasilocal charge-mass plane of Fig.~\ref{FIGe1.854:MassCharge}, this black disk curve extends from the origin $O$, developing cusps $A,B,C \cdots$. The discussion presented for $e=0.23$ applies {\it mutatis mutandis} to the present case. As a minor observation, note however that there is now a window of quasilocal charges (around the Chandrasekhar point $A$) for which the main family has a lower mass than an extremal RN BH with the same charge. 

Unlike for $e<e_{\gamma}$, for $e = 1.854$ (and any $e_{\gamma}<e<e_c$) there is `secondary family' of solitons. This is the magenta triangle curve $\beta'abc\cdots$ in Figs.~\ref{FIGe1.854:MassCharge}-\ref{FIGe1.854:phiK}.
In Fig.~\ref{FIGe1.854:MassCharge}, we see that the main soliton family achieves a maximum charge (the Chandrashekhar point $A$) which is $\textit{less than}$ the minimum charge (at point $a$) reached by the secondary soliton family.  
From this plot, we also infer that the secondary family of solitons extends all the way up to $\beta'$ (which is a point  in the red dashed ``wall" that describes the maximum quasilocal charge that a solution confined in the box of radius $R$ can have). In terms of the $\epsilon$ parameter that we use to construct these solutions, this $\beta'$ configuration is only reached in the limit $\epsilon\to \infty$ and this is why it is difficult to extend the magenta curve all the way till it hits the dashed red line (but when we analyse the $e=2.3$ case we will see that solitons indeed extend all the way to the dashed red line).
Still in Fig.~\ref{FIGe1.854:MassCharge},  we also see that the secondary family of solitons has a  cusp/zig-zagged structure $abc \cdots$ that is very similar to the one found in the main family but this time around the regular cusps $a, b, c, \cdots$. In particular, this structure also translates into damped oscillations or spirals in the quantities plotted in Figs.~\ref{FIGe1.854:fA}-\ref{FIGe1.854:phiK}. 
As we move along the cusps $a,b,c,\cdots$, $f_0\equiv f(0)$ is approaching zero and the Kretschmann curvature is growing unbounded. This suggests that this secondary soliton family also ends in a naked singularity after, possibly, going though an infinite sequence of cusp/oscillations/spirals. On the other hand $f_0$ and $K\big|_{R=0}$ remain finite as we approach $\beta'$. In particular, $f_0\big|_{\beta'} \ll 1$.

The gap $Aa$ between the main and the secondary families of solitons in the $\mathcal{Q}-\Delta\mathcal{M}$ plane of Fig.~\ref{FIGe1.854:MassCharge} is a quantity that deserves special attention. We find that this gap decreases as $e$ increases, approaching zero when $e\to e_c$ from below. A similar gap  $Aa$ present in the several plots of  Figs.~\ref{FIGe1.854:fA}-\ref{FIGe1.854:phiK} also decreases and approaches zero as $e\to e_c$. In the next subsection we will further discuss what happens for values of $e$ at $e_c$ and above.

Note that we can have hairy solitons with higher quasilocal charge than the maximum charge that a RN confined inside a box can have. Indeed, the maximum quasilocal charge that a RN confined inside a box (\ie with horizon radius $R_+\leq 1$) can have is the one of an extremal RN BH with horizon radius $R_+=1$, namely $\mathcal{Q}/L=2^{-1/2}$. This is identified by the vertical dashed grey line in Fig.~\ref{FIGe1.854:MassCharge}. But this figure also demonstrates that we can have secondary solitons with a charge that is above this value.

Other examples of solitons with scalar charge in the window $e_{\gamma}<e<e_c$ can be found in the right panel of Fig.~\ref{FIGseveral:MassCharge}. From the several cases considered, we conclude that secondary solitons (that extend to the red dashed line) exist only for $e_{\gamma}<e<e_c$ and inside the region bounded by the closed auxiliary dashed line $a_c\beta_c\gamma$ of Fig.~\ref{FIGseveral:MassCharge} (these auxiliary lines are inserted to guide reader's eye). As $e\to e_c$, the secondary soliton approaches the auxiliary line $a_c\beta_c$, \ie the equivalent of point $a$ and $\beta'$ of Fig.~\ref{FIGe1.854:MassCharge} approach point $a_c$ and $\beta_c$, respectively, of Fig.~\ref{FIGseveral:MassCharge}. On the opposite end of the window, as $e\to e_{\gamma}$, one finds that the equivalent to points $a$ and $\beta'$ of Fig.~\ref{FIGe1.854:MassCharge} collapse to a single point denoted as $\gamma$ in Fig.~\ref{FIGseveral:MassCharge}.

\subsection{$e_c<e<e_{\hbox{\tiny S}}=\frac{\pi }{\sqrt{2}}\sim 2.221$ \label{subsec:CandS}}

\begin{figure}[t]
\centerline{
\includegraphics[width=.49\textwidth]{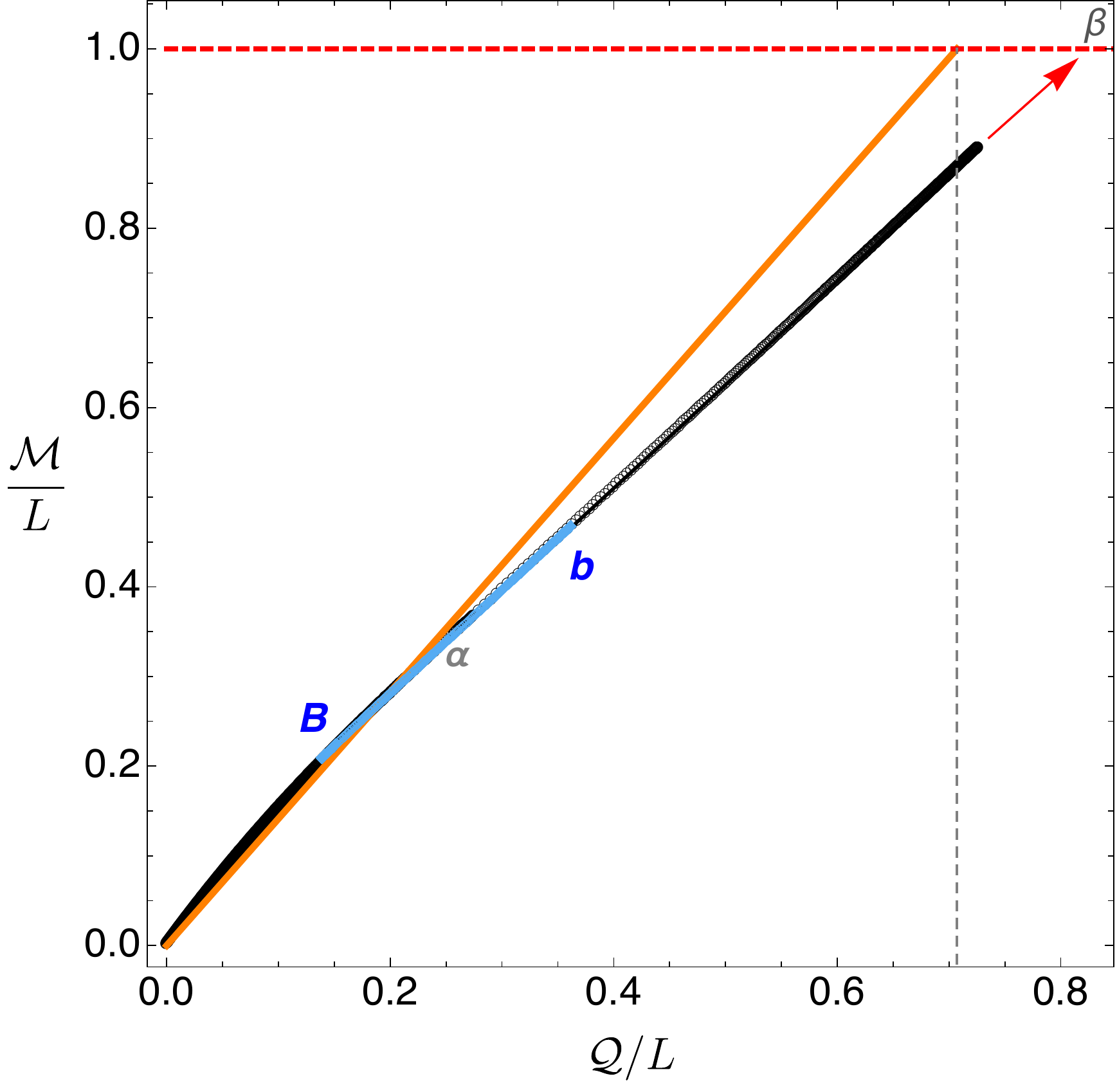}
}
\centerline{
\includegraphics[width=.505\textwidth]{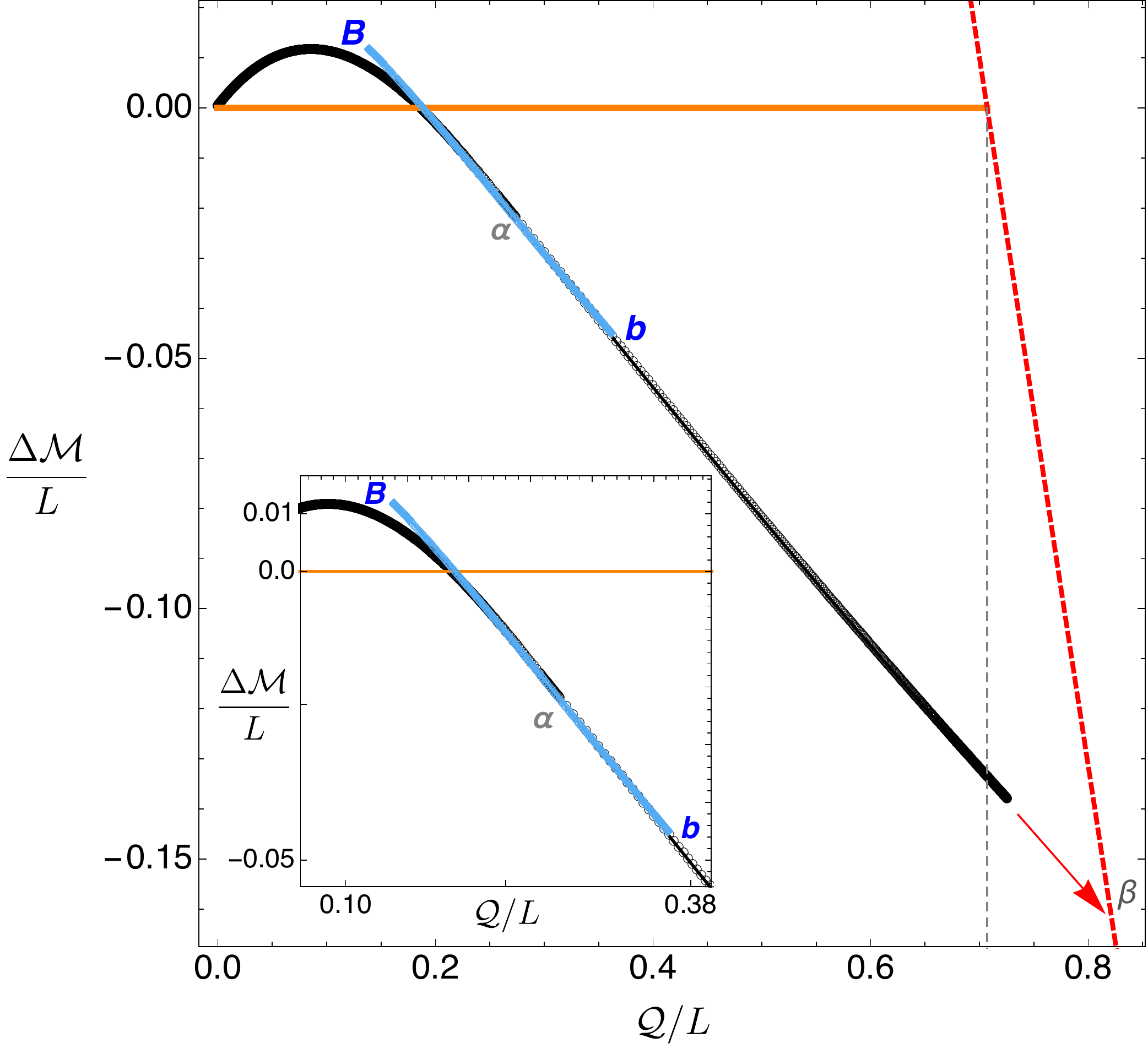}
\hspace{0.3cm}
\includegraphics[width=.505\textwidth]{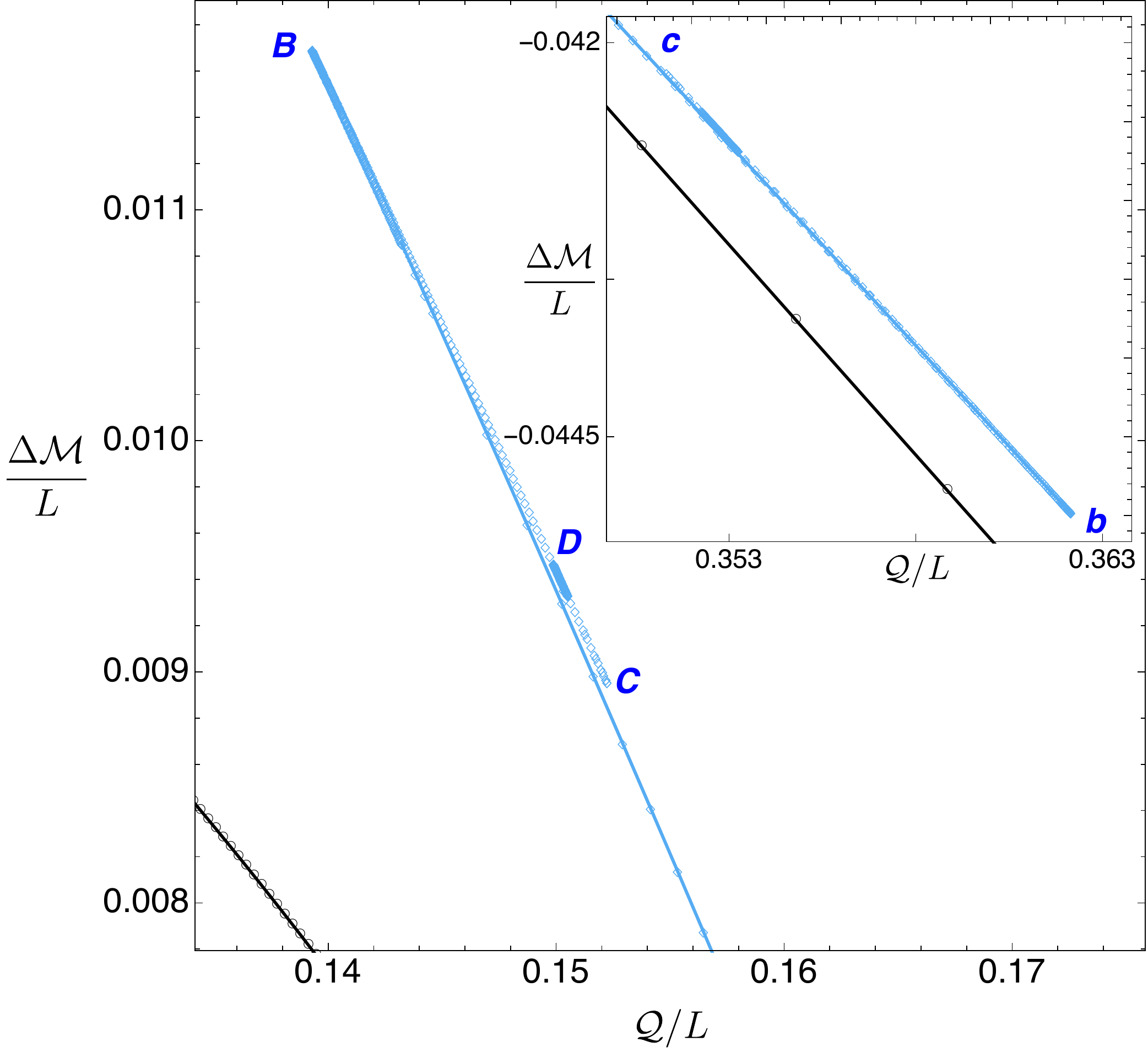} }
\caption{Soliton families with $e=1.855$ ($e_c<e<e_{\hbox{\tiny S}}$). {\bf Top panel:}  the quasilocal charge-mass diagram is now much different from the $e<e_c$ case as better seen in the bottom plots.  {\bf Bottom-left panel:}  for $e>e_c$ the main soliton curve (black disks) no longer has a Chandrasekhar limit; instead it extends from the origin all the way down to point $\beta$ on the red dashed line. On the other hand, the secondary soliton family (blue diamonds curve) is now confined in a window ($Bb$) of quasilocal charge and with mass above the main soliton.  {\bf Bottom-right panel:} zoom of the left plot to amplify the cusp structures $BCD\cdots$ and  $bc \cdots$ that appear on both sides of the secondary soliton family.}
\label{FIGe1.855:MassCharge}
\end{figure}

\begin{figure}[t]
\centerline{
\includegraphics[width=.49\textwidth]{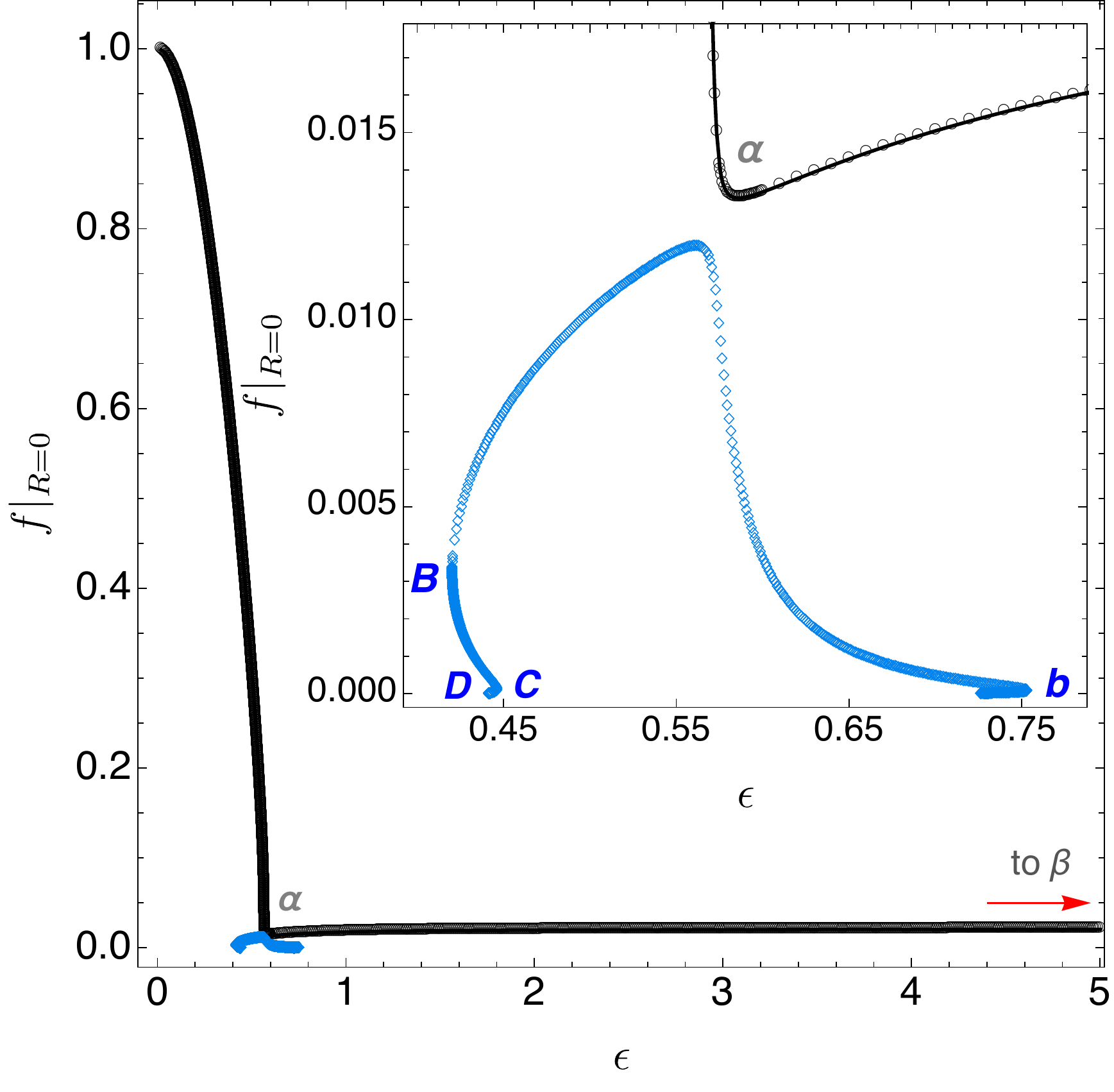}
\hspace{0.3cm}
\includegraphics[width=.49\textwidth]{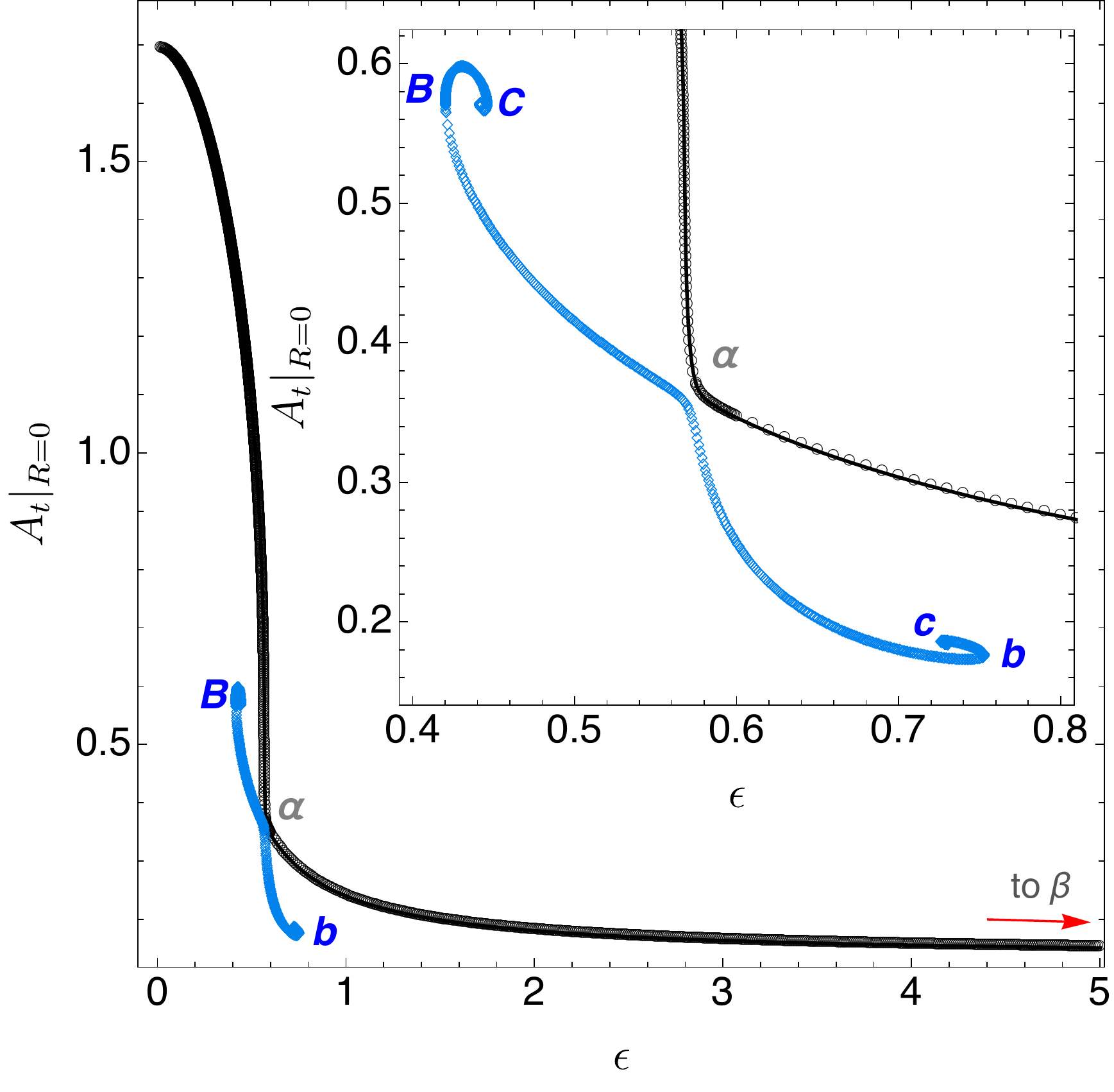}
}
\caption{Soliton families with $e=1.855$ ($e_c<e<e_{\hbox{\tiny S}}$). Comparing with the $e\lesssim e_c$ case of Fig.~\ref{FIGe1.854:fA} we identify main differences: the black curve is still the main soliton family but it now extends to arbitrarily large $\epsilon$. On the other hand, the  secondary soliton family $-$ blue diamond curve  $-$ is now confined to a small window of  scalar condensate amplitude $\epsilon$. Note that as  $e\to e_c^+$, the gap between the two curves, in the region vaguely signalled as $\alpha$, decreases until it vanishes precisely at $e=e_c$. {\bf Left panel:} $f(0)$ as a function of the amplitude $\epsilon$. {\bf Right panel:} The gauge field evaluated at the origin $A_t(0)$ as a function of $\epsilon$. This time the secondary family only exists for a small range of scalar field amplitudes and develops a `double-spiral' structure, \ie a cusp structure develops at either end of the curve, in the trenches $Abc \cdots$ and $BCd \cdots$. }
\label{FIGe1.855:fA}
\end{figure} 

\begin{figure}[t]
\centerline{
\includegraphics[width=.48\textwidth]{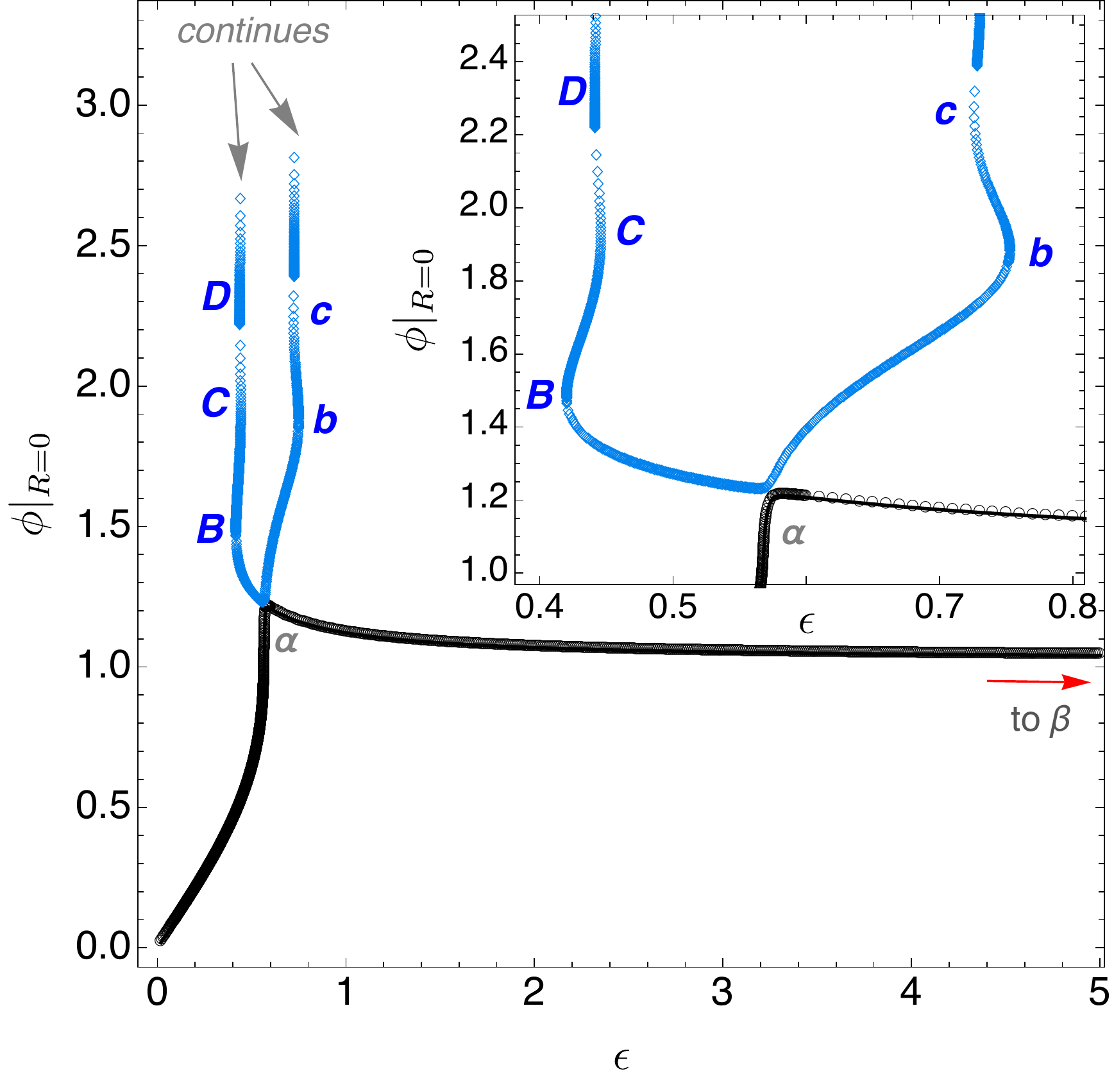}
\hspace{0.3cm}
\includegraphics[width=.515\textwidth]{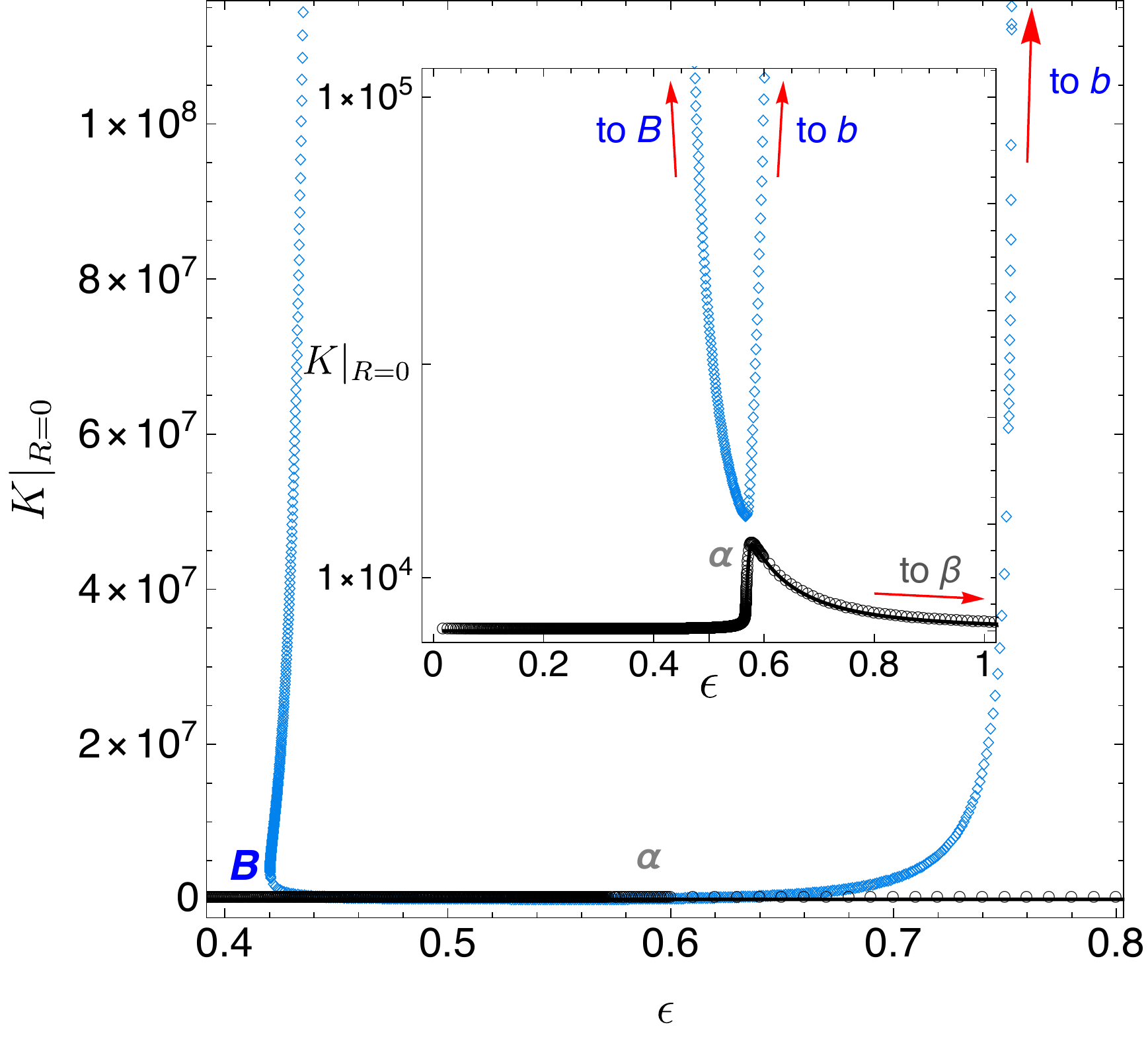}
}
\caption{Soliton families with $e=1.855$ ($e_c<e<e_{\hbox{\tiny S}}$). Alike in the previous figure, comparing the current plots with those of Fig.~\ref{FIGe1.854:phiK}  for the $e\lesssim e_c$ case, we find major differences: the black disk curve is still the main soliton family but it now extends to arbitrarily large $\epsilon$. On the other hand, the  secondary soliton family (blue diamond curve) is now confined to a small window of $\epsilon$. As $e\to e_c^+$, the gap between the two curves, in the region vaguely signalled as $\alpha$, decreases until it vanishes precisely at $e=e_c$. {\bf Left panel:} Core value of the scalar field as a function of $\epsilon$ with damped oscillations seen in both ends of the secondary soliton family.  {\bf Right panel:} The Kretschmann evaluated at the origin as a function of $\epsilon$ for both soliton families. Comparing with Fig.~\ref{FIGe1.854:phiK}  for the $e\lesssim e_c$ case, we  see that now the main soliton family exists for any $\epsilon$ and its curvature at the origin is finite in all its domain of existence. On the other hand, the curvature of the secondary soliton family now diverges at both of its ends.}
\label{FIGe1.855:phiK}
\end{figure}

So, what changes when we cross $e=e_c\sim 1.8545\pm 0.0005$? To address this question we compare the phase diagram for the charge $e=1.854$ just below $e_c$ (Fig.~\ref{FIGe1.854:MassCharge})  with the  one of Fig.~\ref{FIGe1.855:MassCharge} for a charge $e=1.855$ just above $e_c$. From this comparison we can infer the following. Approaching $e_c$ from below,  precisely at $e=e_c$, the cusp $A$ of the main soliton (black disk curve)  merges with the cusp $a$ of the secondary soliton (magenta triangle curve) in Fig.~\ref{FIGe1.854:MassCharge}.  That is to say, at $e=e_c$ the black and magenta curves of Fig.~\ref{FIGe1.854:MassCharge} merge at $A\equiv a$, and this becomes a bifurcation point that irradiates a total of four branches. 

But as $e$ grows above $e_c$, two new independent families emerge. Old point $A\equiv a$ and two of the four branches that irradiate from it at $e=e_c$ are now absorbed in the main soliton family (around the region that we vaguely identify as $\alpha$ in our plots  of Fig.~\ref{FIGe1.855:MassCharge}) that is described by the black disk curve $O\beta$ of Fig.~\ref{FIGe1.855:MassCharge}. This extends from the origin all the way down to point $\beta$ in the dashed red line (which is reached as $\epsilon\to\infty$). That is to say,  for $e>e_c$, the main soliton family no longer has a Chandrashekhar mass limit neither cusps: the Kretschmann curvature is now always finite along this family as it extends from the origin all the way to $\beta$, where the family terminates because it reaches the maximum quasilocal charge that can fit inside the box with dimensionless radius $R=1$. Also note that for $e>e_c$, hairy solitons within the main family can have a quasilocal charge that is higher than the maximum charge $\mathcal{Q}/L=2^{-1/2}$  (identified by the vertical dashed grey line in Fig.~\ref{FIGe1.855:MassCharge}) that a RN confined inside a box can have.

On the other hand, the system now has a novel secondary soliton family (the blue diamond curve $\cdots DCBbc\cdots$ in Fig.~\ref{FIGe1.855:MassCharge}) that exists only in a window of quasilocal charge $\mathcal{Q}/L$ bounded from below by cusp $B$ (that used to belong to the main family for $e<e_c$) and from above by the cusp $b$ (that used to belong to the old secondary magenta triangle family for $e<e_c$). (The distinction between the main black disk family and the secondary blue diamond family is better seen in the plots of Figs.~\ref{FIGe1.855:fA}-\ref{FIGe1.855:phiK}).
This new secondary family absorbs the old point $A\equiv a$ and the two (out of four) branches that irradiate from it at $e=e_c$ that were not embodied in the main family. 
Further confirmation that this new secondary family of soliton derives from connecting the old {\it trench}  $ABCD\cdots$ of the main black disk branch with the old {\it trench} $abc\cdots$ of the secondary magenta triangle branch of Fig.~\ref{FIGe1.854:MassCharge}
can be obtained as follows. Starting at a point in the old trench $ABCD\cdots$ or $abc\cdots$ of  Fig.~\ref{FIGe1.854:MassCharge}, we can run a code where we fix $\epsilon$ or $f_0$ and, {\it this time}, we march the {\it electric scalar charge} $e$ in very small steps from $e=1.854<e_c$ all the way up to $e=1.855>e_c$. Doing this exercise we find that we indeed end on solutions of the blue diamond curve of Figs.~\ref{FIGe1.855:MassCharge}-\ref{FIGe1.855:phiK}. 

Both around $B$ and $b$, the  blue diamond secondary soliton family of Fig.~\ref{FIGe1.855:MassCharge} displays a complex zig-zagged structure that we are already familiar with. In particular, on both sides of this curve, as we move along the several cusps $B,C, D,\cdots$ or $b, c,\cdots$, we find that the Kretschmann curvature is growing unbounded (see Fig.~\ref{FIGe1.855:phiK}) and $f_0\equiv f(0)$ is approaching zero (see Fig.~\ref{FIGe1.855:fA}). Altogether, this suggests that the secondary family of solitons undergoes a sequence of possibly infinite damped oscillations/spirals before it reaches a naked singularity at {\it both} sides of its domain. 
    
The property that, at $e=e_c$, the black and magenta curves of Figs.~\ref{FIGe1.854:MassCharge}-\ref{FIGe1.854:phiK} merge at $A\equiv a$ and then two distinct new soliton families (the black and blue curves of Figs.~\ref{FIGe1.855:MassCharge}-\ref{FIGe1.855:phiK}) emerge for $e>e_c$ is inferred not only from the quasilocal charge-mass plots. Indeed we arrive to the same conclusion  analysing the plots for the fields $f,A_t,\phi$ and Kretschmann when we compare Figs.~\ref{FIGe1.855:fA}-\ref{FIGe1.855:phiK} for $e=1.854\lesssim e_c$ with  Figs.~\ref{FIGe1.855:fA}-\ref{FIGe1.855:phiK} for $e=1.855\gtrsim e_c$ (see in particular their inset plots).

As we keep increasing $e$ above $e_c$ we find that the window of quasilocal charge $\mathcal{Q} \in [0,\mathcal{Q}_0]$ where $\Delta\mathcal{M}\geq 0$, \ie where solitons have a quasilocal mass higher than the extremal RN solution with the same $\mathcal{Q}/L$, decreases and approaches zero as $e\to e_{\hbox{\tiny S}}$ (see next section \ref{subsec:higherS}). We also find that the distance of closest approach between the secondary blue diamond family of solitons and the main black disk family increases in all the plots Figs.~\ref{FIGe1.855:MassCharge}-\ref{FIGe1.855:phiK}. 
There is a scenario that is highly probable: one should expect further critical charges $e_{c_i}>e_c$ ($i\geq 2$) where extra splits in the secondary blue diamond curve  will occur. Indeed, at $e=e_c$ we found that the cusps $A$ and $a$ of Fig.~\ref{FIGe1.854:MassCharge} merge. Similarly, at a new  critical charge $e_{c_2}>e_c$ we might expect that the cusps $C$ and $c$ in the secondary blue curve $\cdots DCBbcd\cdots$ of Fig.~\ref{FIGe1.855:MassCharge} will also merge. If so, above this new merger, \ie for $e>e_{c_2}$, we should have the main family of solitons, then a {\it closed} curve secondary family of solitons $CBbc$ with $C\equiv c$ and then a third family of solitons with the ``leftover" $\cdots Dd\cdots$ that should have an infinite sequence of cusps. We could then have a third critical charge $e_{c_3}> e_{c_2}$ where a new merger of cusps would occur. Then, above it we would end with the main family of solitons, then {\it two closed} families of solitons and a {\it fourth open} family of solitons with an infinite sequence of solitons and so on for   $e_{c_i}, i>4$ (additionally, as discussed before there is also an infinite tower of excited families of solitons besides this ground state families). It is very hard to numerically confirm this scenario. But, starting with our solutions with $e=1.855$, fixing $f_0$ or $\epsilon$ and varying $e$ to march the solutions in the blue curve of Fig.~\ref{FIGe1.855:MassCharge}, we found evidence that this scenario is very plausible. Further evidence in favour of this scenario is provided by the fact that in the context of hairy AdS solitons, closed families of solitons were found (see \eg Figs.~5 and 7 of \cite{Gentle:2011kv}).

\subsection{$e>e_{\hbox{\tiny S}}=\frac{\pi }{\sqrt{2}}\sim 2.221$ \label{subsec:higherS}}

Strictly speaking the only critical charge that marks a substantial qualitative difference on the phase diagram of solitons is $e=e_c$. In this sense, for our discussion it would be enough to provide the illustrative cases $e=1.854<e_c$ (Figs.~\ref{FIGe1.854:MassCharge}-\ref{FIGe1.854:phiK})  and  $e=1.855>e_c$ (Figs.~\ref{FIGe1.855:MassCharge}-\ref{FIGe1.855:phiK}). However, in this section we also discuss a change that occurs in the main soliton family of solitons when $e$ increases above $e_{\hbox{\tiny S}}=\frac{\pi }{\sqrt{2}}\sim 2.221$. Although from a pure soliton perspective this difference seems to be minor, it turns out to have a large impact when we consider the phase diagram of the theory that includes not only the solitons and RN BHs but also the hairy BHs.

\begin{figure}[t]
\centerline{
\includegraphics[width=.49\textwidth]{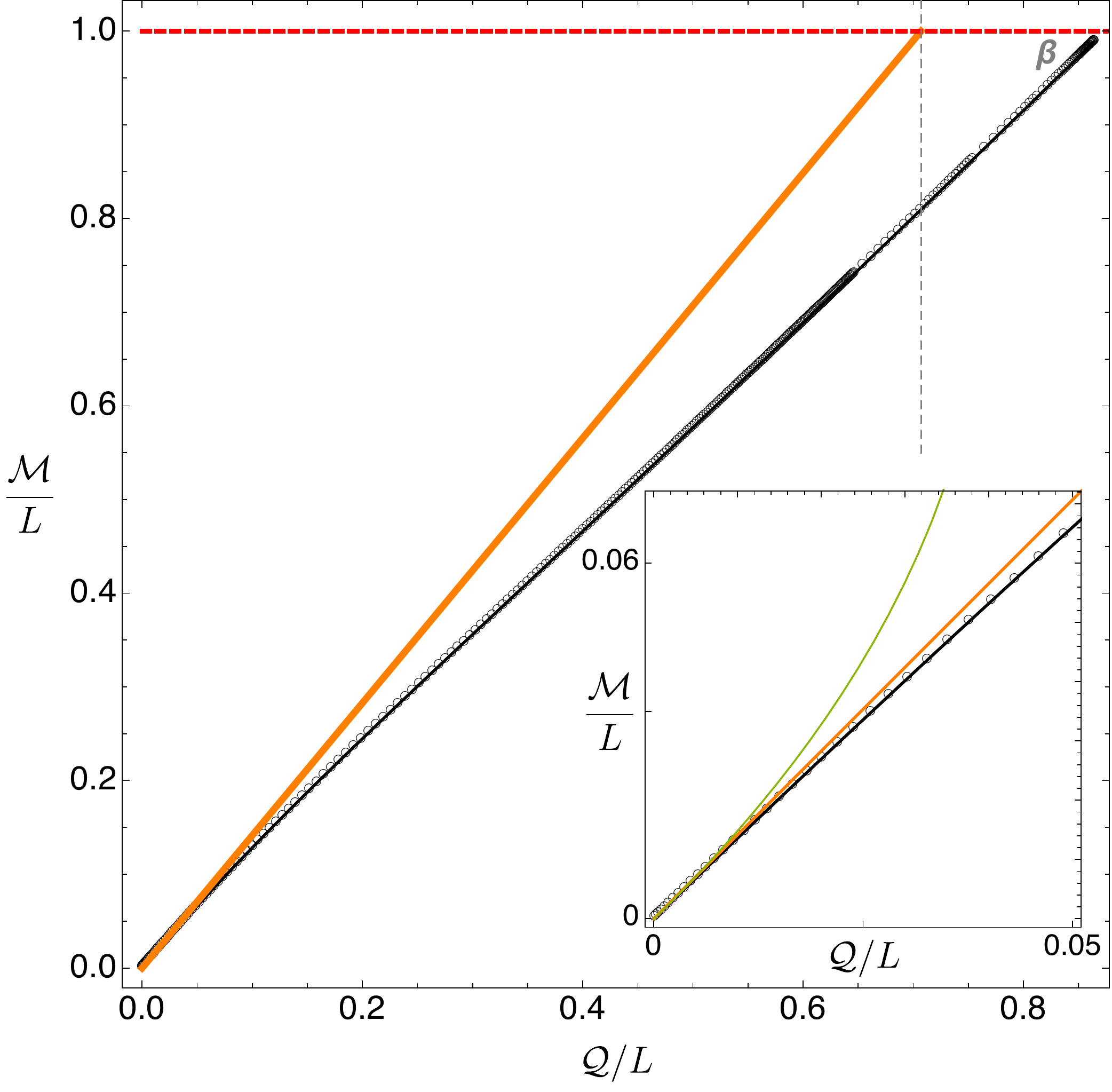}
\hspace{0.3cm}
\includegraphics[width=.505\textwidth]{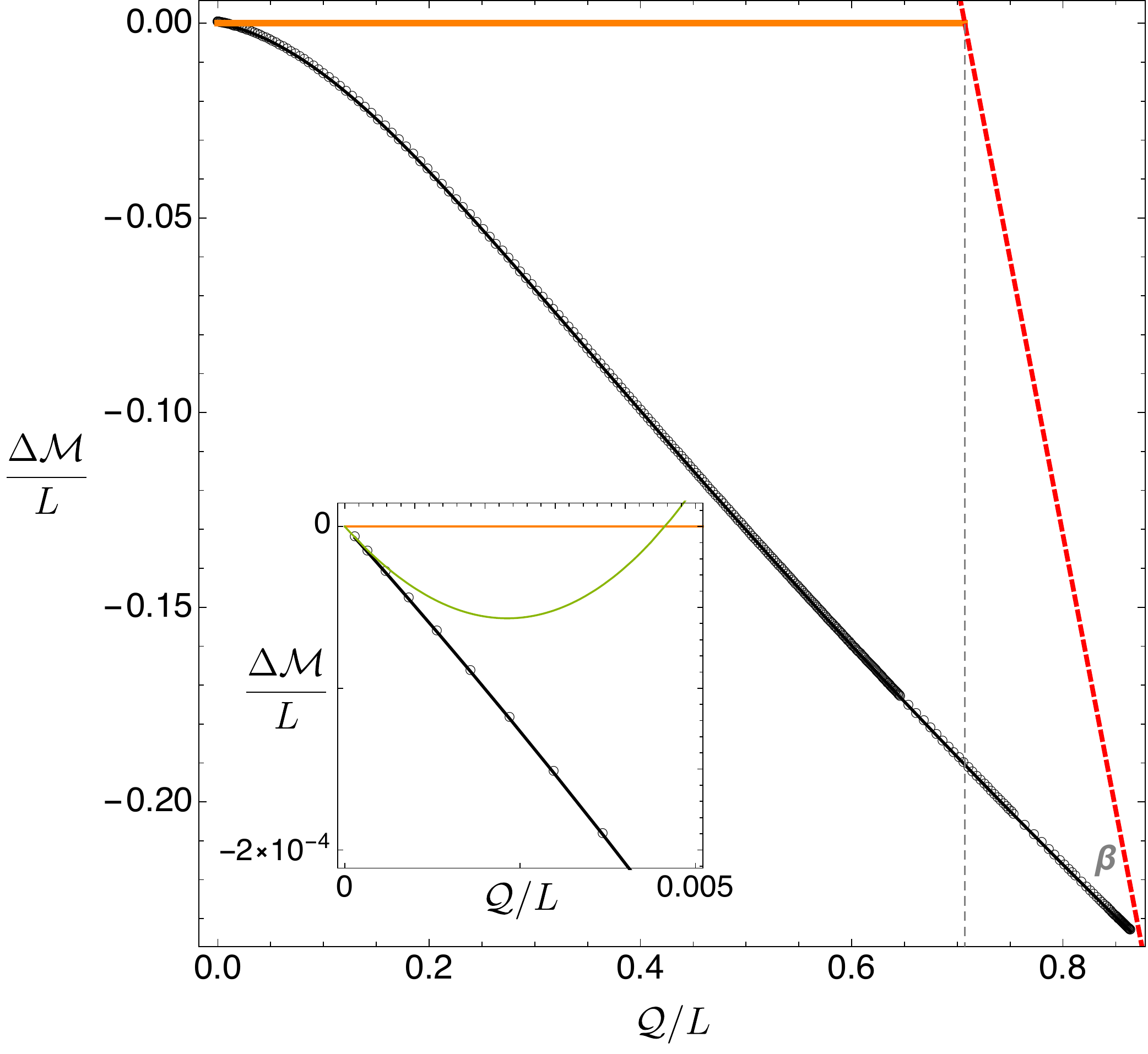}
}
\caption{Main soliton family with $e=2.3$ ($e>e_{\hbox{\tiny S}}$). The soliton extends smoothly from the origin  to $\beta$ and, unlike for $e<e_{\hbox{\tiny S}}$,  the main soliton is now {\it always} less massive than its extremal RN counterpart. This case also demonstrates better a claim that we made for other cases $e$: as the scalar condensate amplitude grows unbounded, $\epsilon \to \infty$, the main soliton approaches point $\beta$ in the red dashed line.
As in other plots, the inset plots show the good agreement between the full result and the analytic prediction (green curve) at small charges.}
\label{FIGe2.3:MassCharge}
\end{figure}

We choose scalar field charge $e = 2.3$ to illustrate the properties of solitons with $e>e_{\hbox{\tiny S}}$. The associated quasilocal charge-mass phase diagram for the main soliton is displayed in Fig.~\ref{FIGe2.3:MassCharge}. Unlike for $e<e_{\hbox{\tiny S}}$, the main solitons now have less quasilocal mass than the extremal RN BH with the same quasilocal charge, \ie one always has $\Delta\mathcal{M}<0$. For $e<e_{\hbox{\tiny S}}$, in a neighbourhood of the origin one has $\Delta\mathcal{M}>0$ but precisely at $e=e_{\hbox{\tiny S}}$, one finds that the slope $\Delta\mathcal{M}/\mathcal{Q}$ vanishes at the origin and then becomes negative for  $e>e_{\hbox{\tiny S}}$. Why is this feature relevant? Well, precisely for $e\geq e_{\hbox{\tiny S}}$ extremal RN BHs are unstable for $\mathcal{M}\geq 0$, \ie the instability onset curve extends all the way down to $\mathcal{M}=0$ (while for $e< e_{\hbox{\tiny S}}$, this onset curve starts at the extremal RN BH at a {\it finite} value of  $\mathcal{M}$). 

These two properties (solitons have $\Delta\mathcal{M}<0$ and RN BHs are unstable all the way down to $\mathcal{M}=0$) strongly suggests that for $e>e_{\hbox{\tiny S}}$, in the charge-mass phase diagram, the hairy charged BHs that emerge from the instability onset of the RN solution ---which is a curve in Fig.~\ref{FIGe2.3:MassCharge} with positive $\Delta\mathcal{M}$ that approaches $\Delta\mathcal{M}=0$ at $\mathcal{Q}=0$ and $\mathcal{Q}=2^{-1/2}$ (see. Fig.~3 of \cite{Dias:2018zjg}) --- should extend all the way down to the soliton (black curve $O\beta$ in Fig.~\ref{FIGe2.3:MassCharge}) when the hairy BH horizon radius goes to zero. This was confirmed for small charges in the perturbative analysis of \cite{Dias:2018yey} and this expectation will also prove correct for higher quasilocal charges (in an ongoing project where the hairy BHs of the theory are constructed at full nonlinear level \cite{DaveyDiasRodgers:2021}). On the other hand, for $e< e_{\hbox{\tiny S}}$ hairy BHs should only exist above a critical quasilocal charge and the main family of hairy solitons is not the zero entropy limit of the hairy BHs of the theory, at least for small charges (and thus these hairy BHs cannot be constructed within perturbation	 theory for small charges \cite{Dias:2018yey}; one needs a full nonlinear numerical analysis that will be provided in \cite{DaveyDiasRodgers:2021}).   

In Fig.~\ref{FIGe2.3:MassCharge} we also find strong evidence to a claim we made previously: as $\epsilon\to\infty$ the main soliton extends all the way to point $\beta$. This point is along the red dashed line that describes the maximal quasilocal  charge that solutions confined inside a box of dimensionless radius $R=1$ can have. For smaller $e$, it is much harder to construct solutions with higher $\epsilon$ (compare the maximum values of $\epsilon$ in the several plots). This is why in the plots for smaller $e$ that we have shown previously, the `last' solution is not so close to $\beta$ as in Fig.~\ref{FIGe2.3:MassCharge}.

Overall, the reader can find a broad overview of the  evolution (as $e$ increases from a small value $e\ll e_{\hbox{\tiny S}}$ all the way above $e_{\hbox{\tiny S}}$) of the phase diagram for the main soliton family in Fig.~\ref{FIGseveral:MassCharge}.

\begin{figure}[t]
\centerline{
\includegraphics[width=.49\textwidth]{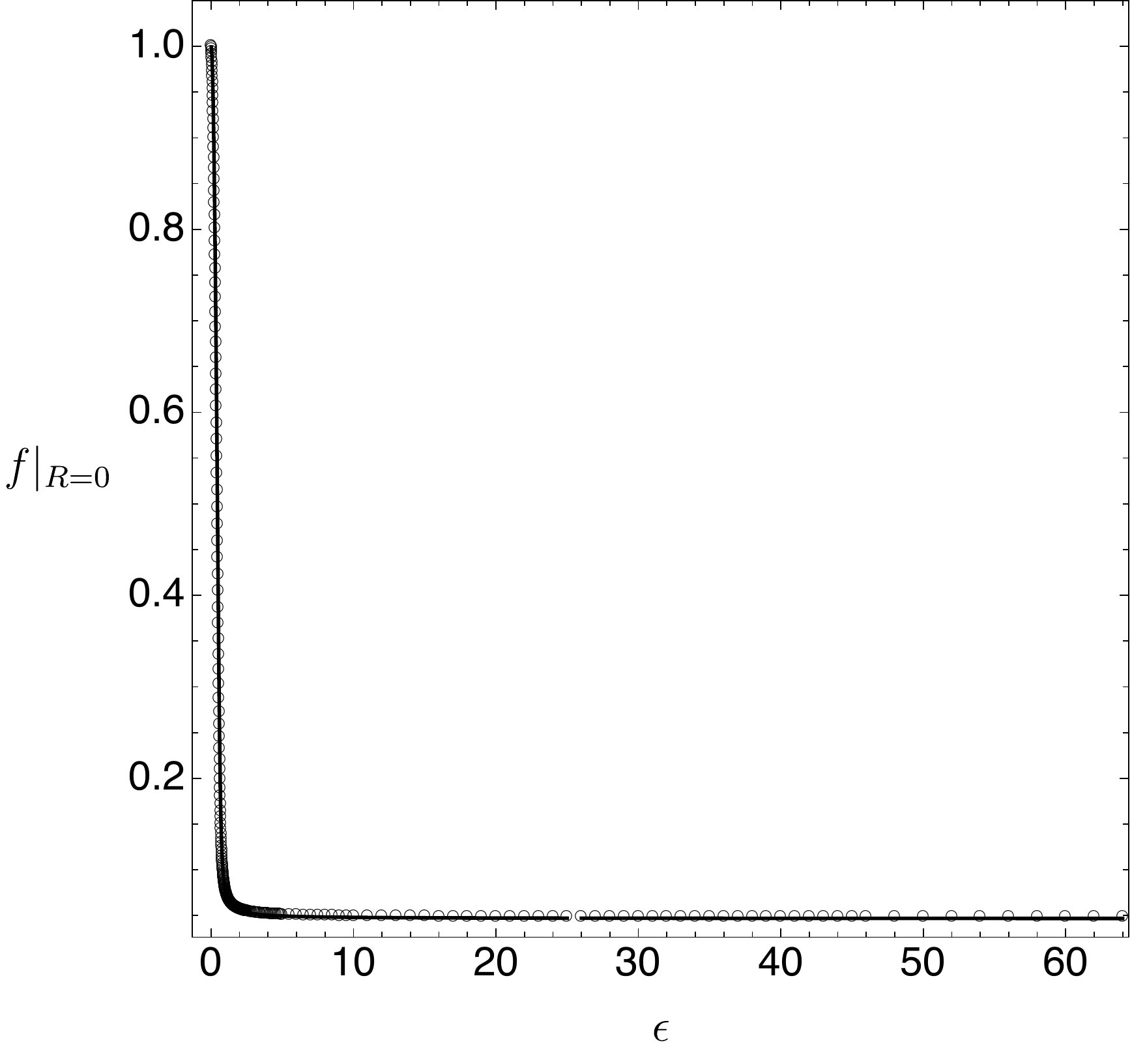}
\hspace{0.3cm}
\includegraphics[width=.49\textwidth]{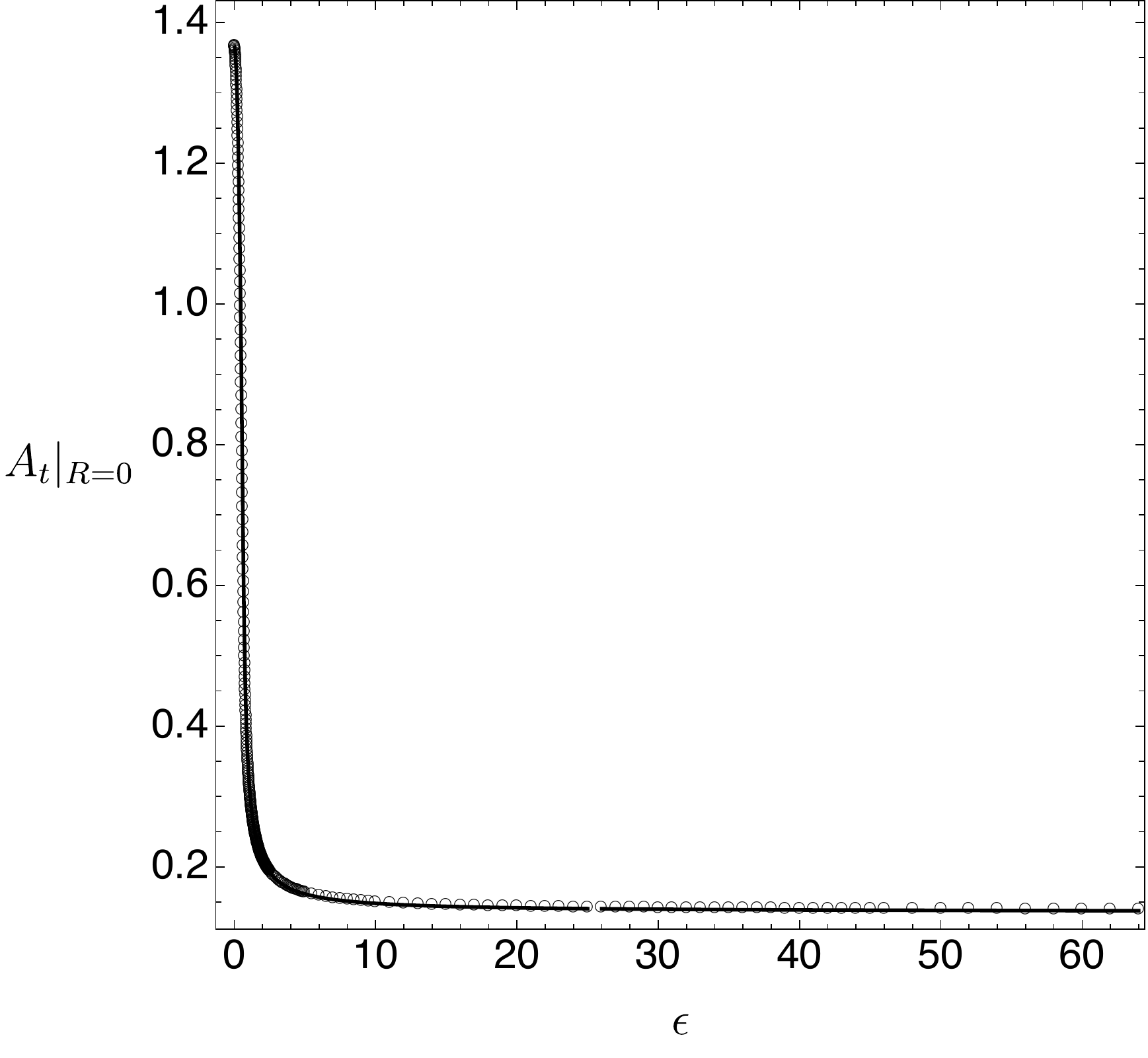}
}
\caption{Main soliton family with $e=2.3$ ($e>e_{\hbox{\tiny S}}$).}
\label{FIGe2.3:fA}
\end{figure}

\begin{figure}[t]
\centerline{
\includegraphics[width=.48\textwidth]{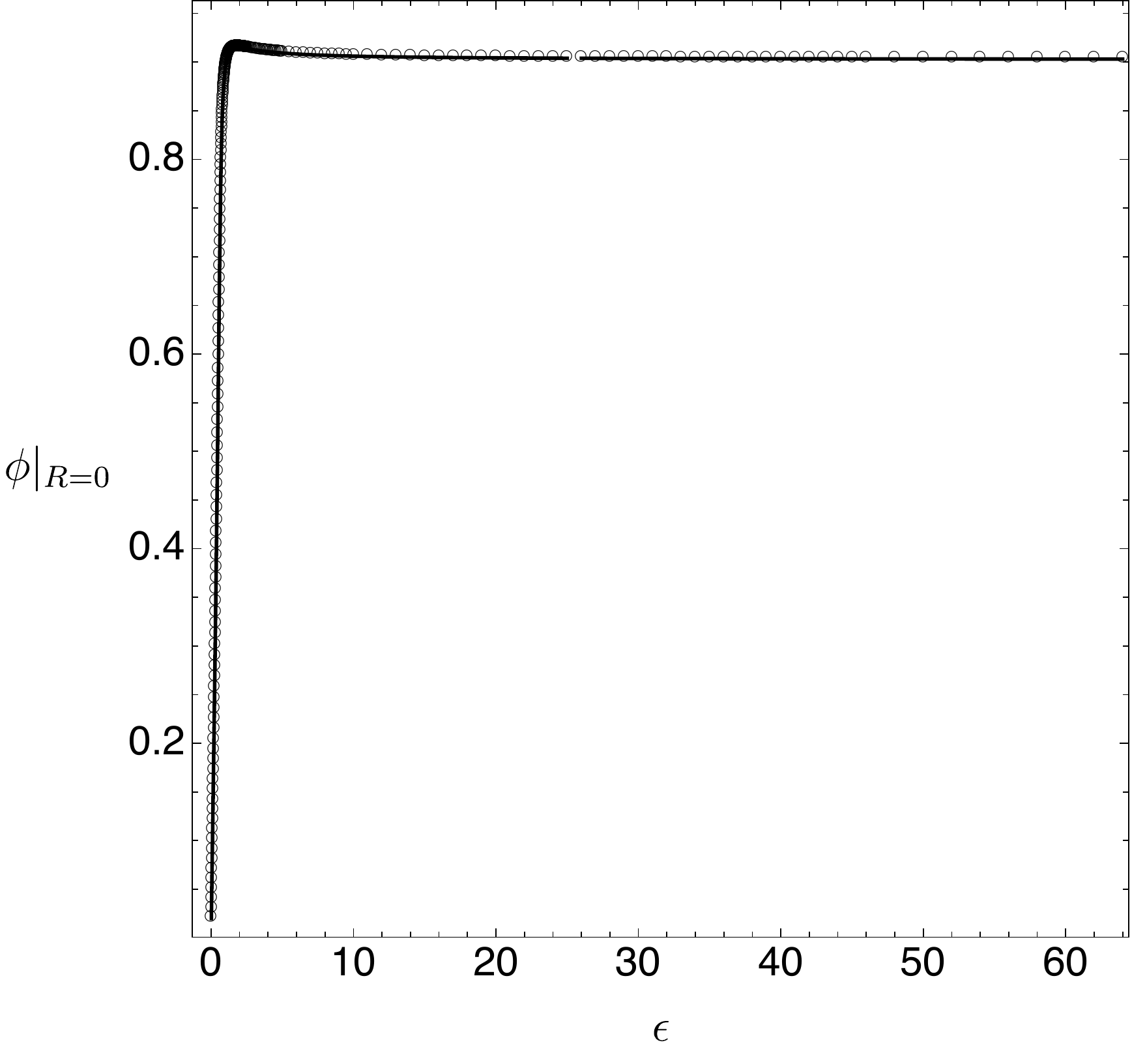}
\hspace{0.3cm}
\includegraphics[width=.515\textwidth]{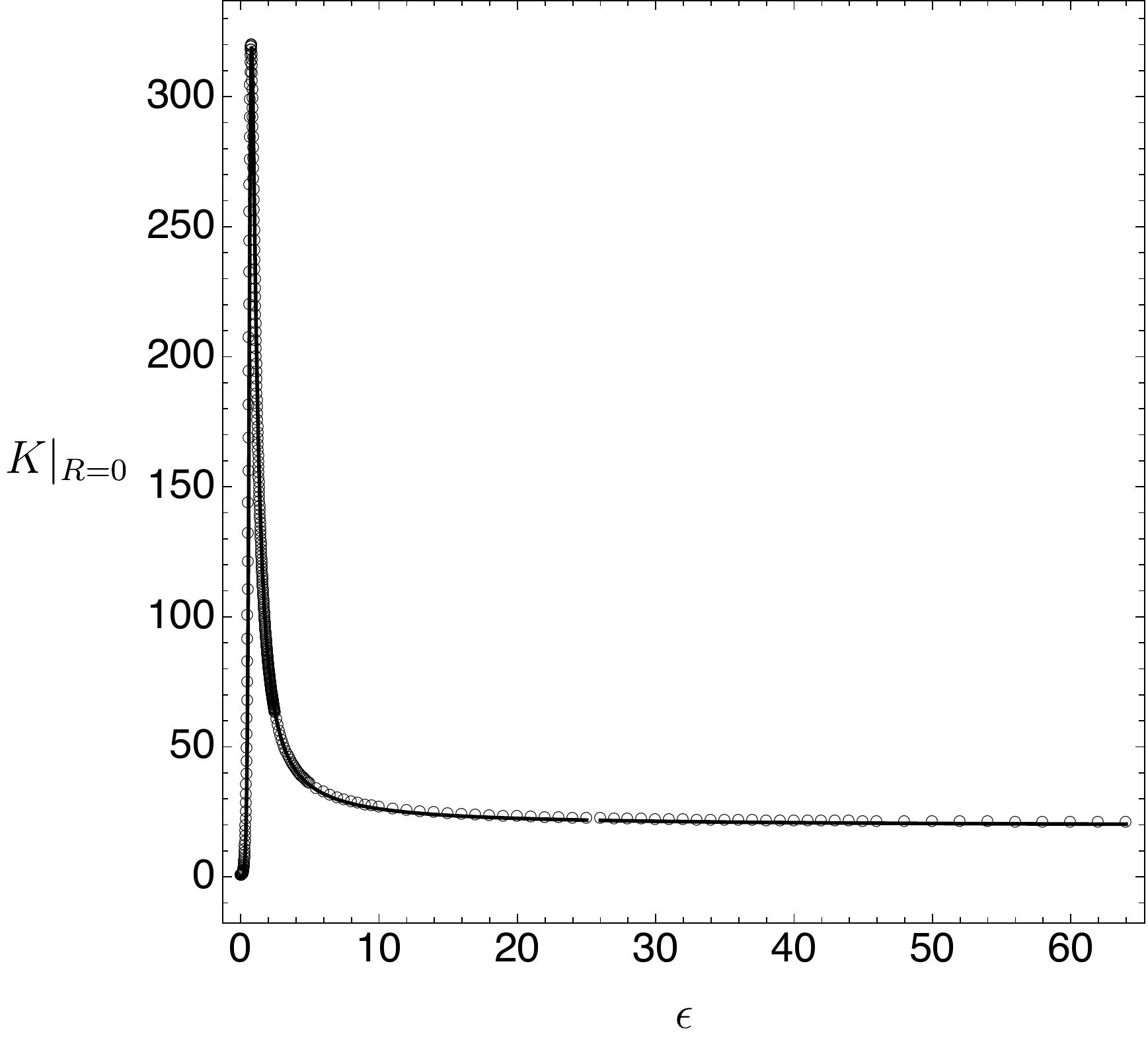}
}
\caption{Main soliton family with $e=2.3$ ($e>e_{\hbox{\tiny S}}$).}
\label{FIGe2.3:phiK}
\end{figure}

For completeness, we also display the evolution of the fields $f,A_t,\phi$ and of the Kretschmann scalar with the scalar condensate amplitude in Figs.~\ref{FIGe2.3:fA}-\ref{FIGe2.3:phiK}. We see that the gradients, observed in the similar plots of Figs.~\ref{FIGe1.855:fA}-\ref{FIGe1.855:phiK} for smaller $e$, become more pronounced.

\section{Israel surface stress tensor and energy conditions}\label{sec:Boxstructure}
 
So far, we were able to construct the phase diagram of hairy solitons inside a box without needing to detail the matter content of the box. Indeed, to get the quasilocal phase diagrams of the previous section, we just had to integrate numerically the equations of motion in the domain $R\in [0,1]$ subject to regular boundary conditions and vanishing scalar field at the box. For this, no further details are required. But the description of the solution is only complete once we discuss the full solution all the way up to the asymptotically Minkowski boundary.
 
In the literature one already has some studies of scalar fields confined in a Minkowski cavity: 1) at the linear level \cite{Herdeiro:2013pia,Hod:2013fvl,Degollado:2013bha,Hod:2014tqa,Li:2014gfg,Hod:2016kpm,Fierro:2017fky,Li:2014xxa,Li:2014fna,Li:2015mqa,Li:2015bfa}, 2) within a higher order perturbative analysis of the elliptic problem \cite{Dias:2018zjg,Dias:2018yey}, 3) as a nonlinear elliptic problem (although without having asymptotically flat boundary conditions \cite{Dolan:2015dha,Ponglertsakul:2016wae,Ponglertsakul:2016anb} or without discussing the exterior solution \cite{Basu:2016srp}), and 4) as an initial-value problem \cite{Sanchis-Gual:2015lje,Sanchis-Gual:2016tcm,Sanchis-Gual:2016ros}. 
However, except in the perturbative analysis of \cite{Dias:2018yey}, the properties of the ``internal structure" of the cavity or surface layer that is necessary to confine the scalar field and its contribution to the ADM mass and charge that ultimately describe, by Birchoff's theorem, the exterior RN solution are not discussed. 

With the interior hairy soliton fields found in the previous section, we can now compute the Lanczos-Darmois-Israel surface stress tensor  \eqref{eq:inducedT} which describes the energy-momentum content of the box $\Sigma$. If this satisfies relevant energy conditions, in practice, we can indeed build a physical box that confines the scalar fields and impedes that it disperses to infinity. 
In our construction we imposed the three Israel junction conditions  \eqref{eq:Israeljunction1}-\eqref{eq:Israeljunction3}  on the gravitoelectric fields on the surface layer $\Sigma$. With these conditions, the fields are continuous across $\Sigma$ and the component of the electric field orthogonal to $\Sigma$ is also continuous across $\Sigma$. This means that we are allowed to choose a surface layer that confines the charged scalar field without needing to have a surface electric charge density. Altogether, the three conditions  \eqref{eq:Israeljunction1}-\eqref{eq:Israeljunction3} match the interior and exteriors solutions, \ie they determine the parameters $M_0,c_A,\rho$ in  \eqref{BCinfinity} as a function of the reparametrization freedom parameter  $N$  introduced in \eqref{SigmaOut}:
\begin{equation}\label{Kab}
M_0=\frac{1}{N^2}\left(1-\frac{A_t'(1)^2}{2}\right)-1,\qquad c_A=\frac{A_t'(1)+A_1(1)}{N}\,, \qquad \rho =-\frac{A_t'(1)}{N}\,.
\end{equation} 
 That is to say, these junction conditions fix the exterior RN solution as a function of the interior field content but also as a function of the box energy-momentum content. And we have a 1-parameter freedom ($N$) to choose the box content that is able to confine the scalar condensate.

How can we fix $N$? Well, ideally we would choose it so that the gravitational field was also differentiable across the box, \ie so that the fourth junction condition \eqref{eq:Israeljunction4} were also obeyed and thus the extrinsic curvature
\begin{equation}\label{Kab}
K^{t}_{\phantom{t}t}=-\frac{f'(R)}{2f(R)\sqrt{g(R)}}\,,
 \qquad K^{i}_{\phantom{i}j}=\frac{1}{R\sqrt{g}}\,\delta^{i}_{\phantom{i}j} \,,\quad (i,j)=(\theta,\varphi)\,,\\
\end{equation} 
 were also continuous across the box. But for our system there is no choice of $N$ that simultaneously makes $[K_t^t]=0$ and $[K_i^i]=0$, unless the scalar field vanishes everywhere. But we can fix $N$ requiring that $[K_t^t]=0$ (at the expense of having $[K_i^i]\neq 0$) or vice-versa, or any other combination.

Our choice of $N$ fixes the energy density and pressure of the box since the surface tensor  of the box can be written in the perfect fluid form, ${\cal S}_{(a)(b)}= \mathcal{E} u_{(a)}u_{(b)}+ \mathcal{P}(h_{(a)(b)}+u_{(a)}u_{(b)})$, with $u=f^{-1/2}\partial_t$ and local energy density $ \mathcal{E}$ and pressure $ \mathcal{P}$ given by
\begin{equation}\label{densitiesbox}
\mathcal{E}=-S^t_{\ t}\,,\qquad \mathcal{P}=S^x_{\ x}=S^{\phi}_{\ \phi}\,.
 \end{equation}
Our choice should be such that  relevant energy conditions are obeyed in order to be physically acceptable. Different versions of these energy conditions read $(i=\theta,\varphi)$ \cite{Wald:106274}:
\begin{eqnarray}\label{energyconditions}
\hbox{Weak energy condition:} && {\cal E}\geq 0 \quad \land \quad {\cal E}+{\cal P}_{i}\geq 0  \,;\\
\hbox{Strong energy condition:} & &{\cal E}+{\cal P}_{i}\geq 0 \quad \land \quad {\cal E}+\sum_{i=1}^2 {\cal P}_{i}\geq 0\,;\\
\hbox{Null energy condition:} & &{\cal E}+{\cal P}_{i}\geq 0\,;\\
\hbox{Dominant energy condition:}& & {\cal E}+|{\cal P}_{i}|\geq 0\,.
\end{eqnarray}

\begin{figure}[th]
\centerline{
\includegraphics[width=.505\textwidth]{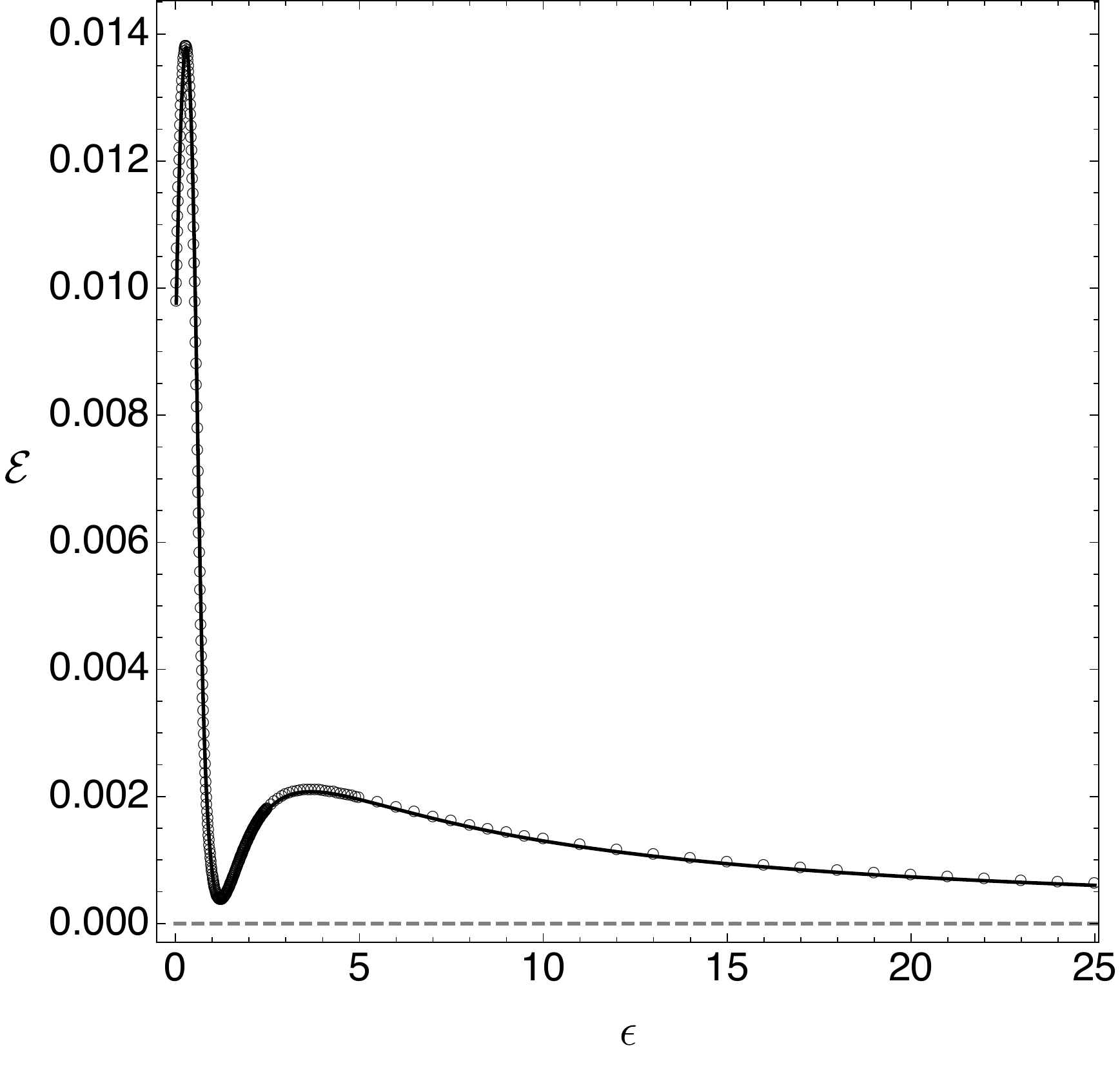}
\hspace{0.3cm}
\includegraphics[width=.48\textwidth]{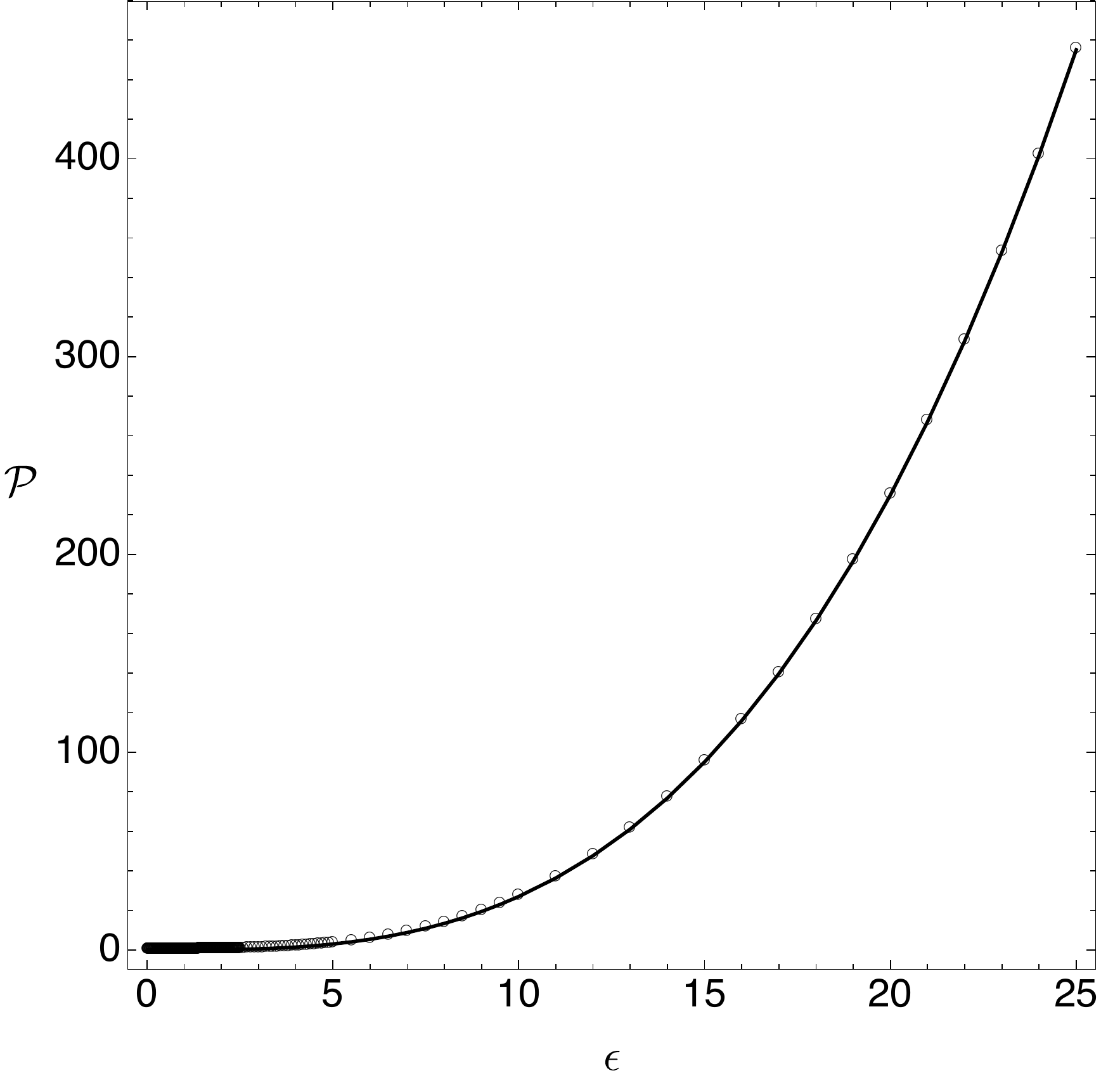}
}
\centerline{
\includegraphics[width=.505\textwidth]{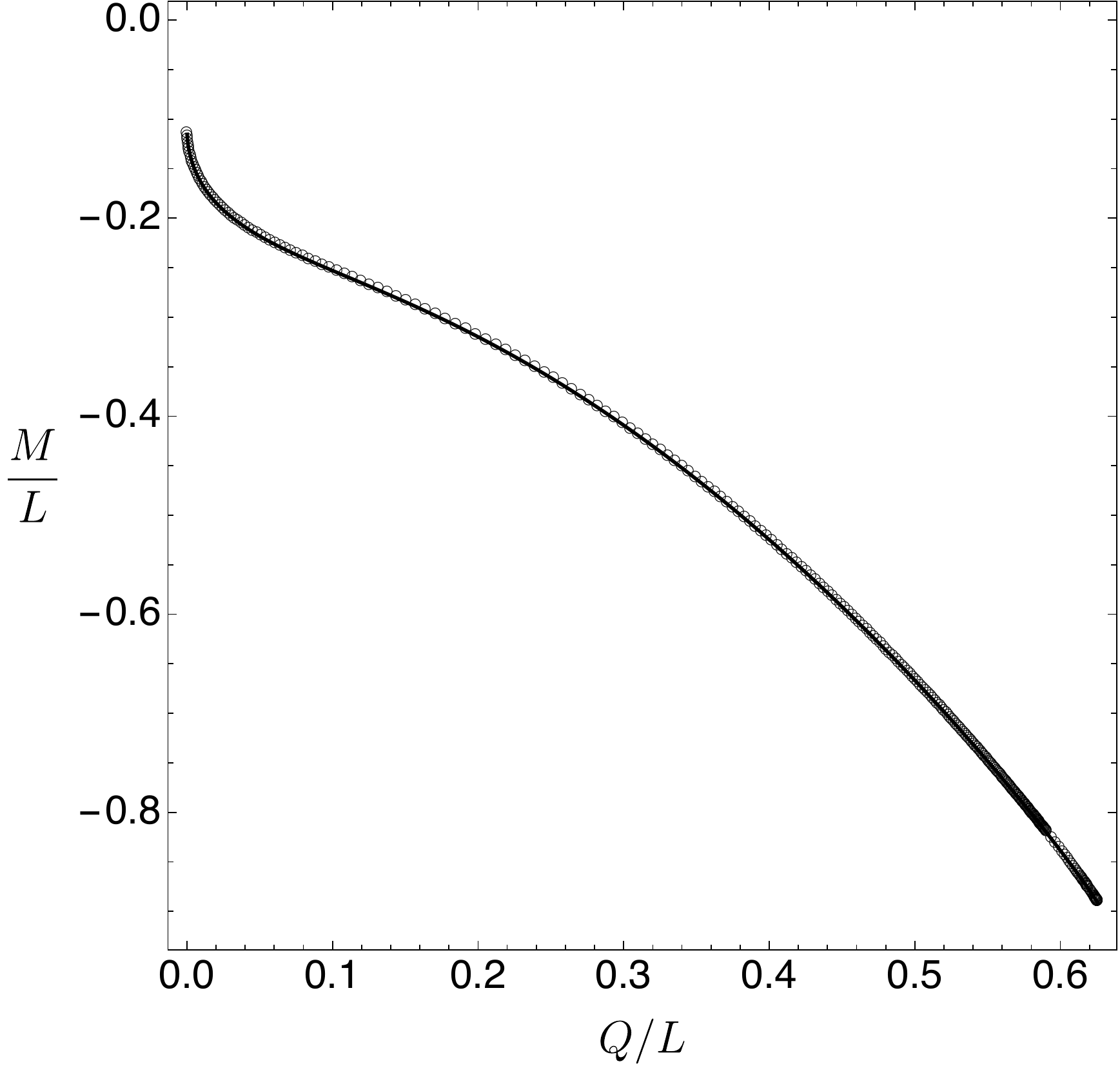} }
\caption{Properties of a box  for a choice of reparametrization normalization $N$ such that the Israel stress tensor satisfies the energy conditions (system with $e=2.3$). {\bf Top panel:} Energy density $\mathcal{E}$ and pressure $\mathcal{P}$. {\bf Bottom panel:} ADM mass $M$ {\it vs} ADM charge $Q$ of the soliton solution (measured at the asymptotic region) includes the contribution from the scalar condensate confined inside the box, the energy-momentum content of the box layer and the exterior RN solution. 
}
\label{FIGe2.3:Israel}
\end{figure} 

To illustrate that it is definitely possible to build boxes that obey these energy conditions, let us take the main soliton family with $e=2.3$ constructed in section~\ref{subsec:higherS}. Let us further choose $N$ such that one has no pressure, $N=N\big|_{\mathcal{P}=0}$. Then the evolution of box's energy density $\mathcal{E}$ as we move along the soliton family is always negative.\footnote{This pressure vanishes in the limit $\epsilon\to 0$ because in this case the interior solution is simply Minkowski space with no scalar hair and thus the box is not necessary.} 
So the energy conditions are not obeyed for this choice of normalization $N$. Let us now insist that we want that all energy conditions \eqref{energyconditions} are obeyed. This is achieved, for example, if we choose $N$ to be $N=N\big|_{\mathcal{P}=0}+\frac{13}{100}$. Then the energy density and pressure of the box are always positive along the whole soliton family, as displayed in the top panel of Fig.~\ref{FIGe2.3:Israel}. Given this choice, we can also compute the ADM energy and charge of our solutions as read at the asymptotic region. These quantities are displayed in the bottom panel of Fig.~\ref{FIGe2.3:Israel}. This case illustrates a feature that we found to be common in similar exercises we did (for other choices of $N$ and $e$): the ADM mass of hairy solitons can be negative, although our solutions are regular everywhere. For reference, recall that  Schwarzschild and RN solutions with negative ADM mass are singular solutions. This might well be a general feature of hairy solutions confined by generic gravitational potentials (e.g. hairy solutions, not necessarily with spin 0, that are confined by more realistic potentials like the one of a massive scalar field or an accretion disk wall).

\vskip .5cm
\centerline{\bf Acknowledgements}
\vskip .2cm
OD acknowledges financial support from the STFC Ernest Rutherford grant ST/K005391/1 and from the STFC ``Particle Physics Grants Panel (PPGP) 2016" Grant No.~ST/P000711/1 and the STFC ``Particle Physics Grants Panel (PPGP) 2018" Grant No.~ST/T000775/1. The authors acknowledge the use of the IRIDIS High Performance Computing Facility, and associated support services at the University of Southampton, in the completion of this work.

\begin{appendix}
\section{Further properties of solitons}

\begin{figure}[t]
\centerline{
\includegraphics[width=.33\textwidth]{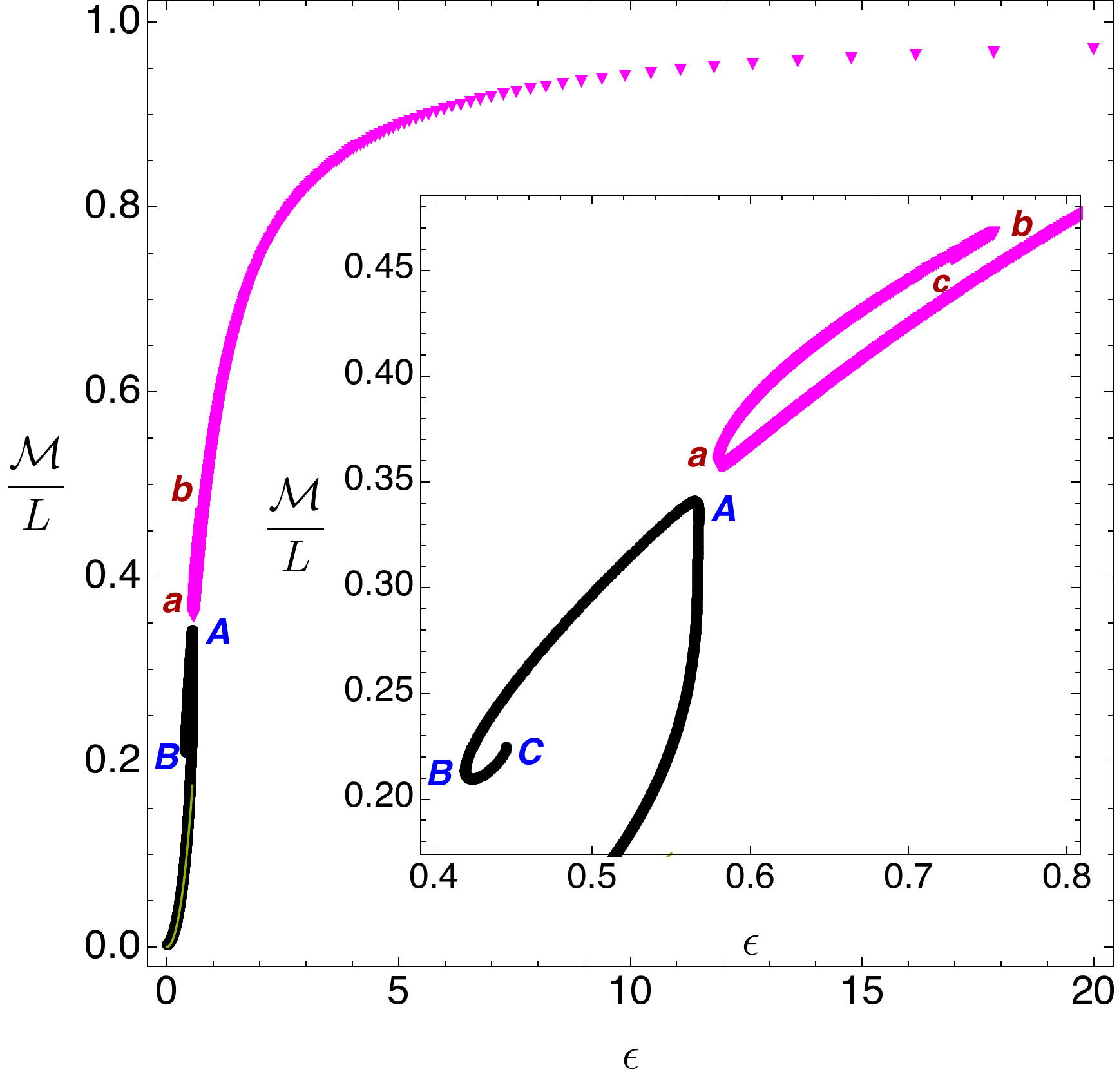}
\hspace{0.2cm}
\includegraphics[width=.33\textwidth]{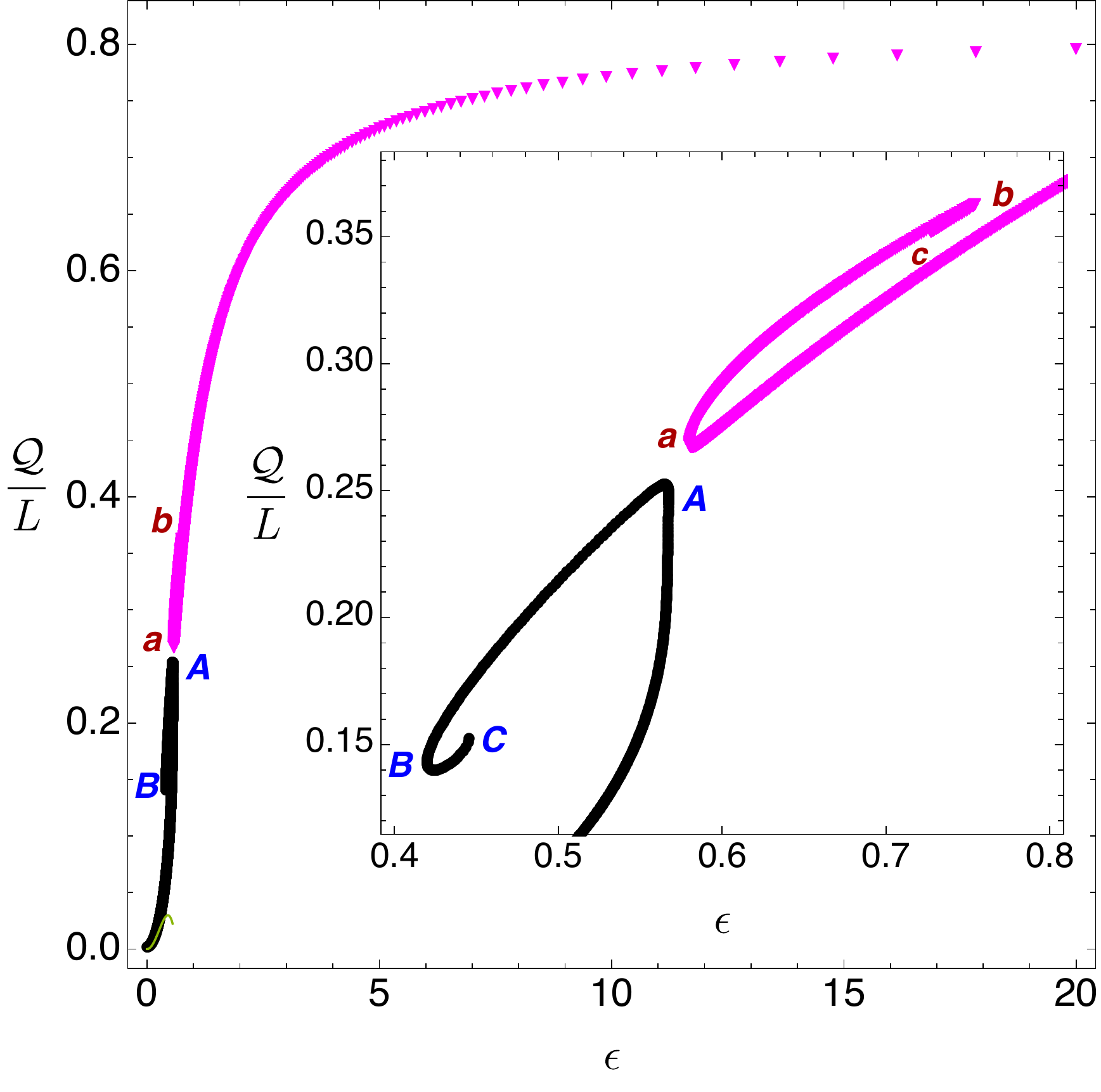}
\hspace{0.2cm}
\includegraphics[width=.33\textwidth]{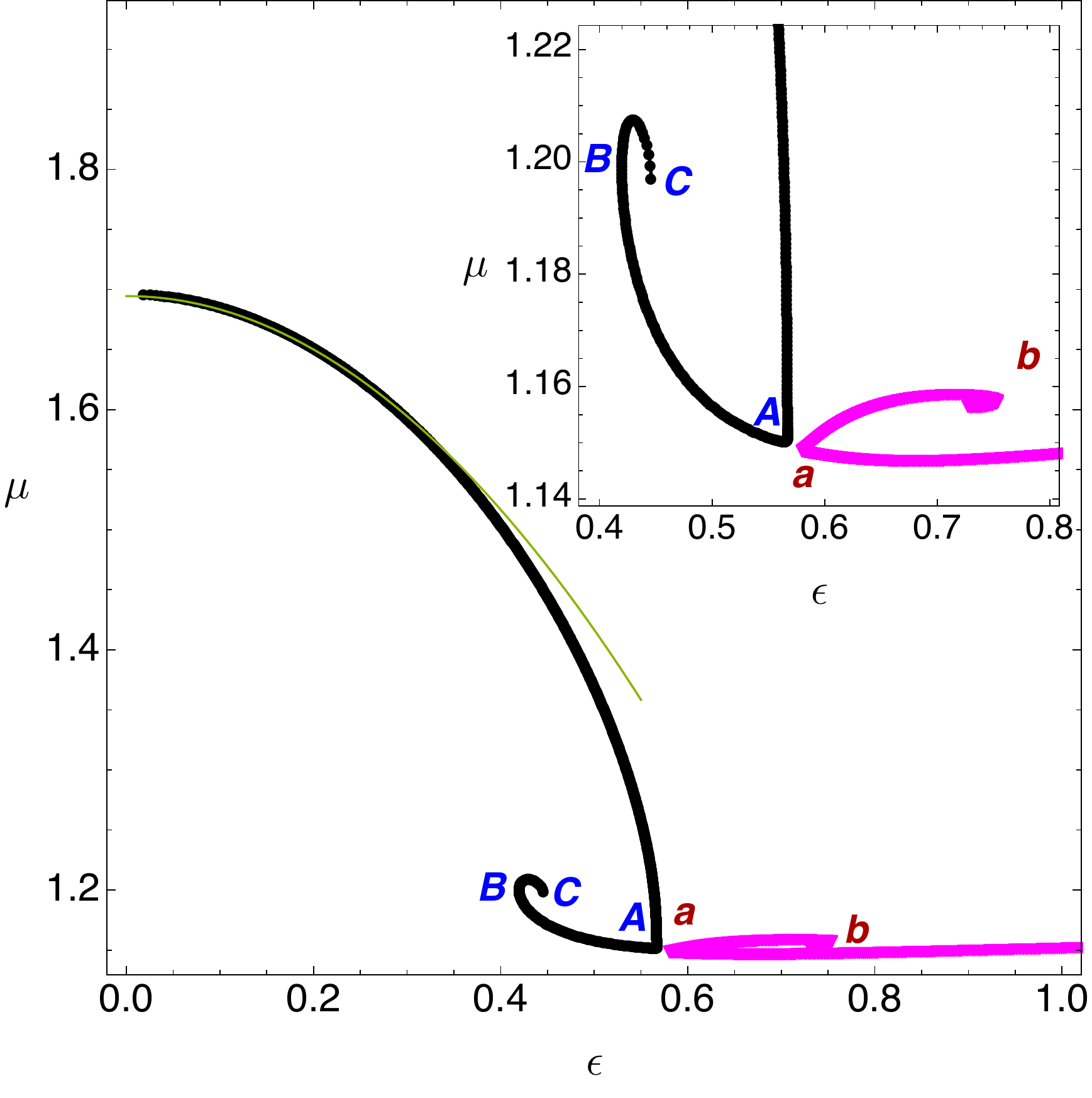}
}
\centerline{
\includegraphics[width=.335\textwidth]{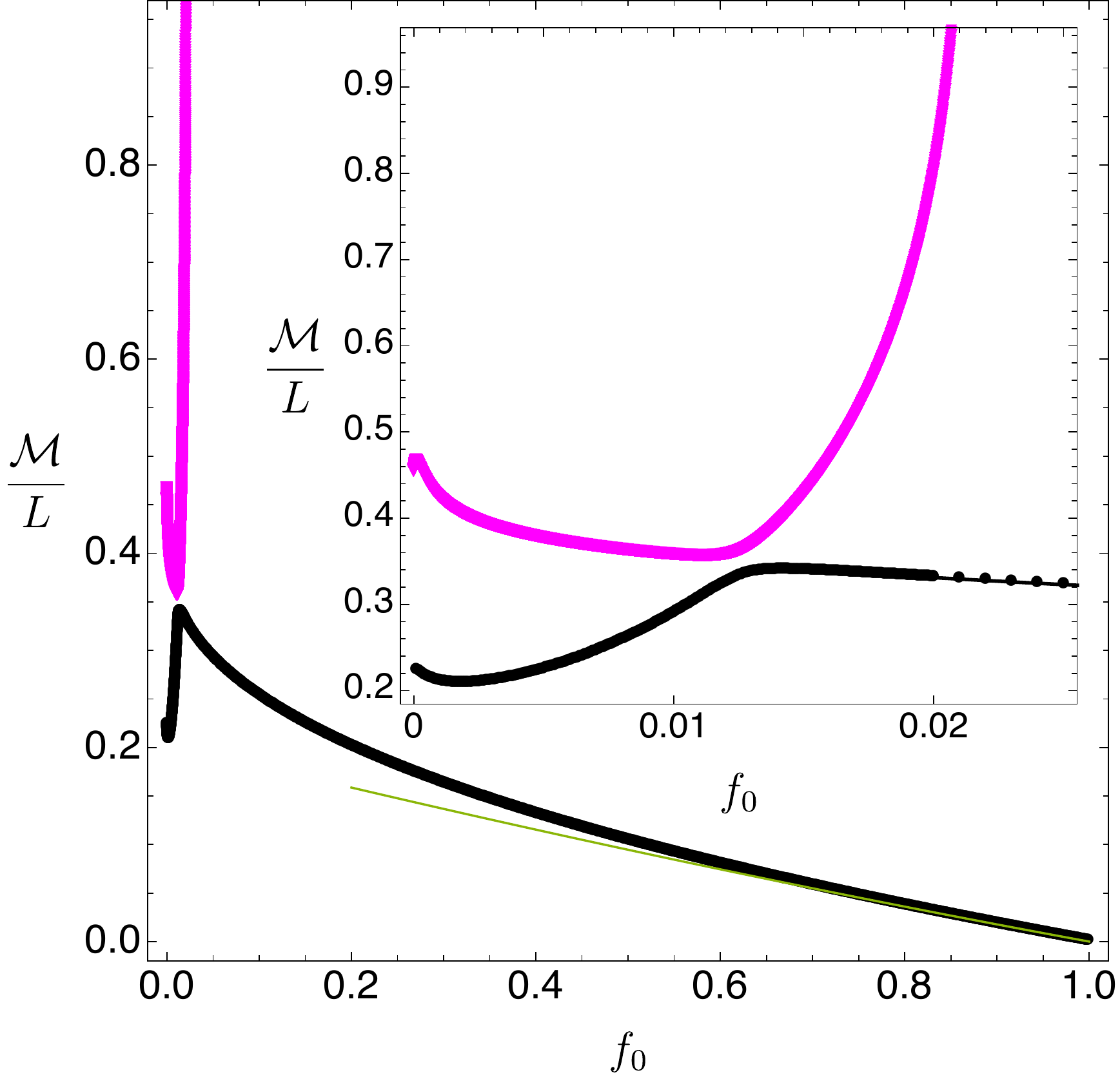}
\hspace{0.2cm}
\includegraphics[width=.33\textwidth]{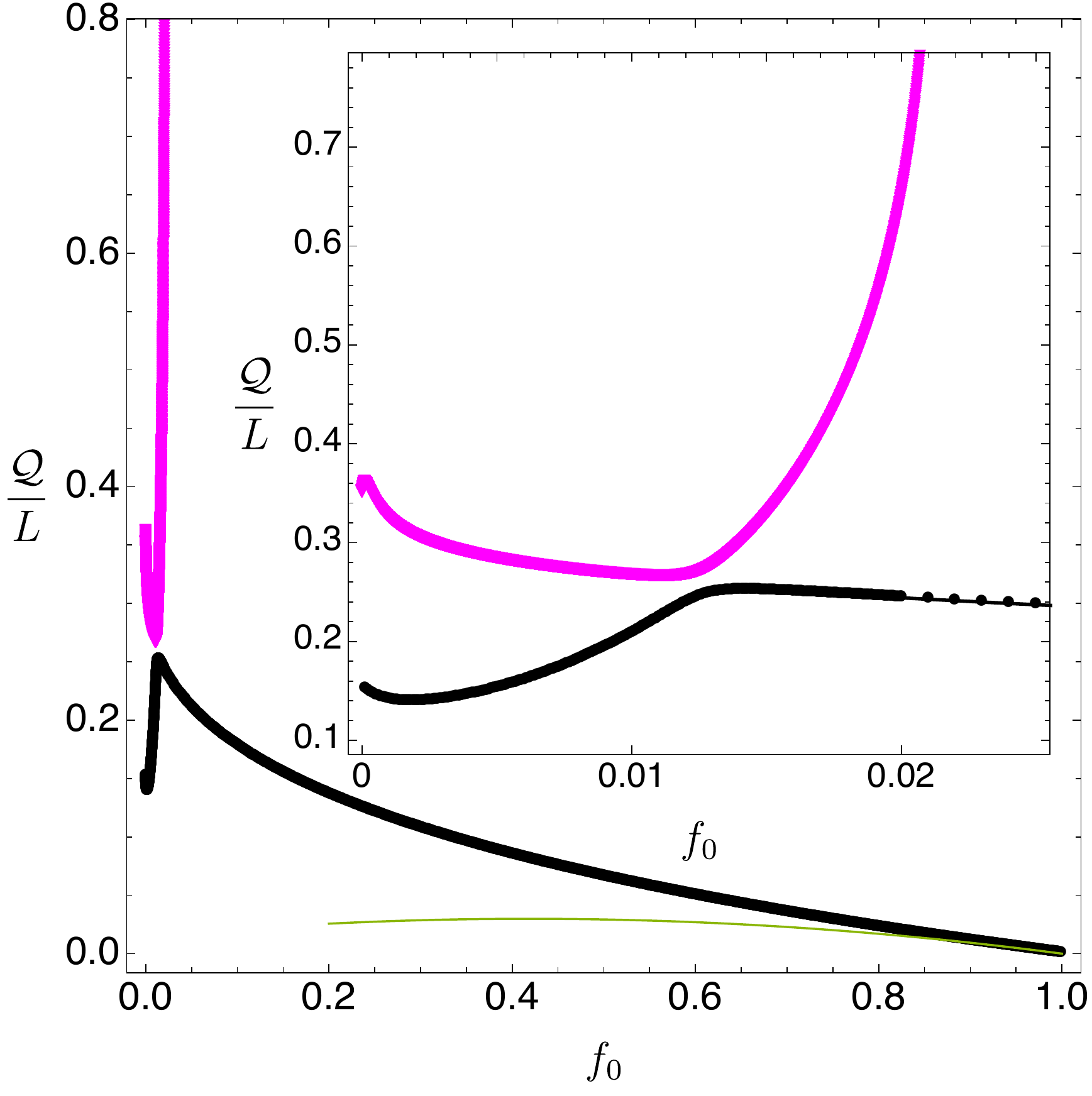}
\hspace{0.2cm}
\includegraphics[width=.325\textwidth]{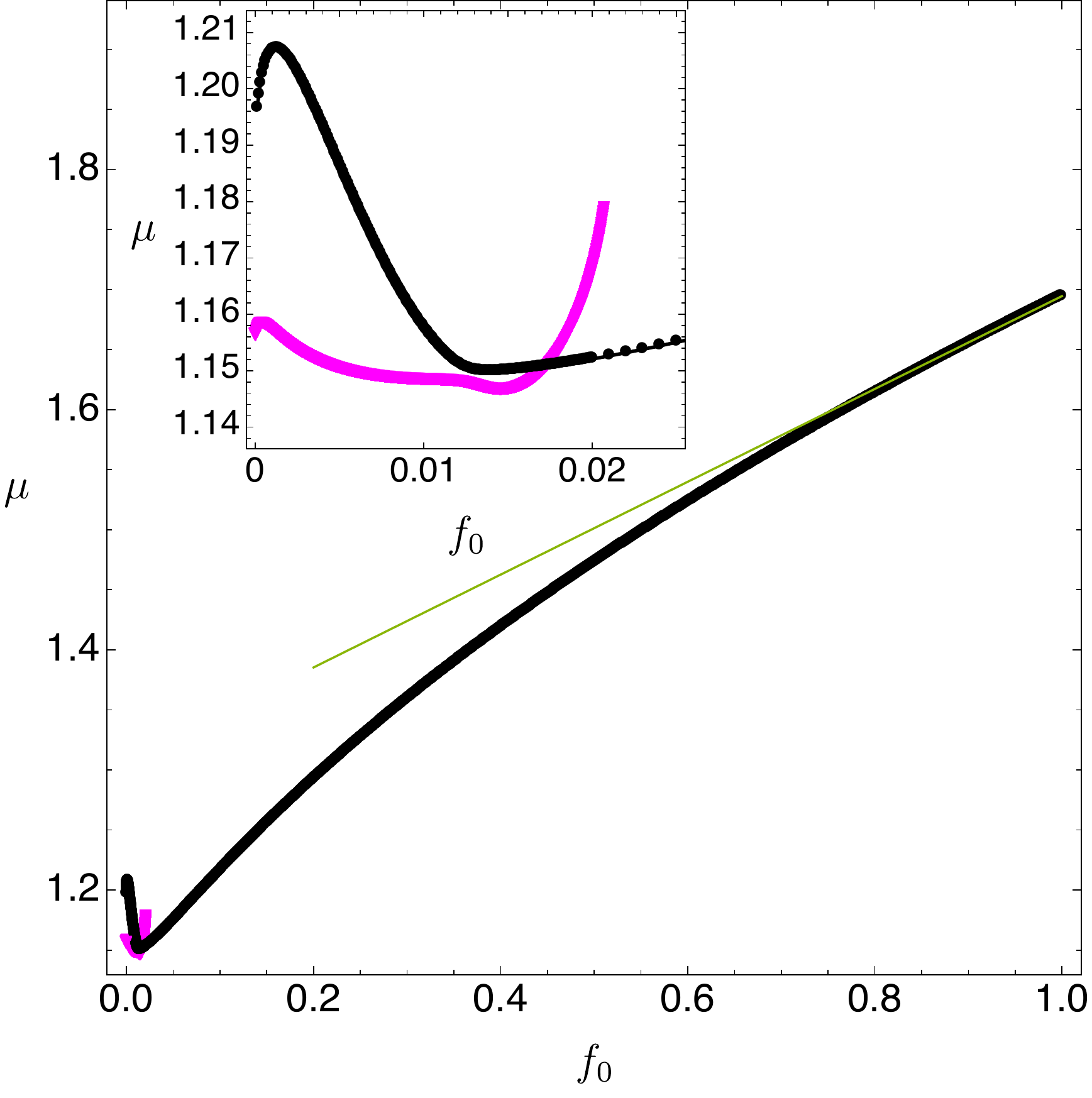}
}
\caption{Soliton families with $e=1.854$ ($e_{\gamma}<e<e_c$). 
{\bf Top panel:} The quasilocal thermodynamic quantities, namely mass, charge and chemical potential of the main (black disks) and secondary (magenta triangles) soliton families are shown as functions of the scalar field amplitude $\epsilon$. 
{\bf Bottom panel:} This time the mass, charge and chemical potential are plotted as functions of $ f_0\equiv f(0)$. }
\label{FigAPP:e1.854}
\end{figure}

\begin{figure}[t]
\centerline{
\includegraphics[width=.33\textwidth]{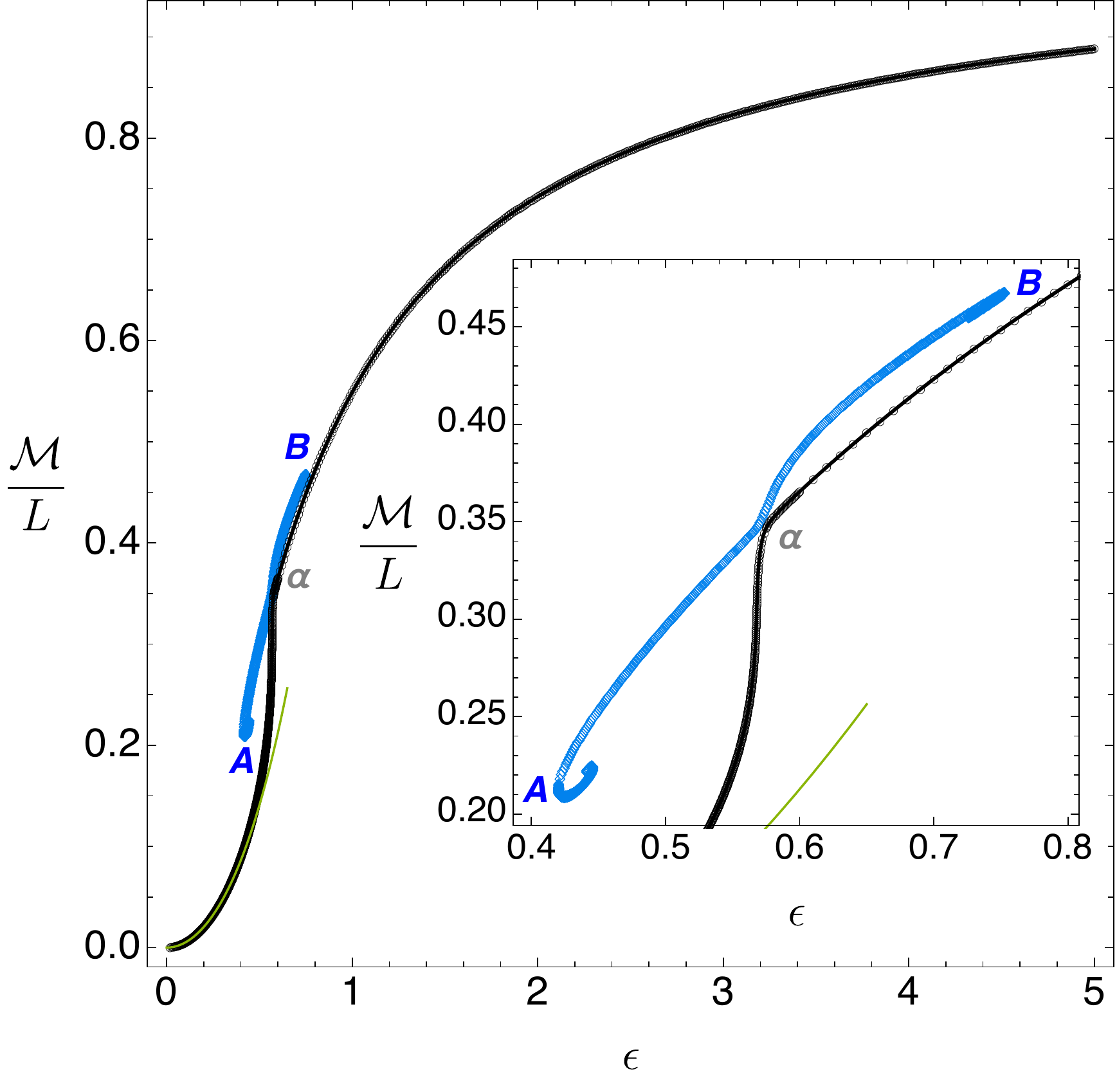}
\hspace{0.2cm}
\includegraphics[width=.33\textwidth]{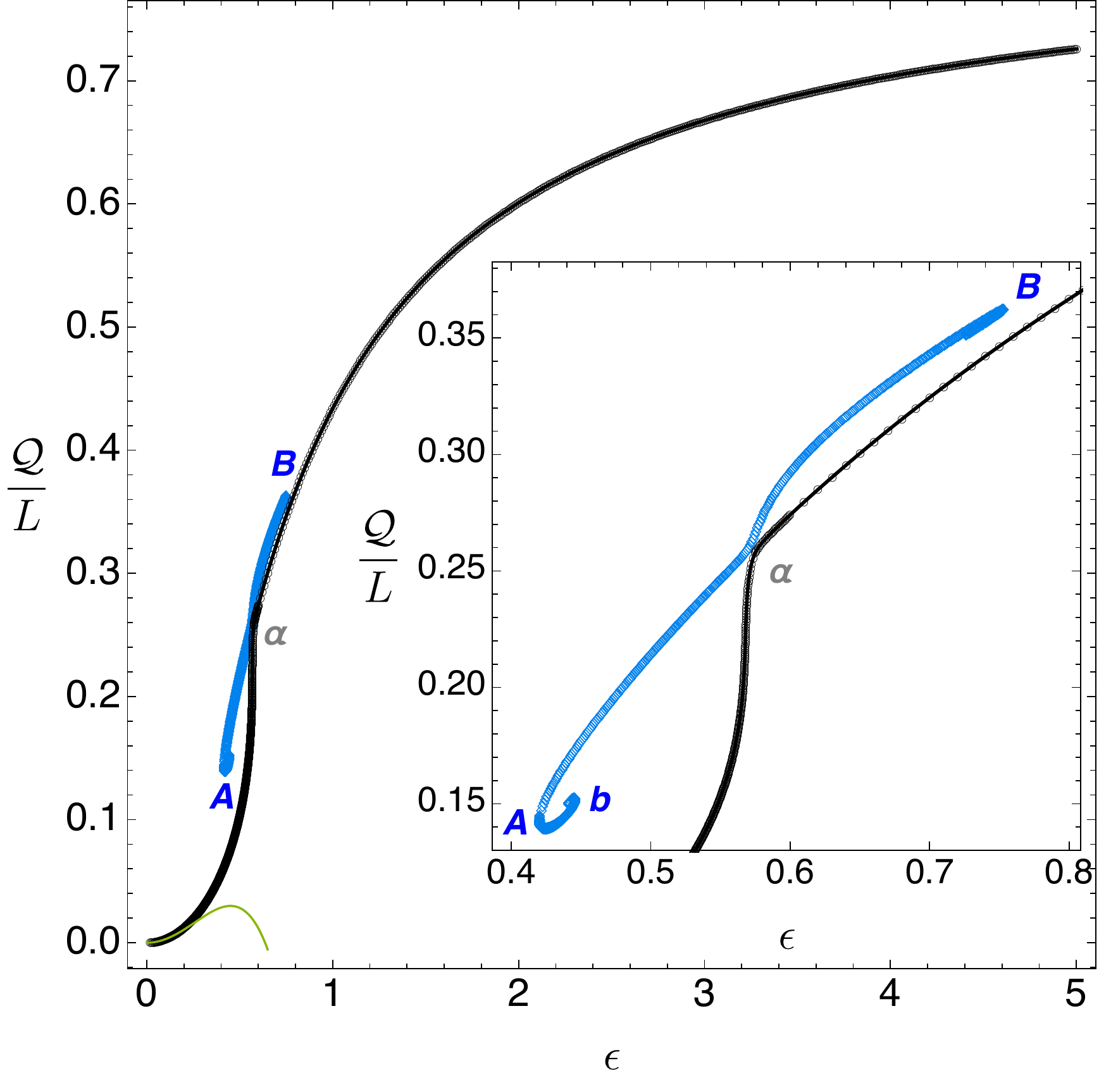}
\hspace{0.2cm}
\includegraphics[width=.33\textwidth]{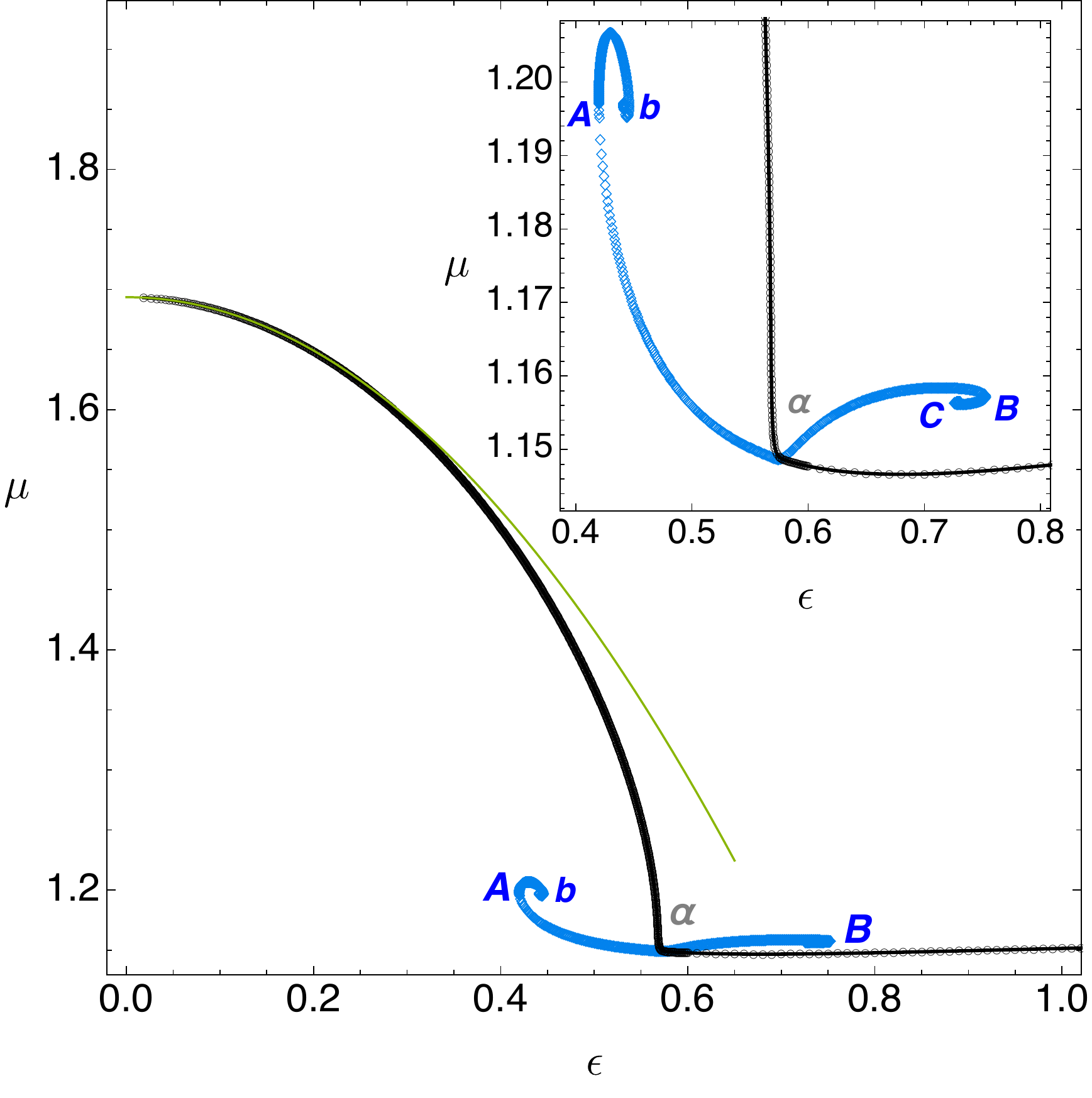}
}
\centerline{
\includegraphics[width=.34\textwidth]{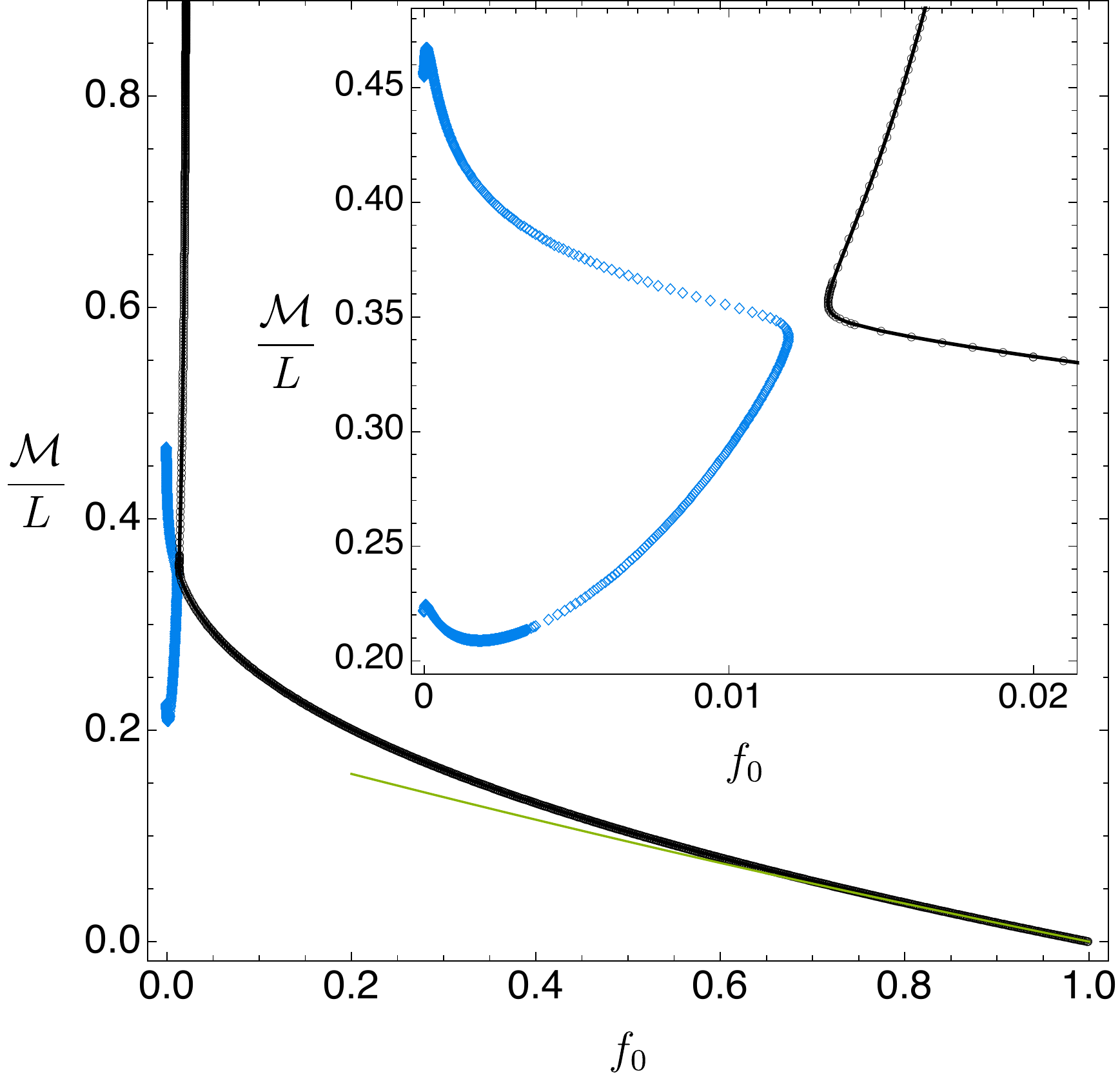}
\hspace{0.2cm}
\includegraphics[width=.33\textwidth]{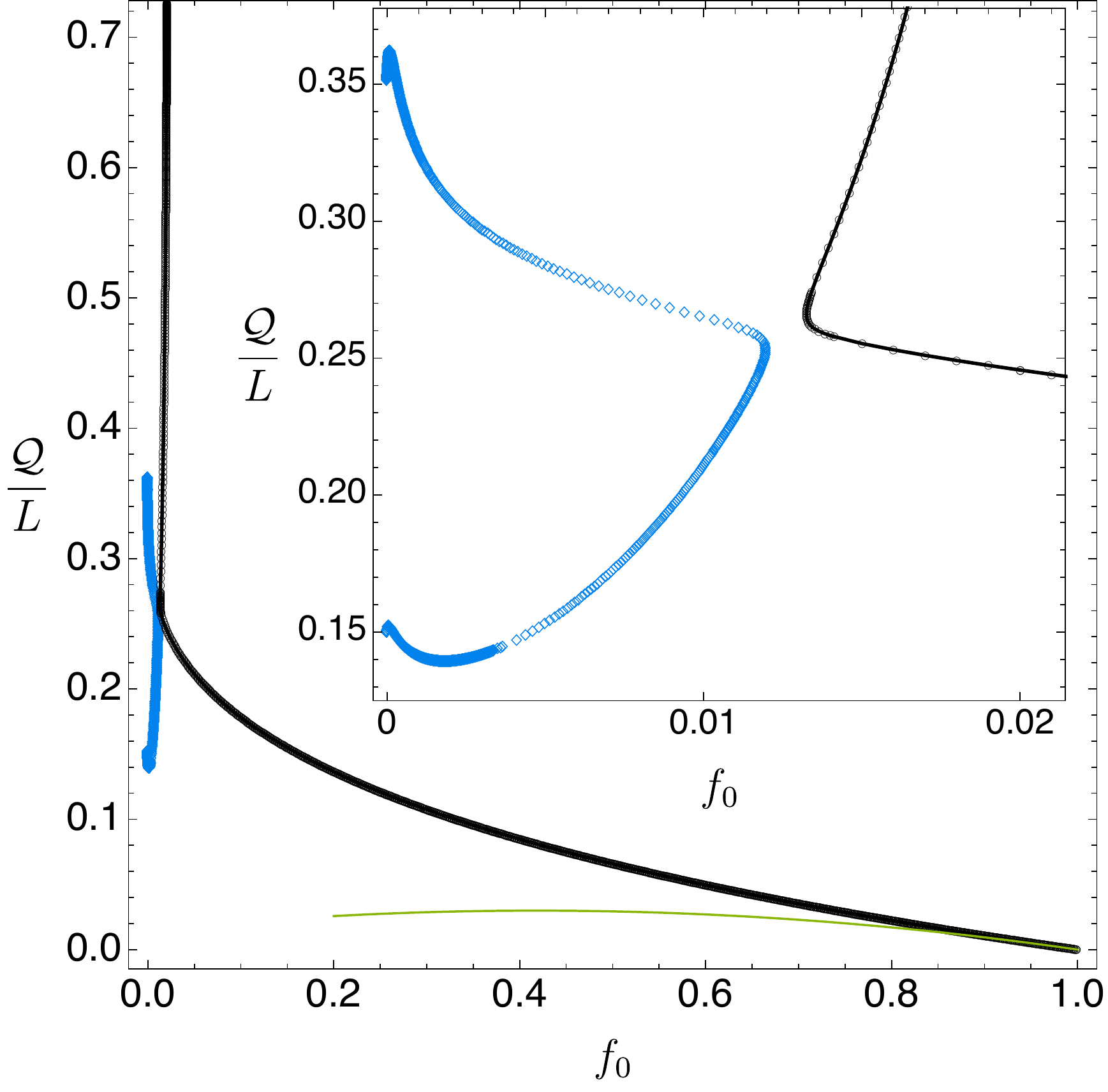}
\hspace{0.2cm}
\includegraphics[width=.32\textwidth]{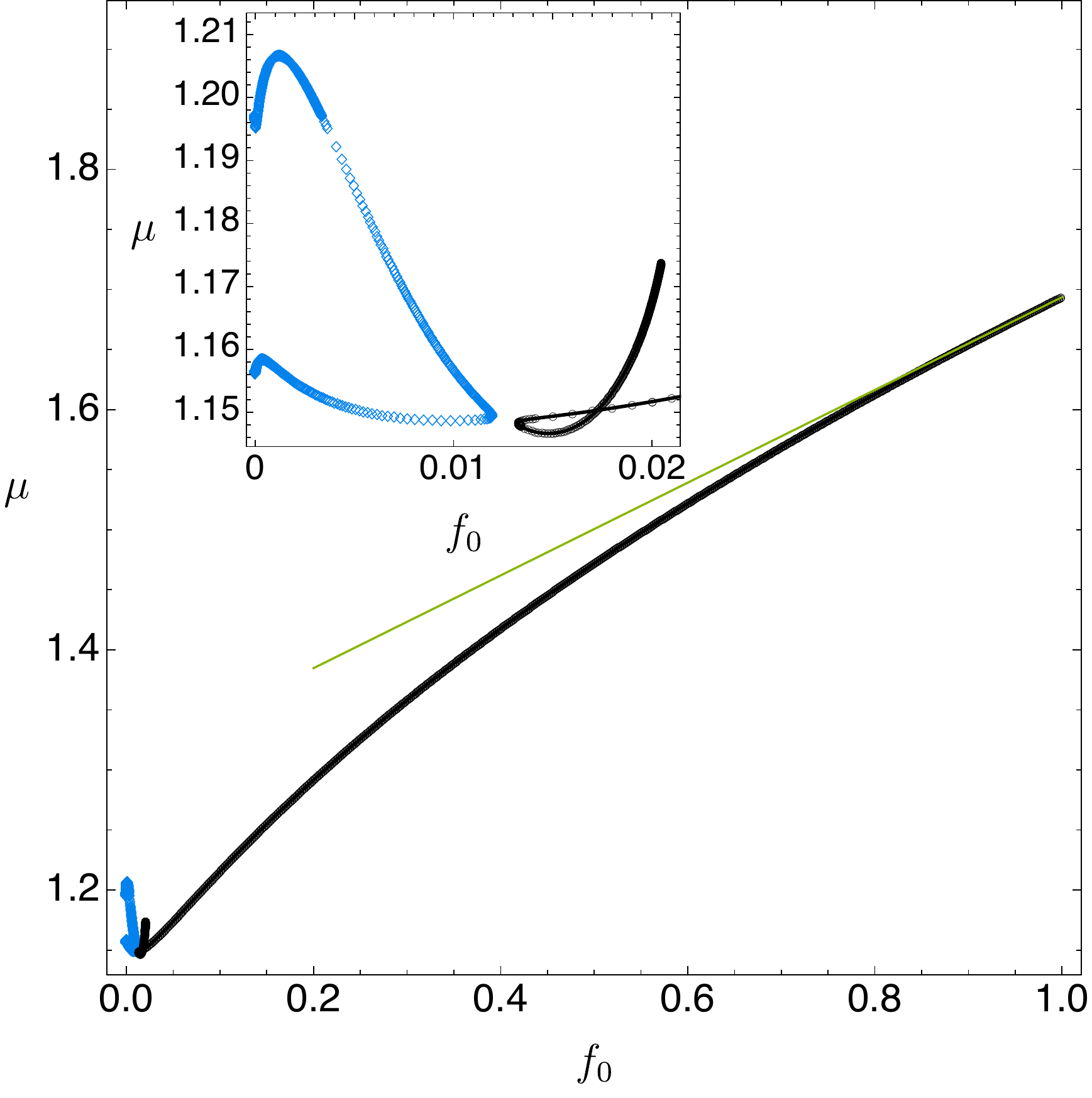}
}
\caption{Soliton families with $e=1.855$ ($e_c<e<e_{\hbox{\tiny S}}$). 
{\bf Top panel:} The quasilocal thermodynamic quantities, namely mass, charge and chemical potential of the main (black disks) and secondary (blue diamons) soliton families are shown as functions of the scalar field amplitude $\epsilon$.
{\bf Bottom panel:} Quasilocal mass, charge and chemical potential plotted as functions of $f_0$.}
\label{FigAPP:e1.855}
\end{figure}

For completeness, in this appendix we plot the soliton quasilocal mass $\mathcal{M}$, charge $\mathcal{Q}$ and chemical potential $\mu$ as a function of the marching parameters $\epsilon=-\phi'(1)$ or $f_0=f(0)$. We do this  for the two most representative cases studied in the main text. Namely, for the solitons with $e=1.854 \lesssim e_c$ (in Fig.~\ref{FigAPP:e1.854}) and $e=1.855\gtrsim e_c$ (in Fig.~\ref{FigAPP:e1.855}).

These quantities display the spiral behaviour already described in the main text. In particular, the turning points of these spirals translate into the regular cusps in the quasilocal charge plots: compare \eg  Fig.~\ref{FigAPP:e1.854} with  Fig.~\ref{FIGe1.854:MassCharge} or   Fig.~\ref{FigAPP:e1.855} with  Fig.~\ref{FIGe1.855:MassCharge}. Complementing the discussion given in the main text, these plots also  illustrate how points $A$  of the main (black) curve and  point $a$ of the secondary (magenta) curve) approach each other as $e$ tends to $e_c$ from below (Fig.~\ref{FigAPP:e1.854}) and then give origin to a new branch of main (black) and secondary (blue) soliton families (Fig.~\ref{FigAPP:e1.855})

\end{appendix}

\bibliography{refs_Box}{}
\bibliographystyle{JHEP}

\end{document}